\batchmode

\documentclass[11pt,times]{article}

\usepackage{times}
\usepackage{graphicx}
\usepackage{amsfonts}
\usepackage{amsmath}
\usepackage{amssymb}
\usepackage{amstext}
\usepackage{amsthm}

\usepackage{epsfig}
\usepackage{xcolor}
\usepackage{latexsym}
\usepackage{rotating}

\usepackage {subfigure}

\begin{document}

\baselineskip 6mm


\newcommand{\nc}{\newcommand}
\newcommand{\rnc}{\renewcommand}


\rnc{\baselinestretch}{1.24}    
\setlength{\jot}{6pt}       
\rnc{\arraystretch}{1.24}   

\makeatletter
\rnc{\theequation}{\thesection.\arabic{equation}}
\@addtoreset{equation}{section}
\makeatother



\nc{\be}{\begin{equation}}

\nc{\ee}{\end{equation}}

\nc{\bea}{\begin{eqnarray}}

\nc{\eea}{\end{eqnarray}}

\nc{\xx}{\nonumber\\}

\nc{\ct}{\cite}

\nc{\la}{\label}

\nc{\eq}[1]{(\ref{#1})}

\nc{\newcaption}[1]{\centerline{\parbox{6in}{\caption{#1}}}}

\nc{\fig}[3]{

\begin{figure}
\centerline{\epsfxsize=#1\epsfbox{#2.eps}}
\newcaption{#3. \label{#2}}
\end{figure}
}


\def\CA{{\cal A}}
\def\CC{{\cal C}}
\def\CD{{\cal D}}
\def\CE{{\cal E}}
\def\CF{{\cal F}}
\def\CG{{\cal G}}
\def\CH{{\cal H}}
\def\CK{{\cal K}}
\def\CL{{\cal L}}
\def\CM{{\cal M}}
\def\CN{{\cal N}}
\def\CO{{\cal O}}
\def\CP{{\cal P}}
\def\CR{{\cal R}}
\def\CS{{\cal S}}
\def\CU{{\cal U}}
\def\CV{{\cal V}}
\def\CW{{\cal W}}
\def\CY{{\cal Y}}
\def\CZ{{\cal Z}}


\def\IB{{\hbox{{\rm I}\kern-.2em\hbox{\rm B}}}}
\def\IC{\,\,{\hbox{{\rm I}\kern-.50em\hbox{\bf C}}}}
\def\ID{{\hbox{{\rm I}\kern-.2em\hbox{\rm D}}}}
\def\IF{{\hbox{{\rm I}\kern-.2em\hbox{\rm F}}}}
\def\IH{{\hbox{{\rm I}\kern-.2em\hbox{\rm H}}}}
\def\IN{{\hbox{{\rm I}\kern-.2em\hbox{\rm N}}}}
\def\IP{{\hbox{{\rm I}\kern-.2em\hbox{\rm P}}}}
\def\IR{{\hbox{{\rm I}\kern-.2em\hbox{\rm R}}}}
\def\IZ{{\hbox{{\rm Z}\kern-.4em\hbox{\rm Z}}}}


\def\a{\alpha}
\def\b{\beta}
\def\d{\delta}
\def\ep{\epsilon}
\def\ga{\gamma}
\def\k{\kappa}
\def\l{\lambda}
\def\s{\sigma}
\def\t{\theta}
\def\w{\omega}
\def\G{\Gamma}


\def\half{\frac{1}{2}}
\def\dint#1#2{\int\limits_{#1}^{#2}}
\def\goto{\rightarrow}
\def\para{\parallel}
\def\brac#1{\langle #1 \rangle}
\def\curl{\nabla\times}
\def\div{\nabla\cdot}
\def\p{\partial}


\def\Tr{{\rm Tr}\,}
\def\det{{\rm det}}


\def\vare{\varepsilon}
\def\zbar{\bar{z}}
\def\wbar{\bar{w}}
\def\what#1{\widehat{#1}}


\def\ad{\dot{a}}
\def\bd{\dot{b}}
\def\cd{\dot{c}}
\def\dd{\dot{d}}
\def\so{SO(4)}
\def\bfr{{\bf R}}
\def\bfc{{\bf C}}
\def\bfz{{\bf Z}}

\begin{titlepage}


\hfill\parbox{3.7cm} {{\tt arXiv:2206.08108}}

\vspace{5mm}

\begin{center}

{\Large \bf  Algebraic Properties of Riemannian Manifolds}

\vspace{10mm}

Youngjoo Chung ${}^{a}$\footnote{ychung@gist.ac.kr}, Chi-Ok Hwang ${}^{b}$\footnote{chwang@gist.ac.kr}
and Hyun Seok Yang ${}^c$\footnote{Corresponding Author:hsyang@gist.ac.kr}
\\[10mm]

${}^a$ {\sl School of Electrical Engineering and Computer Science, \\ Gwangju Institute of Science and Technology, Gwangju 61005, Korea}

${}^b$ {\sl Division of Liberal Arts and Sciences, \\ Gwangju Institute of Science and Technology, Gwangju 61005, Korea}

${}^c$ {\sl Department of Physics and Photon Science, \\ Gwangju Institute of Science and Technology, Gwangju 61005, Korea}

\end{center}

\thispagestyle{empty}

\vskip1cm


\centerline{\bf ABSTRACT}
\vskip 4mm
\noindent

Algebraic properties are explored for the curvature tensors of Riemannian manifolds,
using the irreducible decomposition of curvature tensors.
Our method provides a powerful tool to analyze the irreducible basis as well as an algorithm
to determine the linear dependence of arbitrary Riemann polynomials.
We completely specify 13 independent basis elements for the quartic scalars and explicitly
find 13 linear relations among 26 scalar invariants.
Our method provides several completely new results, including some clues to identify 23 independent basis elements 
from 90 quintic scalars, that are difficult to find otherwise.
\\



\vspace{1cm}

\today

\end{titlepage}

\renewcommand{\thefootnote}{\arabic{footnote}}
\setcounter{footnote}{0}

\section{Introduction}

The scalar invariants of the Riemann tensor $R_{abcd}$ are important in general relativity since they
allow a manifestly coordinate invariant characterization of certain geometrical properties
of space-times. For example, they are important in studying curvature singularities
\cite{c-sing1,c-sing2,c-sing3}, the classification of curvature tensors such as the Petrov type of the Weyl tensor $W_{abcd}$ \cite{petrov,penrose}, and
the Segre type of the trace-free Ricci tensor $S_{ab} \equiv R_{ab} -
\frac{1}{4} \delta_{ab} R$ \cite{pleban,mcin-2,zak-car}, and more generally in studying the equivalence problem, i.e.,
the question of whether two space-time metrics are equivalent \cite{eq-st1,eq-st2,cohepe}.
See also \cite{weinberg,stepetal}. The scalar invariants are also important
in understanding the structure of the diffeomorphism group and the effective action
for quantum fields with gravitational interaction \cite{gibbons,dewitt,vilk,amsber}.

The invariant characterization of a curved space must be in terms of scalars constructed from
curvature tensors $R_{\mu\nu\rho\sigma}$ and a metric tensor $g_{\mu\nu}$.
A simple argument implies (see \S 6.7 in \ct{weinberg}) that there are fourteen algebraically independent curvature invariants in four dimensions. Since the Riemann curvature tensor $R_{abcd}$  has 20 independent components, the space of tensors $R_{abcd}$ is 20-dimensional. But these twenty pieces of information are not independent because there are six degrees of freedom represented by
the Lorentz transformation under which the metric is invariant. Hence there are $20-6=14$
independent invariants (see \S 8.6 in \ct{penrose}) that can be locally constructed by curvature tensors. However, the problem of
obtaining a `complete set' is not an easy problem since it requires to know all possible algebraic relations (known as {\it syzygies}) for the members of the set to any higher orders. This issue was one of the driving forces in the early development
of computer algebra \cite{mac-com,port-com}.

The classification of the Riemann tensor in general relativity depends on properties of
the algebraic invariants of the Weyl tensor and the Ricci tensor.
For several decades, many attempts have been directed towards understanding these invariants and their relationships. In 1902 Haskins showed \cite{haskins} that there were 14 independent differential invariants constructed from the metric tensor and the Riemann curvature tensor. Forty-five years later Narlikar and Karmarkar produced such a set of scalars \cite{narl-kar,harvey2}. Sometime thereafter, G\'eh\'eniau and Debever also presented a set \cite{gehe-dede1,gehe-dede2,gehe-dede3}. Various sets have been constructed by several
authors \cite{l-witten,safko-witten,greenberg,sobc,sned-0} and have been claimed to be independent. However it was pointed out \cite{carmal,sned-1,mcin,bona} that the number of the sets proposed was deficient in some sense and none of these sets can be algebraically complete.
Carminati and McLenaghan showed \cite{carmal} that the 14 previously suggested invariants fail to describe solutions admitting a perfect fluid and came up with 16 independent invariants to cover all subspaces including the degenerate cases such as the Einstein-Maxwell theory and the perfect fluid.
Later, Zahkary and McIntosh constructed the first complete set of algebraic invariants for all possible type of metrics \cite{mcin}.
This is an indication that the problem is not as straightforward as it may seem.
The main difficulty is that the relationships between different invariants are not at all well understood.

The problem of linear dependences between the monomials (or polynomials) of the Riemann curvature tensor is not a simple question to answer.
Because of the subtleties of tensor symmetry, terms may be linearly dependent in non-obvious ways.
This problem was partially solved in \cite{ten-poly} by applying some representation theory of the symmetric, general linear and orthogonal groups to the set formed from the Riemann tensor by covariant differentiation, multiplication and contraction.
But the results of Ref. \cite{ten-poly} are incomplete because possible algebraic relations between the set of tensor polynomials were not addressed. Moreover only parity even polynomials have been included, so parity odd polynomials, i.e.,  pseudo-scalars or pseudo-tensors, are missing.
In this paper we will develop a powerful tool to analyze the irreducible basis as well as an algorithm to determine the linear dependence of arbitrary Riemann polynomials
using the irreducible decomposition of curvature tensors \cite{oh-yang}. See also \cite{yang-col1,yang-col2,yang-col3,yang-col4}.

A special feature of four-dimensional Riemannian manifolds is the fact that the Lie group $Spin(4)$ splits into a product of two groups:
\be \la{spin4}
Spin (4) = SU(2)_+ \times SU(2)_-.
\ee
The group $Spin (4)$ is a double cover of the four-dimensional Euclidean Lorentz group $SO(4)$,
i.e., $SO(4) =  SU(2)_+ \times SU(2)_-/\mathbb{Z}_2$. It also leads to the splitting of the Lie algebra
\be \la{lie-so4}
so(4) = su_+ (2) \oplus su(2)_-.
\ee
The splitting of the vector space $\mathfrak{g} = so(4)$ is isomorphically  related to the decomposition of the 2-forms on a four-manifold $M$. Using the Hodge $*$ operator acting on exterior 2-forms,
one can split the 2-forms into self-dual and anti-self-dual 2-forms \cite{book-dokr}
\be \la{split-2}
\Omega^2 \equiv \Lambda^2 T^*M = \Omega^2_+ \oplus \Omega^2_-,
\ee
where $\Omega^2_\pm$ is the $\pm 1$ eigenspaces of the Hodge star operator $*: \Omega^2 \to \Omega^2$.

The splitting of the vector spaces, Eqs. \eq{spin4} and \eq{split-2}, can be applied to the curvature form of any bundle with connection over an oriented
four-manifold. The canonical splitting leads to the irreducible decomposition of
Riemann curvature tensor $R \in C^\infty (\mathfrak{g} \otimes \Omega^2)$ as \cite{s-duality,book-besse,egh-report}
\be \la{riem-dec}
R = R_{(++)} \oplus R_{(+-)} \oplus R_{(-+)} \oplus R_{(--)},
\ee
where the subscript $(\pm, \pm)$ refers to the splitting of the vector spaces $\mathfrak{g} = so(4) = su_+ (2) \oplus su(2)_-$ and $\Omega^2 = \Lambda^2 T^*M = \Omega^2_+ \oplus \Omega^2_-$.
This splitting of the vector spaces occupies a central position of the Donaldson theory of four-manifolds and permeates four-dimensional geometry (see, for example, Chaps. 1.G in Ref. \cite{book-besse} and Secs. 1.1 and 2.1 in Ref. \cite{book-dokr}).
Essentially the same decomposition also appears in \cite{l-witten} (Eq. (10))
and \cite{penrose} (Eq. (4.6.38)) using the spinor formalism and in \cite{buchdahl1,buchdahl2,cahenetal}
using complex self-dual bivectors. In particular, our approach corresponds to the real version of the latter, which clearly reveals an elegant structure for the Riemann polynomial under parity transformation.

This paper is organized as follows. In section 2, the index symmetries of the Riemann curvature tensor are reviewed which show how they reduce the number of independent index orders.
In section 3, we review the results of \cite{oh-yang}. We present the explicit form of the decomposition \eq{riem-dec} for a general Riemannian manifold. The essentially same decomposition
also appeared in the spinor approach for general relativity \cite{penrose,l-witten}, which has been used by most of the authors cited above. Since our method uses only vector indices rather than spinor indices \cite{buchdahl1,buchdahl2,cahenetal}, it is much simpler than the spinor method and is much more convenient for a practical calculation, even for developing computer algorithms.
In section 4, we illustrate how the irreducible decomposition of curvature tensors \eq{riem-dec}
provides a powerful tool to analyze the irreducible basis as well as an algorithm to determine the linear dependence of arbitrary Riemann polynomials. After a brief application to the algebraic relation of some Riemann polynomials and the generalization of the Gauss-Bonnet relation
to illustrate the power of our methodology, we completely specify 13 independent basis elements for the quartic scalars and explicitly find 13 linear relations among 26 scalar invariants.
Our method also provides some clues to identification of 23 independent basis elements from
90 quintic scalars. We also clarify the algebraic properties of the second-rank cubic tensors,
which was shown in \cite{ten-poly} to have 16 basis elements.
In section 5, we discuss future directions and suggest several generalizations.
In particular, we address an issue of how to use our results in understanding the complete set of Riemann tensor invariants proposed in \cite{carmal,mcin}.

All detailed results related to Chapter 4 are deferred to the appendices.
Appendix A contains some useful identities of the ’t Hooft symbols which are extensively used in our calculations. Appendix B is a warming-up calculation to show how effective our method is. We show {\it by hand} the vanishing of some quintic and quartic Riemann polynomials which were the examples presented in \cite{portugal} to illustrate why the use of a computer is necessary since they are difficult to show by hand. Appendix C contains detailed results of the cubic scalars and second-rank tensors classified in \cite{ten-poly}. We identify the independent basis elements of 8 cubic scalars by expanding them using the irreducible decomposition \eq{riem-dec}. We confirm that there are 6 independent cubic scalars and precisely reproduce the two algebraic relations found by Xu \cite{dxu} and Harvey \cite{harvey}. We extend the analysis to second-rank cubic tensors and get several new results. Appendix D contains our main results. We show in detail how the 26 quartic scalars in  \cite{ten-poly} can be expanded using the decomposition \eq{riem-dec} and explicitly identify
14 basis elements. We find that these 14 basis elements are not completely independent and there is one more linear relation among them. This implies there must be 13 linear relations known as syzygies
among the 26 quartic scalars. We found all of them. This completes the syzygies of Riemann curvature invariants in cubic and quartic orders.

\section{Algebraic symmetry of the Riemann curvature tensor}

A solution of gravitational field equations is given by the line element
\begin{equation}\label{metric-vier}
  ds^2 = g_{\mu\nu} (x) dx^\mu dx^\nu = e^a \otimes e^a
\end{equation}
with the symmetric, covariant metric tensor $g_{\mu\nu}$ or
four basis covectors $e^a = e^a_\mu (x) dx^\mu$ in a locally inertial frame.
The basic objects of a metric are the Christoffel symbol and Riemann tensor
which are defined as follows:
\begin{eqnarray} \label{christoffel}
&& \Gamma^\mu_{\nu\lambda} = \frac{1}{2} g^{\mu\rho} \left(\frac{\partial g_{\rho\nu}}{\partial x^\lambda}
+ \frac{\partial g_{\rho\lambda}}{\partial x^\nu} - \frac{\partial g_{\nu\lambda}}{\partial x^\rho} \right)
= \Gamma^\mu_{\lambda\nu}, \\
\label{riemann}
&& {R^\mu}_{\nu\rho\sigma} = \frac{\partial \Gamma^\mu_{\nu\sigma}}{\partial x^\rho}
- \frac{\partial \Gamma^\mu_{\nu\rho}}{\partial x^\sigma} + \Gamma^\mu_{\rho\lambda}\Gamma^\lambda_{\nu\sigma} - \Gamma^\mu_{\sigma\lambda}\Gamma^\lambda_{\nu\rho},
\end{eqnarray}
and $R_{\mu\nu\rho\sigma} = g_{\mu\lambda}{R^\lambda}_{\nu\rho\sigma}$.
The Ricci tensor and Ricci scalar are defined by
\begin{equation}\label{ricci-ten}
  R_{\mu\nu} = {R^\rho}_{\mu\rho\nu} = g^{\rho\sigma} R_{\rho\mu\sigma\nu}, \qquad
  R = g^{\mu\nu} R_{\mu\nu}.
\end{equation}
Throughout this paper, we will use the Riemann curvature tensor expressed in a locally inertial frame:
\begin{equation}\label{curv-inerf}
  R_{abcd} = \mathfrak{e}_a^\mu \mathfrak{e}_b^\nu \mathfrak{e}_c^\rho \mathfrak{e}_d^\sigma R_{\mu\nu\rho\sigma}.
\end{equation}
where $\mathfrak{e}_a = \mathfrak{e}_a^\mu (x) \partial_\mu$ are basis vectors defined by the conditions
$\mathfrak{e}_a^\mu e^b_\mu = \delta^b_a$ and $\mathfrak{e}_a^\mu e^a_\nu = \delta^\mu_\nu$.

From Eq. \eq{riemann}, one may read off the algebraic properties of the curvature tensor \cite{weinberg}:

(A) Antisymmetry

\begin{equation} \label{a-symm}
    R_{abcd} = - R_{bacd} = - R_{abdc} = R_{badc}.
\end{equation}

(B) The first Bianchi identity

\begin{equation} \label{bianchi}
    R_{abcd} + R_{acdb} + R_{adbc} = 0.
\end{equation}

(C) Pair symmetry

\begin{equation} \label{symm}
    R_{abcd} = R_{cdab}.
\end{equation}

Indeed the last property (\ref{symm}) can be derived from (A) and (B).
From these symmetries of the curvature tensor, it is easy to show \cite{ten-poly} that
of the $4!=24$ different possible orderings of the indices, only two are independent.
A convenient choice of them is
\begin{equation} \label{2-riem}
    R_{abcd} \quad \textrm{and} \quad R_{acbd}.
\end{equation}

\section{Irreducible decomposition of curvature tensors}

On an orientable Riemannian four-manifold, Eq. \eq{split-2} shows that the 2-forms in $\Omega^2 (M) = \Lambda^2 T^* M$ decompose into
the space of self-dual and anti-self-dual 2-forms.
Another important feature, which permeates four-dimensional geometry, is the fact that
the Lorentz group $Spin(4)$ splits into a product of two groups as expressed by Eq. \eq{spin4}.
The group $Spin(4)$ is a double cover of the four-dimensional Euclidean Lorentz group $SO(4)$,
 i.e., $SO(4) \cong SU(2)_+ \times SU(2)_-/\mathbb{Z}_2$.
The product structure \eq{spin4} also leads to the splitting of the Lie algebra \eq{lie-so4}.
The splitting of the vector space $\mathfrak{g} = so(4)$ is defined by the chiral operator
$\gamma_5 = - \gamma_1 \gamma_2 \gamma_3 \gamma_4$ in the Clifford algebra which obeys $\gamma_5^2 =1$.
Indeed, the splitting of vector spaces is induced by the existence of the projection operators \ct{yang-col2}
\begin{equation}\label{proj-2}
  P_\pm = \frac{1}{2} (1 \pm *), \qquad P_\pm = \frac{1}{2} (1 \pm \gamma_5)
\end{equation}
acting on the vector spaces $\Omega^2$ and the $so(4)$ generators $J_{ab}
= \frac{1}{4} [\gamma_a, \gamma_b ]$, respectively.

The explicit realization of the splitting \eq{lie-so4} reads as
\begin{equation}\label{split-so4}
J_{ab} = J_{ab}^+ \oplus J_{ab}^-
\end{equation}
where $J_{ab}^\pm  \equiv P_\pm J_{ab}$ and each part consists of three $(2 \times 2)$ anti-Hermitian traceless matrices.
So they can be expanded in the basis of the Pauli matrices $\tau^i \; (i=1,2,3)$ as
\begin{equation}\label{split-su2}
J_{ab}^+ = \frac{i}{2} \eta^i_{ab} \tau^i \in  su(2)_+, \qquad
J_{ab}^- = \frac{i}{2}\bar{ \eta}^i_{ab} \tau^i \in su(2)_-,
\end{equation}
where the expansion coefficients, the so-called 't Hooft symbols, are given by\footnote{The 't Hooft symbols, dubbed by physicists, were first introduced in \cite{buchdahl1,buchdahl2}
as quantities (dubbed ``tensor rotors") connecting a self-dual bivector to a vector in, so-called, ``rotor space" and an anti-self-dual bivector to a complex conjugated rotor. See also \cite{cahenetal}. The 't Hooft symbols have played an important role for instanton physics in Yang-Mills gauge theory \cite{raja-inst}.}
\begin{equation}\label{t-symbol}
 \eta^i_{ab} = - i \mathrm{Tr} \big( J_{ab}^+ \tau^i \big), \qquad
 \bar{\eta}^i_{ab} = - i \mathrm{Tr} \big( J_{ab}^- \tau^i \big).
\end{equation}
An explicit representation of the ’t Hooft symbols is shown up in Appendix A where
we also list some useful identities of the ’t Hooft symbols.

The decomposition \eq{split-2} is that the six-dimensional vector space of two-forms canonically splits
into the sum of three-dimensional vector spaces of self-dual and anti-self-dual two-forms.
Canonical basis elements of self-dual and anti-self-dual two forms are given by
\begin{equation}\label{sd-2form}
  \zeta^i_+ = \frac{1}{2} \eta^i_{ab} e^a \wedge e^b \in  \Omega^2_+, \qquad
  \zeta^i_- = \frac{1}{2} \bar{\eta}^i_{ab}  e^a \wedge e^b \in \Omega^2_-,
\end{equation}
where the canonical basis elements satisfy the Hodge-duality equation
\begin{equation}\label{hodge-bais}
  * \zeta^i_\pm = \pm \zeta^i_\pm.
\end{equation}
They also satisfy the intersection relation
\begin{equation}\label{intersection}
\zeta^i_\pm \wedge \zeta^j_\pm = \pm 2 \delta^{ij} \sqrt{g} d^4 x, \qquad
\zeta^i_\pm \wedge \zeta^j_\mp = 0.
\end{equation}

The Riemann curvature tensor $R_{ab} = \frac{1}{2} R_{ab\mu\nu} dx^\mu \wedge dx^\nu
= \frac{1}{2} R_{abcd} e^c \wedge e^d \in C^\infty(\mathfrak{g} \otimes \Omega^2)$ is $so(4)$-valued 2-forms.
Thus it is involved with two vector spaces $\mathfrak{g} = so(4)$ and $\Omega^2$
in \eq{lie-so4} and \eq{split-2}.
Let us apply the decompositions of the vector spaces $\mathfrak{g} = so(4)$ and $\Omega^2$ to Riemann curvature tensors.
The first decomposition is that the Riemann tensor can be split into a pair of $SU(2)_+$ and $SU(2)_-$ curvatures
according to the Lie algebra splitting \eq{split-so4}:
\begin{equation}\label{split-riemann}
  R_{ab} = F^{(+)i} \eta^i_{ab}  + F^{(-)i} \bar{\eta}^i_{ab},
\end{equation}
where $SU(2)_\pm$ field strengths are 2-forms on $M$ defined by
\begin{eqnarray}\label{sd-2f}
  F^{(\pm)i} &=& \frac{1}{2} F^{(\pm)i}_{cd} e^c \wedge e^d \nonumber \\
             &=& dA^{(\pm)i} - \varepsilon^{ijk} A^{(\pm)j} \wedge A^{(\pm)k}.
\end{eqnarray}
The second decomposition is that the $SU(2)_\pm$ field strengths in Eq. \eq{split-riemann} can be decomposed as
\begin{equation}\label{decop-2f}
 F^{(+)i} =  f^{ij}_{(++)}\zeta^j_{+} + f^{ij}_{(+-)}\zeta^j_{-}, \qquad
  F^{(-)i} =  f^{ij}_{(-+)}\zeta^j_{+} + f^{ij}_{(--)}\zeta^j_{-}.
\end{equation}
Combining the two decompositions \eq{split-riemann} and \eq{decop-2f} leads to an irreducible decomposition of
the general Riemann curvature tensor \cite{oh-yang,yang-col2}:
\begin{equation}\label{decom-riem}
  R_{abcd} = f^{ij}_{(++)}\eta^i_{ab}\eta^j_{cd} + f^{ij}_{(+-)}\eta^i_{ab} \bar{\eta}^j_{cd}
  +  f^{ij}_{(-+)} \bar{\eta}^i_{ab} \eta^j_{cd} + f^{ij}_{(--)} \bar{\eta}^i_{ab}\bar{\eta}^j_{cd},
\end{equation}
which is the explicit form of the decomposition \eq{riem-dec}.
Eq. (\ref{symm}) imposes the symmetry property
\begin{equation}\label{symm-fij}
f^{ij}_{(++)} = f^{ji}_{(++)}, \quad f^{ij}_{(--)}=f^{ji}_{(--)},
\quad  f^{ij}_{(+-)} = f^{ji}_{(-+)}.
\end{equation}
The first Bianchi identity (\ref{bianchi}), being totally 16 conditions, imposes an additional constraint
\begin{equation}\label{1-more}
  f^{ij}_{(++)} \delta^{ij} = f^{ij}_{(--)} \delta^{ij}
\end{equation}
that is equivalently written as
\begin{equation}\label{pseudo-r}
  \varepsilon^{abcd} R_{abcd} = 0.
\end{equation}

The above results can be applied to the Ricci tensor $R_{ab} \equiv R_{acbc}$ and
the Ricci scalar $R \equiv R_{aa}$ to yield
\begin{eqnarray}\label{ricci}
  R_{ab} &=& \big(f^{ij}_{(++)} \delta^{ij} + f^{ij}_{(--)} \delta^{ij} \big) \delta_{ab}
  + 2 f^{ij}_{(+-)}\eta^i_{ac} \bar{\eta}^j_{bc}, \nonumber \\
  R &=& 4 \big(f^{ij}_{(++)} \delta^{ij} + f^{ij}_{(--)} \delta^{ij} \big).
\end{eqnarray}
The ``trace-free part" of the Riemann curvature tensor is called the Weyl tensor defined by
\begin{equation}\label{weyl}
  W_{abcd} = R_{abcd} - \frac{1}{2} (\delta_{ac} R_{bd} - \delta_{ad} R_{bc} - \delta_{bc} R_{ad} + \delta_{bd} R_{ac} )
  + \frac{1}{6} (\delta_{ac} \delta_{bd} - \delta_{ad} \delta_{bc}) R.
\end{equation}
The Weyl tensor shares all the symmetry structures of the curvature tensor and all its traces with the
metric are zero. Then the Weyl tensor can be decomposed as
\begin{eqnarray}\label{weyl-riem}
  W_{abcd} &=& f^{ij}_{(++)}\eta^i_{ab}\eta^j_{cd} + f^{ij}_{(--)} \bar{\eta}^i_{ab}\bar{\eta}^j_{cd}
  - \frac{1}{3} \Big(  f^{ij}_{(++)} \delta^{ij} + f^{ij}_{(--)} \delta^{ij} \Big)
  (\delta_{ac} \delta_{bd} - \delta_{ad} \delta_{bc}) \xx
  &=& {\widetilde f}^{ij}_{(++)} \eta^i_{ab}\eta^j_{cd} + {\widetilde f}^{ij}_{(--)} \bar{\eta}^i_{ab}\bar{\eta}^j_{cd},
\end{eqnarray}
where ${\widetilde f}^{ij}_{(++)} \equiv f^{ij}_{(++)} - \frac{1}{3} \delta^{ij} \big(  f^{kl}_{(++)} \delta^{kl} \big)$ and ${\widetilde f}^{ij}_{(--)} \equiv f^{ij}_{(--)} - \frac{1}{3} \delta^{ij} \big(  f^{kl}_{(--)} \delta^{kl} \big)$ are symmetric, traceless $3 \times 3$ matrices.
In the end the Riemann tensor is decomposed as
\begin{equation}\label{dec-riem}
  R_{abcd} = W_{abcd} + \frac{1}{12} R (\delta_{ac} \delta_{bd} - \delta_{ad} \delta_{bc})
  + f^{ij}_{(+-)} \big( \eta^i_{ab} \bar{\eta}^j_{cd} + \bar{\eta}^j_{ab} \eta^i_{cd} \big)
\end{equation}
where the last one corresponds to the traceless Ricci tensor
$S_{ab} \equiv R_{ab}-\frac{1}{4}\delta_{ab} R = 2 f^{ij}_{(+-)} \eta^i_{ac} \bar{\eta}^j_{bc}$
or, conversely, $f^{ij}_{(+-)} = \frac{1}{8} S_{ab} \eta^i_{ac} \bar{\eta}^j_{bc}$.

Using the decomposition (\ref{decom-riem}) and (\ref{ricci}), it is easy to calculate quadratic (pseudo-)scalars of curvature tensors
\begin{eqnarray}\label{quadratic-r}
  R_{abcd} R_{abcd} &=& 16 \Big(f^{ij}_{(++)} f^{ij}_{(++)} + 2 f^{ij}_{(+-)}f^{ij}_{(+-)}
  + f^{ij}_{(--)}f^{ij}_{(--)} \Big), \nonumber \\
  R_{ab} R_{ab} &=& 4 \big(f^{ij}_{(++)} \delta^{ij} + f^{ij}_{(--)} \delta^{ij} \big)^2  + 16 f^{ij}_{(+-)} f^{ij}_{(+-)}, \nonumber \\
  R^2 &=& 16 \big(f^{ij}_{(++)} \delta^{ij} + f^{ij}_{(--)} \delta^{ij} \big)^2, \\
  \varepsilon^{abcd}\varepsilon^{efgh} R_{abef}R_{cdgh} &=& 64 \Big(f^{ij}_{(++)} f^{ij}_{(++)} - 2 f^{ij}_{(+-)}f^{ij}_{(+-)}
   + f^{ij}_{(--)}f^{ij}_{(--)}\Big), \nonumber \\
  \varepsilon^{cdef} R_{abcd}R_{abef} &=& 32 \Big(f^{ij}_{(++)} f^{ij}_{(++)} -  f^{ij}_{(--)}f^{ij}_{(--)}\Big). \nonumber
\end{eqnarray}
One can immediately see that
\begin{eqnarray}\label{gb-relation}
  R_{abcd} R_{abcd} - 4 R_{ab} R_{ab} + R^2 = \frac{1}{4}
  \varepsilon^{abcd}\varepsilon^{efgh} R_{abef}R_{cdgh}.
\end{eqnarray}
This is the famous Gauss-Bonnet relation (see Appendix B in \cite{thooft-velt}) where the right-hand side is the well-known Euler density
$\rho_\chi = \frac{1}{128 \pi^2} \varepsilon^{abcd}\varepsilon^{efgh} R_{abef}R_{cdgh}$ such that
$\chi = \int_M \rho_\chi \sqrt{g} d^4 x = n \geq 0$.
The last one in Eq. (\ref{quadratic-r}) corresponds to the Hirzebruch density $\rho_\tau = \frac{1}{96 \pi^2}
\varepsilon^{cdef} R_{abcd}R_{abef}$ in which the Hirzebruch signature is defined by
$\tau = \int_M \rho_\tau \sqrt{g} d^4 x = m \in \mathbb{Z}$ \cite{egh-report,gh-cmp79}.
However, since the Hirzebruch density $\rho_\tau$ is a pseudo-scalar, $\rho_\tau$ cannot be written in a similar way
as the Euler density $\rho_\chi$.

Using the decomposition \eq{weyl-riem}, it is easy to get
\begin{eqnarray*}
    W_{acde} W_{bcde} &=&  4 \Big(f^{ij}_{(++)} f^{ij}_{(++)} + f^{ij}_{(--)}f^{ij}_{(--)} \Big) \delta_{ab} - \frac{1}{24} R^2 \delta_{ab} \\
    &=& 4 \Big({\widetilde f}^{ij}_{(++)} {\widetilde f}^{ij}_{(++)} +
    {\widetilde f}^{ij}_{(--)} {\widetilde f}^{ij}_{(--)} \Big) \delta_{ab}.
\end{eqnarray*}
Using this result, one can get the important identity \ct{ten-poly}
\begin{eqnarray} \label{quad-weyl}
     & W_{cdea} W_{cdeb} = \frac{1}{4} \delta_{ab} W_{cdef} W_{cdef}.
\end{eqnarray}
Using the definition \eq{weyl}, it can be written as
\begin{eqnarray} \label{2r-weyl}
 R_{cdea} R_{cdeb} = \frac{1}{4} \delta_{ab} R_{cdef} R_{cdef}
+ 2 R_{acbd} R_{cd} + 2 R_{ac} R_{bc} - \delta_{ab} R_{cd}^2
- R \left(R_{ab} - \frac{1}{4} \delta_{ab} R \right).
\end{eqnarray}

Under the parity transformation which flips the orientation of four-manifolds,
the self-dual and anti-self-dual 't Hooft symbols exchange their role and it induces the interchange
\begin{equation}\label{exchange}
f^{ij}_{(++)} \leftrightarrow f^{ij}_{(--)}, \qquad f^{ij}_{(+-)} \leftrightarrow f^{ij}_{(-+)} = f^{ji}_{(+-)}.
\end{equation}
Therefore the normal scalars and tensors which are invariant under the parity transformation must be symmetric
for the interchange \eq{exchange}. The quadratic scalars in Eq. \eq{quadratic-r} exhibits this symmetry.
But, pseudo-scalars and pseudo-tensors change the sign under the parity transformation and
the Hirzebruch density shows this behavior.

For Einstein manifolds satisfying the equations, $R_{\mu\nu} = \lambda g_{\mu\nu}$ (or $R_{ab} = \lambda \delta_{ab}$), with $\lambda$ a cosmological constant, one can show \cite{oh-yang,yang-col4} that
\begin{equation} \label{einstein}
  f^{ij}_{(+-)} = f^{ji}_{(-+)} = 0.
\end{equation}
Therefore the curvature tensor for Einstein manifolds takes the simple form
\begin{equation}\label{em-riem}
  R_{abcd} = f^{ij}_{(++)}\eta^i_{ab}\eta^j_{cd} + f^{ij}_{(--)} \bar{\eta}^i_{ab}\bar{\eta}^j_{cd}.
\end{equation}

\section{Algebraic relations of curvature tensors}

We have already illustrated in Eq. \eq{quadratic-r} how efficient our method using the irreducible decomposition of
curvature tensors \eq{decom-riem} is to find algebraic relations among the curvature tensors such as \eq{gb-relation}.
To further illustrate the power of our methodology, let us quote the following sentence in \cite{portugal}:
\begin{quotation}
It is difficult to show by hand that
\begin{equation}\label{dif-5}
  R_{a_1 a_2 b_1 b_2} R_{c_1 c_2 d_1 d_2} R_{e_1 a_1 e_2 c_1} R_{a_2 b_1 b_2 e_1} R_{c_2 d_1 d_2 e_2} = 0.
\end{equation}
This example does not require the use of the cyclic identity of the Riemann tensor. The next example
is more difficult to prove:
\begin{eqnarray}\label{dif-4}
 && 2 R_{a_1 a_2 b_1 b_2} R_{c_1 c_2 a_1 d_1} R_{d_2 b_1 d_1 a_2} R_{b_2 d_2 c_1 c_2}
 + 4 R_{a_1 a_2 b_1 b_2} R_{c_1 c_2 d_1 a_1} R_{d_2 a_2 b_1 c_1} R_{b_2 d_1 c_2 d_2}  \\
&& \qquad - R_{a_1 a_2 b_1 b_2} R_{c_1 c_2 d_1 a_1} R_{d_2 a_2 c_2 c_1} R_{b_1 b_2 d_1 d_2}
  + 4 R_{a_1 a_2 b_1 b_2} R_{d_2 d_1 b_2 c_2} R_{c_1 c_2 d_1 a_1} R_{d_2 a_2 b_1 c_1} = 0. \nonumber
\end{eqnarray}
\end{quotation}
In Appendix B, we have shown these equations \textit{by hand} using the decomposition (\ref{em-riem}).

Now we will extend the calculation in Eq. \eq{quadratic-r} to general cases: Rank-$r$ tensors
(with $r$ free indices) formed from the Riemann tensor by multiplication and contraction. 
Furthermore, we will generalize our calculations to general Riemannian manifolds
after a brief warm-up with the Einstein manifold.
Incidentally, our method can also be applied to a Lorentzian manifold although it is based on the decompositions
\eq{lie-so4} and \eq{split-2} which are characteristic properties established in the Euclidean signature.
The signature of the metric is irrelevant in studying the algebraic relationship
of the curvature tensors of a Riemannian (or Lorentzian) manifold.\footnote{There may
be a subtlety for pseudo-tensors since the Wick rotation of the Levi-Civita tensor will introduce
an extra imaginary factor $i=\sqrt{-1}$. However, since the pseudo-tensor contains only an odd number of the Levi-Civita tensors, such a subtlety can be easily dealt with.}
Indeed we will see that our results are consistent with the previous results for Lorentzian manifolds.

\subsection{Quadratic order}

The quadratic monomials of Riemann tensors are completely classified in \cite{ten-poly}.
The scalar monomials have three independent basis elements listed in Eq. \eq{quadratic-r}.
The independent basis elements are given by the product of $f^{ij}_{(\pm\pm)}$ and $f^{ij}_{(\pm\mp)}$
\begin{eqnarray} \label{quadratic-b}
f^{ij}_{(++)} f^{ij}_{(++)} + f^{ij}_{(--)}f^{ij}_{(--)},
\quad f^{ij}_{(+-)} f^{ij}_{(+-)}, \quad
\big(f^{ij}_{(++)} \delta^{ij} + f^{ij}_{(--)} \delta^{ij} \big)^2 = \frac{R^2}{16}.
\end{eqnarray}
Considering the parity symmetry \eq{exchange}, the expression clearly verifies why the quadratic scalars
have only three independent basis elements.
The Gauss-Bonnet relation in \eq{quadratic-r} is not a linear relation among the quadratic scalars
but a simple consequence of the formula \cite{thooft-velt}
$$ \varepsilon^{abcd} \varepsilon^{efgh} = \delta^a_{[e} \delta^b_f \delta^c_g \delta^d_{h]}$$
where $[efgh]$ means anti-symmetrization over all four indices.

Although we will consider a general Riemannian manifold later, let us restrict,
for the time being, our calculation to Einstein manifolds
whose curvature tensors are given by (\ref{em-riem}).
Let us point out why this restriction does not lose generality while making the calculation simple.
The main reason is that both the Ricci tensor and Ricci scalar are non-zero in the case of Einstein manifolds
so that the index structure of rank-$r$ tensors is exactly the same as the general case.
Furthermore, the Weyl tensor for an Einstein manifold takes the same form as a general Riemannian manifold,
as shown in \eq{weyl-riem}.\footnote{Therefore, the result for an Einstein manifold can easily be converted to the result for the Weyl tensor of a general Riemannian manifold using the relation \eq{weyl-riem} which can be written as $W_{abcd} = R^E_{abcd} - \frac{1}{12} R (\delta_{ac} \delta_{bd} - \delta_{ad} \delta_{bc})$ where $R^E_{abcd}$ is the Riemann tensor given by \eq{em-riem}. See, for example, Eqs. \eq{quad-weyl} and \eq{2r-weyl}.} However, in the conventional approach of dealing directly with the Riemann curvature tensor, the restriction to the Einstein manifold is of little help.

Since determining the linear dependence of the monomials of curvature tensors at quadratic order
is a trivial problem, we will instead generalize the Gauss-Bonnet relation to the case with four free indices.
Let us illuminate how to find a generalized Gauss-Bonnet relation
among homogeneous monomials of the Riemann tensor with explicit examples.
For example, using the result (\ref{em-riem}), it is easy to get
\begin{eqnarray}\label{2-1}
  R_{abef}R_{efcd} = 4 \big( f^{ik}_{(++)}  f^{kj}_{(++)} \eta^i_{ab} \eta^j_{cd}
  + f^{ik}_{(--)}  f^{kj}_{(--)} \bar{\eta}^i_{ab} \bar{\eta}^j_{cd} \big).
\end{eqnarray}
Now we want to express the right-hand side of Eq. \eq{2-1} in terms of the product of the Riemann tensors.
We will use the relation
\begin{eqnarray}\label{2-2}
  f^{ij}_{(++)} = \frac{1}{16} R_{abcd} \eta^i_{ab} \eta^j_{cd}, \qquad
  f^{ij}_{(--)} = \frac{1}{16} R_{abcd} \bar{\eta}^i_{ab} \bar{\eta}^j_{cd}, \qquad
  f^{ij}_{(+-)} = \frac{1}{16} R_{abcd} \eta^i_{ab} \bar{\eta}^j_{cd} =  f^{ji}_{(-+)}.
\end{eqnarray}
Plugging these expressions into Eq. \eq{2-1} and using various identities of the 't Hooft symbols in Appendix A
lead to the result
\begin{eqnarray}\label{2-3}
R_{abef}R_{efcd} &=& \frac{1}{12} \big( \varepsilon_{a_1 a_2 ab} \varepsilon_{b_1 b_2 c_1 c_2}
  R_{a_1 a_2 b_1 b_2} R_{c_1 c_2 cd} + \varepsilon_{a_1 a_2 ab} \varepsilon_{c_1 c_2 c d}
  R_{a_1 a_2 b_1 b_2} R_{b_1 b_2 c_1 c_2} \nonumber \\
 && \qquad  + \varepsilon_{a_1 a_2 b_1 b_2} \varepsilon_{c_1 c_2 c  d}
  R_{a b a_1 a_2} R_{b_1 b_2 c_1 c_2} \big).
\end{eqnarray}
This is a generalization of the Gauss-Bonnet relation \eq{gb-relation} in the sense that
contracting $[ab]$ and $[cd]$ leads to the relation. (Since we have assumed the Einstein manifold, the last two terms
on the left-hand side of Eq. \eq{gb-relation} cancel each other.)
The cyclic symmetry \eq{bianchi} leads to the relation
\begin{eqnarray}\label{4-1}
  R_{abef}R_{cedf} &=& R_{aebf} R_{cdef} = \frac{1}{2}  R_{abef}R_{efcd}.
\end{eqnarray}

Similarly, we get
\begin{eqnarray}\label{2-4}
 R_{aebf}R_{cedf} &=& \big( f^{ij}_{(++)}  f^{ij}_{(++)} + f^{ij}_{(--)}  f^{ij}_{(--)} \big)  \delta_{ac} \delta_{bd}
  + 2  f^{ij}_{(++)} f^{kl}_{(--)} \big( \eta^i_{ae} \bar{\eta}^k_{ce} \big) \big( \eta^j_{bf} \bar{\eta}^l_{df} \big) \nonumber \\
 &&  + \varepsilon^{ikm} \varepsilon^{jln} f^{ij}_{(++)} f^{kl}_{(++)}
  \eta^m_{ac} \eta^n_{bd} + \varepsilon^{ikm} \varepsilon^{jln} f^{ij}_{(--)} f^{kl}_{(--)}
  \bar{\eta}^m_{ac} \bar{\eta}^n_{bd}.
\end{eqnarray}
Using the identity $\varepsilon^{ikm} \varepsilon^{jln} = \delta^i_{[j} \delta^k_l \delta^m_{n]}$ where $[jln]$ means anti-symmetrization over all three indices, and the relations,
\begin{eqnarray} \label{f-sd-asd}
&& f^{ij}_{(++)} \eta^i_{ab} \eta^j_{cd} = \frac{1}{2} \Big( R_{abcd} + \frac{1}{2} \varepsilon_{abef} R_{efcd} \Big), \nonumber \\
&& f^{ij}_{(--)} \bar{\eta}^i_{ab} \bar{\eta}^j_{cd} = \frac{1}{2} \Big( R_{abcd} - \frac{1}{2} \varepsilon_{abef} R_{efcd} \Big),
\end{eqnarray}
we obtain, after a little algebra, the result
\begin{eqnarray}\label{4-3}
 R_{aebf}R_{cedf} &=& R_{acef} R_{efbd} + \frac{1}{16} R^2 \big( \delta_{ab} \delta_{cd} - \delta_{ad} \delta_{bc} \big)
 - \frac{1}{2} R R_{acbd} \nonumber \\
 && - \frac{1}{8} R_{a_1 a_2 b_1 b_2} R_{a_1 a_2 b_1 b_2}
 \big( \delta_{ab} \delta_{cd} - \delta_{ad} \delta_{bc} - \delta_{ac} \delta_{bd}  \big) \nonumber \\
 && - \frac{1}{4} \varepsilon_{ae a_1 a_2} \varepsilon_{ce b_1 b_2} R_{a_1 a_2 b f} R_{b_1 b_2 d f}
  + \frac{1}{16} \varepsilon_{abcd} \varepsilon_{a_1 a_2 b_1 b_2}
 R_{c_1 c_2 a_1 a_2} R_{c_1 c_2 b_1 b_2}.
\end{eqnarray}
Using the cyclic symmetry \eq{bianchi} leads to the relation
\begin{eqnarray} \label{4-4}
 R_{aebf}R_{cfde} &=& R_{aebf}R_{cedf} - \frac{1}{2} R_{abef}R_{efcd}.
\end{eqnarray}
But, directly plugging \eq{2-2} into Eq. \eq{2-4} leads to a different expression of the same result
\begin{eqnarray}\label{4-3'}
 R_{aebf}R_{cedf} &=& \frac{1}{12} \big( \varepsilon_{a e a_1 a_2} \varepsilon_{b f b_1 b_2} R_{cedf} R_{a_1 a_2 b_1 b_2}
  + \varepsilon_{c e a_1 a_2} \varepsilon_{d f b_1 b_2} R_{aebf} R_{a_1 a_2 b_1 b_2} \big) \nonumber \\
 && + \frac{1}{48} \varepsilon_{ae a_1 a_2} \varepsilon_{bf b_1 b_2}
 \varepsilon_{ce c_1 c_2} \varepsilon_{df d_1 d_2} R_{a_1 a_2 b_1 b_2} R_{c_1 c_2 d_1 d_2} \\
 \label{4-3''}
 &=& - \frac{1}{4} \left( \delta_{ac} R_{b e a_1 a_2}  R_{d e a_1 a_2} + \delta_{bd} R_{a e a_1 a_2}  R_{c e a_1 a_2}
 - \frac{1}{2} \delta_{ac} \delta_{bd} R_{a_1 a_2 b_1 b_2} R_{a_1 a_2 b_1 b_2} \right) \nonumber \\
 && + \frac{1}{8} \big( \varepsilon_{ae a_1 a_2} \varepsilon_{bf b_1 b_2} R_{c e d f} R_{a_1 a_2 b_1 b_2}
 + \varepsilon_{ce a_1 a_2} \varepsilon_{df b_1 b_2}  R_{a e b f} R_{a_1 a_2 b_1 b_2} \big).
\end{eqnarray}

By contracting $c$ and $d$ in Eq. (\ref{4-3}) and $b$ and $d$ in Eqs. (\ref{4-3'}) and (\ref{4-3''}), we get
\begin{eqnarray}\label{4-5}
 R_{acef}R_{bcef} &=& \frac{1}{4} \delta_{ab} R_{a_1 a_2 b_1 b_2} R_{a_1 a_2 b_1 b_2} \nonumber \\
                  &=& \frac{1}{12} \big( \varepsilon_{a c a_1 a_2} \varepsilon_{b_1 b_2 c_1 c_2} R_{bc c_1 c_2} R_{a_1 a_2 b_1 b_2}
  + \varepsilon_{bc a_1 a_2} \varepsilon_{b_1 b_2 c_1 c_2} R_{ac c_1 c_2} R_{a_1 a_2 b_1 b_2} \nonumber \\
  && \qquad + \varepsilon_{ac a_1 a_2} \varepsilon_{bc b_1 b_2} R_{a_1 a_2 c_1 c_2} R_{b_1 b_2 c_1 c_2} \big),
\end{eqnarray}
and
\begin{eqnarray}\label{4-6}
\delta_{ab} R_{a_1 a_2 b_1 b_2} R_{a_1 a_2 b_1 b_2} &=& \frac{1}{2} \big( \varepsilon_{a c a_1 a_2}
  \varepsilon_{b_1 b_2 c_1 c_2} R_{bc c_1 c_2} R_{a_1 a_2 b_1 b_2}
+ \varepsilon_{bc a_1 a_2} \varepsilon_{b_1 b_2 c_1 c_2} R_{ac c_1 c_2} R_{a_1 a_2 b_1 b_2} \big) \nonumber \\
 &=& \varepsilon_{a c a_1 a_2} \varepsilon_{bc b_1 b_2} R_{a_1 a_2 c_1 c_2} R_{b_1 b_2 c_1 c_2}.
\end{eqnarray}

We can also find algebraic relations for pseudo-tensors.
There are four types at quadratic order:
\begin{eqnarray} \label{4-7}
&& \varepsilon_{a_1 a_2 b_1 b_2} R_{ab a_1 a_2} R_{cd b_1 b_2}
  = \frac{1}{3} \big(  \varepsilon_{ab a_1 a_2} R_{cd b_1 b_2}
 + \varepsilon_{cd b_1 b_2} R_{ab a_1 a_2} \big) R_{a_1 a_2 b_1 b_2} \nonumber \\
 && \hspace{4.2cm}  + \frac{1}{12} \varepsilon_{ab a_1 a_2} \varepsilon_{cd c_1 c_2} \varepsilon_{b_1 b_2 d_1 d_2}
 R_{a_1 a_2 b_1 b_2} R_{c_1 c_2 d_1 d_2}, \xx
 && \varepsilon_{a_1 a_2 b_1 b_2} R_{ab a_1 a_2} R_{c b_1 d b_2} =
\frac{1}{2} \varepsilon_{a_1 a_2 b_1 b_2} R_{ab a_1 a_2} R_{cd b_1 b_2},   \nonumber  \\
 && \varepsilon_{a_1 a_2 b_1 b_2} R_{a a_1 b a_2} R_{c b_1 d b_2} =
\frac{1}{4} \varepsilon_{a_1 a_2 b_1 b_2} R_{ab a_1 a_2} R_{cd b_1 b_2},\\
&& \varepsilon_{a a_1 a_2 a_3} R_{b a_1 a_3 a_4} R_{c d a_2 a_4} = - \frac{1}{8} \varepsilon_{cdef} R R_{abef}
+ \frac{1}{4} \varepsilon_{a_1 a_2 b_1 b_2} R_{a b a_1 a_2} R_{c d b_1 b_2}, \nonumber \\
&& \varepsilon_{a a_1 a_2 a_3} R_{b a_1 a_3 a_4} R_{a_2 c a_4 d} =
\frac{1}{32} \varepsilon_{a_1 a_2 b_1 b_2} R_{c_1 c_2 a_1 a_2} R_{c_1 c_2 b_1 b_2}
(\delta_{ac}\delta_{bd} - \delta_{ab}\delta_{cd} + \delta_{bc}\delta_{ad} ) \nonumber \\
&& \hspace{4.2cm}  + \frac{1}{4} \big( \varepsilon_{a_1 a_2 b_1 b_2} R_{ac a_1 a_2} R_{bd b_1 b_2}
 - \frac{1}{2} \varepsilon_{bdef} R R_{efac} \big) \nonumber \\
&& \hspace{4.2cm}  + \frac{1}{16} \varepsilon_{abcd}
 \big( R_{a_1 a_2 b_1 b_2} R_{a_1 a_2 b_1 b_2} - 2 R_{a_1 a_2} R_{a_1 a_2} \big),  \nonumber 
\end{eqnarray} 
\begin{eqnarray} \label{4-8}
&& \varepsilon_{a b a_1 a_2} R_{c d b_1 b_2} R_{a_1 a_2 b_1 b_2} =
\varepsilon_{a_1 a_2 b_1 b_2} R_{a b a_1 a_2} R_{c d b_1 b_2},  \nonumber \\
&& \varepsilon_{a b a_1 a_2} R_{c a_1 b_1 b_2} R_{d b_1 a_2 b_2} =
\frac{1}{2} \varepsilon_{a b a_1 a_2} R_{c a_1 b_1 b_2} R_{d a_2 b_1 b_2}, \\
&& \varepsilon_{a b c a_1} R_{a_1 a_2 a_3 a_4} R_{d a_3 a_4 a_2} =
- \frac{1}{2} \varepsilon_{a b c a_1} R_{a_1 a_2 a_3 a_4} R_{d a_2 a_3 a_4}.  \nonumber
\end{eqnarray}
Contracting $(a, b)$ and $(c, d)$ all leads to the Hirzebruch density in \eq{quadratic-r}
except the last one which identically vanishes.

\subsection{Cubic order}

We also want to generalize the calculation to higher orders. At higher orders, many terms are involved.
So, at the cubic order, we will consider scalars and tensors with 2 free indices, for simplicity.
It is known \cite{ten-poly} that, at the cubic order, there are 8 types of scalar monomials made of
Riemann curvature tensors:
\begin{itemize}
 \item One curvature-scalar term: $R^3$,
 \item Two Ricci terms: $R R_{ab} R_{ab}, \; R_{ab} R_{bc} R_{ca}$,
 \item Five Riemann terms: $R_{ac} R_{bd} R_{abcd}, \; R R_{abcd} R_{abcd}, \;
 R_{ab} R_{acde} R_{bcde}, \; R_{abcd}R_{cdef} R_{efab}, \; R_{acbd} R_{cedf} R_{eafb}.$
\end{itemize}
Among the five Riemann terms, $R_{ab} R_{acde} R_{bcde} = \frac{1}{4} R R_{abcd} R_{abcd}$
and $R_{ac} R_{bd} R_{abcd} = \frac{1}{16} R^3$ for Einstein manifolds
which satisfy $R_{ab} = \frac{1}{4} R \delta_{ab}$.
But there is another relation, which can be found by expanding the last two terms:
\begin{eqnarray*}
&& R_{abcd}R_{cdef} R_{efab} = 64 \left( f^{i_1 i_2}_{(++)} f^{i_1 j_1}_{(++)} f^{i_2 j_1}_{(++)}
  + f^{i_1 i_2}_{(--)} f^{i_1 j_1}_{(--)} f^{i_2 j_1}_{(--)} \right),  \\
  && R_{acbd} R_{cedf} R_{ebfa} = \frac{1}{16} R^3 - 6R \left( f^{i_1 i_2}_{(++)} f^{i_1 i_2}_{(++)}
  + f^{i_1 i_2}_{(--)} f^{i_1 i_2}_{(--)} \right)
 + 16 \left( f^{i_1 i_2}_{(++)} f^{i_1 j_1}_{(++)} f^{i_2 j_1}_{(++)}
  + f^{i_1 i_2}_{(--)} f^{i_1 j_1}_{(--)} f^{i_2 j_1}_{(--)} \right).
\end{eqnarray*}
Therefore, we get the relation
\begin{equation}\label{cubic-riem4}
R_{abcd}R_{cdef} R_{abef} - 4 R_{acbd} R_{cedf} R_{ebfa} = - \frac{1}{4} R^3 + \frac{3}{2} R R_{abcd} R_{abcd},
\end{equation}
which can be written as the relation among the Riemann terms
\begin{equation}\label{cubic-riem5}
R_{abcd}R_{cdef} R_{efab} - 4R_{acbd} R_{cedf} R_{ebfa} + 4 R_{ac} R_{bd} R_{abcd}
- 6 R_{ab} R_{acde} R_{bcde} = 0.
\end{equation}
Thus only three of the five Riemann terms can remain independent \cite{ten-poly}.
Using the relation
\begin{eqnarray} \label{cubic-riem3}
&& R^3 - 12 R R_{ab} R_{ab} + 16 R_{ab} R_{bc} R_{ca} + 3 R R_{abcd} R_{abcd} + 24 R_{ac} R_{bd} R_{abcd}
- 24 R_{ab} R_{acde} R_{bcde} \nonumber \\
  &=& \frac{1}{2} R^3 - 3 R R_{abcd} R_{abcd},
\end{eqnarray}
we obtain the famous identity in the cubic order (see Eq. (3.5) in \cite{ten-poly}):
\begin{eqnarray}\label{q3-fi}
  {R_{[a b}}^{ab} {R_{cd}}^{cd} {R_{e f]}}^{ef}
 &=& \frac{1}{90} \left( R^3 - 12 R R_{ab} R_{ab} + 16 R_{ab} R_{ac} R_{bc} + 3 R R_{abcd} R_{abcd}
 + 24 R_{ac} R_{bd} R_{abcd} \right. \nonumber \\
   && \left. - 24 R_{ab} R_{acde} R_{bcde} + 2R_{abcd}R_{cdef} R_{efab} - 8R_{acbd} R_{cedf} R_{ebfa} \right) = 0,
\end{eqnarray}
where $[\cdots]$ means anti-symmetrization over all six subscripts.

D. Xu had derived other identities related to the cubic scalars \cite{dxu}:
\begin{eqnarray} \label{dxu1}
&&  R_{ab} R_{acde} R_{bcde} = \frac{1}{4} R^3 -2 R R_{ab}R_{ab} + 2 R_{ab} R_{bc} R_{ca}
+ 2 R_{ac} R_{bd} R_{abcd}  + \frac{1}{4}  R R_{abcd} R_{abcd}, \\
\label{dxu2}
&& R_{acbd} R_{cedf} R_{eafb} = - \frac{5}{8} R^3 + \frac{9}{2} R R_{ab}R_{ab} - 4 R_{ab} R_{bc} R_{ca}
-3 R_{ac} R_{bd} R_{abcd}  - \frac{3}{8} R R_{abcd} R_{abcd} \nonumber \\
&& \hspace{3.2cm} + \frac{1}{2} R_{abcd} R_{cdef} R_{efab}.
\end{eqnarray}
The first identity \eq{dxu1} can be derived from Eq. \eq{2r-weyl} by contracting $R_{ab}$
on both sides \cite{ten-poly}.
A. Harvey pointed out \cite{harvey} that the identities \eq{dxu1} and \eq{dxu2} can be derived
from the fact \cite{ten-poly}
that an $(n+1)$ index object anti-symmetrized on an $n$-dimensional manifold vanishes identically.
Indeed, Eqs. \eq{dxu1} and \eq{dxu2} can be derived from the identities, respectively,
\begin{equation}\label{harvey-id}
  \delta^p_{[a} \delta^q_b \delta^r_c \delta^s_d \delta^t_{e]} R_{abpq} R_{cdrs} R_{et} = 0, \qquad
  \delta^p_{[a} \delta^q_b \delta^r_c \delta^s_d \delta^t_{e} \delta^u_{f]} R_{abpq} R_{cdrs} R_{eftu} = 0.
\end{equation}
His argument implies that these two identities are the only linear relations existing in the cubic terms.
In fact, the identity \eq{q3-fi} can be derived by combining the identities \eq{dxu1} and \eq{dxu2} and
using the identity
\begin{equation*}
  R_{acbd} R_{cedf} R_{ebfa} = R_{acbd} R_{cedf} R_{eafb} - \frac{1}{4} R_{abcd} R_{cdef} R_{efab}.
\end{equation*}
In Appendix C, we prove that the identities \eq{dxu1} and \eq{dxu2} are the only linear relations existent
in the cubic terms for a general Riemannian manifold.
This means that the cubic scalars have six independent basis elements.\footnote{\label{parity-odd}Our approach can also be applied to pseudo-tensors. For example, we get
\begin{eqnarray*}\label{q3-1}
  && \varepsilon^{b_1 b_2 b_3 b_4} R_{a a_1 b_1 b_2} R_{b a_3 a_1 a_2}  R_{a_2 a_3 b_3 b_4}
  = - 16 \left( f^{i_1 i_3}_{(++)} f^{i_1 i_2}_{(++)} f^{i_3 i_2}_{(++)}
  - f^{i_1 i_3}_{(--)} f^{i_1 i_2}_{(--)} f^{i_3 i_2}_{(--)} \right) \delta_{ab}.
\end{eqnarray*}
However, pseudo-tensors also have as many basis elements as normal tensors, and systematically classifying them
as in Ref. \cite{ten-poly} is a new problem. We leave this issue as a problem addressed in the future. But, it would be possible to know how to generate a pseudo-scalar basis that is parity odd, as we will discuss later.}

The second-rank tensors at the cubic order (i.e., with 2 free indices) are listed in Table \ref{table:1}
which reproduces the table in Appendix C, denoted by $\mathcal{R}^2_{6,3}$, in \cite{ten-poly}. As in the scalar terms, each term can be expanded using Eq. \eq{decom-riem}  and the result has more complicated expansions compared to the scalar case.
This is because some terms vanishing in the case of a scalar are included. For example,
we have the expansion
\begin{eqnarray}
R R_{ac} R_{bc} = R \left( \frac{R^2}{16} \delta_{ab} - R f^{ij}_{(+-)} \big(\eta^i \bar{\eta}^j \big)_{ab} + 4 f^{ij}_{(+-)}f^{ij}_{(+-)} \delta_{ab}  + 4 \varepsilon^{i_1 i_2 i_3} \varepsilon^{j_1 j_2 j_3} f^{i_1 j_1}_{(+-)}f^{i_2 j_2}_{(+-)} \big(\eta^{i_3} \bar{\eta}^{j_3} \big)_{ab}
\right). \quad
\end{eqnarray}

\begin{table} [h!]
    \centering
\begin{tabular}{|l|l|}
  \hline
  \hline
  A: $R^2 R_{ab}$ &               I: $R_{ab} R_{cdef} R_{cdef}$ \\
  B: $R R_{ac} R_{bc}$ &          J: $R_{ac} R_{bedf} R_{cedf}$ \\
  C: $R_{ab} R_{cd} R_{cd}$  &    K: $R_{aecd} R_{bfcd} R_{ef}$ \\
  D: $R_{ac} R_{bd} R_{cd}$  &    L: $R_{acbd} R_{cedf} R_{ef}$  \\
  E: $R R_{acbd} R_{cd}$ &        M: $R_{acde} R_{bcdf} R_{ef}$ \\
  F: $R_{acbd} R_{ce} R_{de}$ &   N: $R_{agcd} R_{bgef} R_{cdef} $ \\
  G: $R_{ac} R_{becd} R_{de}$ &   O: $R_{aecg} R_{bfdg} R_{cdef}$ \\
  H: $R R_{aecd} R_{becd}$ &      P: $R_{aebg} R_{cdef} R_{cdfg}$  \\
  \hline
  \hline
\end{tabular}
\caption{The 16 second-rank tensors at cubic order}
    \label{table:1}
\end{table}

The expansion of second-rank cubic monomials using Eq. \eq{decom-riem} exhibits some patterns. The second-rank cubic monomials in Table \ref{table:1} are symmetric with respect to $(a \leftrightarrow b)$ except in $G$ and $J$.
It turns out that terms that would appear in the scalar case again appear with the factor $\frac{1}{4} \delta_{ab}$, and novel terms that do not appear in the scalar case (or should disappear in the scalar case) appear with the factor $(\eta^i \bar{\eta}^j)_{ab}$. Indeed,
$\delta_{ab}$ and $(\eta^i \bar{\eta}^j)_{ab}$ are the only symmetric tensors constructed from the product of 't Hooft symbols. The latter is trace-free due to Eq. \eq{eta-etabar}.
The details are listed in Appendix C.

\subsection{Quartic order}

The 26 quartic scalars are listed in Table \ref{table:2}
which reproduces the last table in Appendix B, denoted by $\mathcal{R}^0_{8,4}$, in \cite{ten-poly}.
\begin{table} [h!]
    \centering
\begin{tabular}{|l|l|}
  \hline
  \hline
  A: $R^4$ &                                 N: $R_{ab} R_{cd} R_{aecf} R_{bedf}$ \\
  B: $R^2 R_{ab} R_{ab}$ &                   O: $R R_{abcd} R_{cdef} R_{efab}$ \\
  C: $R R_{ab} R_{bc} R_{ca}$  &             P: $R R_{acbd} R_{aebf} R_{cedf}$ \\
  D: $ \big(R_{ab} R_{ab} \big)^2 $  &       Q: $R_{ab}R_{acbd} R_{efgc} R_{efgd}$  \\
  E: $R_{ab} R_{bc} R_{cd} R_{da}$ &         R: $R_{ab} R_{cdef} R_{agef} R_{bgcd}$ \\
  F: $R R_{ab} R_{cd} R_{acbd} $ &           S: $R_{ab} R_{cedf} R_{egfa} R_{gcbd} $ \\
  G: $R_{ab} R_{ce} R_{ed} R_{acbd}$ &       T: $\big( R_{abcd} R_{abcd} \big)^2 $ \\
  H: $R^2 R_{abcd} R_{abcd}$ &               U: $R_{abcd} R_{abce} R_{fghd} R_{fghe}$  \\
  I: $R R_{ab} R_{acde} R_{bcde}$ &          V: $R_{abcd} R_{cdef} R_{efgh} R_{ghab}$ \\
  J: $R_{ab} R_{ab} R_{cdef} R_{cdef}$ &     W: $R_{abcd} R_{abef} R_{cegh} R_{dfgh}$ \\
  K: $R_{ab} R_{bc} R_{defa} R_{defc} $ &     X: $R_{abcd} R_{efab} R_{gche} R_{gdhf}$ \\
  L: $R_{ab} R_{cd} R_{acef} R_{bdef}$ &     Y: $R_{acbd} R_{cedf} R_{egfh} R_{gahb}$ \\
  M: $R_{ab} R_{cd} R_{aebf} R_{cedf}$ &     Z: $R_{acbd} R_{eafb} R_{gehc} R_{fgdh}$ \\
  \hline
  \hline
\end{tabular}
\caption{The 26 quartic scalars}
    \label{table:2}
\end{table}
A. Harvey found six linear relations \cite{harvey} among the 26 monomials in Table \ref{table:2}
using the fact \cite{ten-poly} that an $(n+1)$ index object anti-symmetrized
on an $n$-dimensional manifold vanishes identically.\footnote{\label{sign-error}We found several errors in Ref. \cite{harvey}. These are corrected in Appendix D. Rectification of such errors played a crucial role in finding the correct independent basis.} Therefore he reduced the number of independent scalars shown in Table \ref{table:2} from $26$ to $20$. He pointed out that his method does not rule out
the possible existence of other identities further reducing this number.
We prove that the quartic scalars in Table \ref{table:2} only have the $13$ independent basis elements.
This result implies that there must be 13 linear relations among the quartic
monomials in Table \ref{table:2}. We explicitly find such linear relations.
Therefore we found that there are seven more linear relations in addition to the 6 relations found by Harvey \cite{harvey}. The details defer to Appendix D.
 Our method pins down that no more linear relation exists.

\subsection{Independent basis elements of quintic scalars}

Our method using the decomposition \eq{decom-riem} reveals an elegant structure for scalar basis elements. Scalar quantities must be symmetric under the parity transformation \eq{exchange}.
This enforces the expansion coefficients $f^{ij}_{(\pm\pm)}$ and $f^{ij}_{(\pm\mp)}$ to appear very symmetrical as shown in Table \ref{table:4}. In order to exhibit such a symmetry more explicitly,
let us introduce $3\times 3$ (real) matrices defined by
\begin{equation} \label{def-metab}
    (A_\pm)_{ij} \equiv f^{ij}_{(\pm \pm)}, \qquad  (B)_{ij} \equiv f^{ij}_{(+-)},
    \qquad  (B^T)_{ij} \equiv f^{ij}_{(-+)}, \qquad i, j = 1,2,3.
\end{equation}
Then the matrices $A_\pm$ are symmetric, i.e. $A_\pm^T = A_\pm$, but the matrix $B$ is
a general $3\times 3$ matrix with no explicit symmetry.
The parity transformation \eq{exchange} denoted by $P$ acts on the matrices as
\begin{equation} \label{parity-mat}
 P: A_\pm  \leftrightarrow A_\mp, \qquad  P: B \leftrightarrow B^T.
\end{equation}

It may be instructive to represent the basis elements in Table \ref{table:4} in terms of
the matrices in \eq{def-metab} to illuminate the symmetry structure enforced by the parity symmetry.
In the matrix notation, the Ricci scalar in \eq{ricci} reads
\begin{equation} \label{matrix-r}
    R = 4 \big( \textrm{Tr} A_+ + \textrm{Tr} A_- \big).
\end{equation}
It is easy to translate the Table \ref{table:4} into the matrix notation.
\begin{table} [h!]
    \centering
\begin{tabular}{|l|l|}
  \hline
  \hline
  I: $R^4$ &
  VIII: $ \left(\textrm{Tr} \big( B B^T \big) \right)^2  $ \\     
  II: $R^2 \textrm{Tr} \left( B B^T \right)$ &
  IX: $ \textrm{Tr} \big( B B^T \big) \textrm{Tr} \left( A_+^2 +  A_-^2 \right)  $ \\
  III: $R^2 \textrm{Tr} \left( A_+^2 +  A_-^2 \right) $ &
  X: $ \textrm{Tr} A_+^2 \textrm{Tr} A_+^2 + \textrm{Tr} A_-^2 \textrm{Tr} A_-^2 $ \\
  IV: $ R \textrm{Tr} \left(  A_+^3 +  A_-^3 \right)$  &
  XI: $ \textrm{Tr} \left( (B B^T)^2 \right) $  \\
  V: $ R \textrm{Tr} \left( B^T A_+ B + B A_- B^T \right)$ &
  XII: $ \textrm{Tr} \left( A_+ B A_- B^T \right) $ \\
  VI: $R  \,  \det B  $ &
  XIII: $ \textrm{Tr} \left( B^T A_+^2 B + B A_-^2 B^T \right) $ \\
  VII: $ \textrm{Tr} A_+^2 \textrm{Tr} A_-^2  $ &
  XIV: $ \textrm{Tr} \left( A_+^4 + A_-^4 \right) $ \\
  \hline
  \hline
\end{tabular}
\caption{Matrix representation of the quartic basis elements}
    \label{table:3}
\end{table}
Surprisingly, it turns out that the 14 basis elements listed in Table \ref{table:3}
or Table \ref{table:4} are not completely independent.
There exists one linear relation among the quartic basis elements given by Eq. \eq{19-exp}.
Therefore, the number of linearly independent basis elements for quartic scalars is 13.

The quintic scalars have 90 basis elements, denoted by ${\cal R }^0_{10,5}$ in
Appendix A of Ref. \cite{ten-poly}, among which the number of Weyl tensor monomials,
denoted by ${\cal C}^0_{10,5}$ in Appendix D, is 19.
But their explicit expression has not been displayed in Ref. \cite{ten-poly}.
Thus we do not know the explicit form of quintic monomials.
Nevertheless, our method gives a hint as to how many independent quintic basis elements can exist. It is necessary to count how many quintic matrix polynomials made of the matrices in \eq{def-metab} which are parity even. To do this, let us first summarize the quadratic and cubic basis elements in \eq{quadratic-b} and \eq{six-cubic}:
\begin{eqnarray}
&& Q_2 = \{ R^2, \textrm{Tr} \left( A_+^2 +  A_-^2 \right), \textrm{Tr} \big( B B^T \big) \}, \xx
&&  Q_3 = \{ R^3, R \, \textrm{Tr} \left( A_+^2 +  A_-^2 \right), R\, \textrm{Tr} \big( B B^T \big), \textrm{Tr} \left( A_+^3 + A_-^3 \right), \textrm{Tr} \left( B^T A_+ B + B A_- B^T \right),  \det B  \}. \qquad
\end{eqnarray}
Denote the basis in \eq{matrix-r} as $Q_1$ and the set of basis elements in Table \ref{table:3} as $Q_4$. Then the quintic basis can be generated by considering the tensor products $Q_1 \otimes Q_4$ and $Q_2 \otimes Q_3$ and newly generated basis $Q_5$ at the quintic order. Since $Q_1$ has the unique element $R$ in \eq{matrix-r}, it is obvious that the tensor products $Q_1 \otimes Q_4$ consists of the set in Table \ref{table:3} multiplied by $R$ which contains 13 elements. The tensor product $Q_2 \otimes Q_3$ generates new elements which are not overlapped with those in $Q_1 \otimes Q_4$. Their list is
\begin{eqnarray} \label{5=2+3}
Q_2 \otimes Q_3 &\supset& \{ \textrm{Tr} A_+^2  \textrm{Tr} A_+^3  + \textrm{Tr} A_-^2  \textrm{Tr} A_-^3,  \; \textrm{Tr} A_+^2  \textrm{Tr}  A_-^3 + \textrm{Tr} A_-^2 \textrm{Tr} A_+^3, \xx
&& \; \textrm{Tr} \left( A_+^2 + A_-^2 \right) \textrm{Tr} \left( B^T A_+ B + B A_- B^T \right), \; \textrm{Tr} B B^T \textrm{Tr} \left( A_+^3 + A_-^3 \right), \xx
&& \; \textrm{Tr} \left( A_+^2 + A_-^2 \right) \det B,  \; \textrm{Tr} B B^T \textrm{Tr} \left( B^T A_+ B + B A_- B^T \right), \; \textrm{Tr} B B^T \det B \}.
\end{eqnarray}
It is not difficult to find the newly generated basis $Q_5$ at the quintic order. They are given by\footnote{There is a subtle element: $Q_5^{\textrm{odd}} = \textrm{Tr} \left( B^T B A_+ B A_-  + B B^T A_- B^T A_+ \right)$. This is parity even in itself but the odd power of the matrix $B$.
If this combination were allowed at the quintic order, a similar term $ \textrm{Tr} \left( B^T B A_+ B  + B^T A_- B^T B \right)$ would also be allowed at the quartic order. But this term did not appear at the quartic order. So we rule out $Q_5^{\textrm{odd}}$. The result in Table \ref{table:3} alludes that a term of odd powers of $B$ is only possible in the form of $\det B$
because $\det B = \det B^T$.}
\begin{eqnarray} \label{quintic-5}
Q_5 &=& \{ \textrm{Tr} \left( A_+^5  + A_-^5 \right),  \; \textrm{Tr} \left( B^T A_+^3 B + B A_-^3 B^T \right), \;
\textrm{Tr} \left( A_+^2 B A_- B^T +  A_-^2 B A_+ B^T \right), \xx
&& \; \textrm{Tr} \left( B B^T  A_+ B B^T + B^T B A_- B^T B \right) \}.
\end{eqnarray}

\begin{table} [h!]
    \centering
\begin{tabular}{|l|l|}
  \hline
  \hline
  $S^5_1  \sim S^5_{13}: \; Q_1 \otimes Q_4 $ &
  $S^5_{19}: \; \textrm{Tr} B B^T \textrm{Tr} \left( B^T A_+ B + B A_- B^T \right)$  \\
  $S^5_{14}:\; \textrm{Tr} A_+^2  \textrm{Tr} A_+^3  + \textrm{Tr} A_-^2  \textrm{Tr} A_-^3$ &
  $S^5_{20}:\; \textrm{Tr} B B^T \det B$ \\
  $S^5_{15}: \; \textrm{Tr} A_+^2  \textrm{Tr}  A_-^3 + \textrm{Tr} A_-^2 \textrm{Tr} A_+^3$ &
  $S^5_{21}:\; \textrm{Tr}  \left( A_+^5  + A_-^5  \right) $ \\
  $S^5_{16}: \; \textrm{Tr} \left( A_+^2 + A_-^2 \right) \textrm{Tr} \left( B^T A_+ B + B A_- B^T \right)$ &
  $S^5_{22}:\;  \textrm{Tr} \left( B^T A_+^3 B + B A_-^3 B^T \right)$  \\
  $S^5_{17} : \; \textrm{Tr} B B^T \textrm{Tr} \left( A_+^3 + A_-^3 \right)$ &
  $S^5_{23}: \; \textrm{Tr} \left( A_+^2 B A_- B^T \right) + \textrm{Tr} \left( A_-^2 B A_+ B^T \right)$  \\
  $S^5_{18} : \; \textrm{Tr} \left( A_+^2 + A_-^2 \right) \det B $ &
  $S^5_{24}:\;  \textrm{Tr} \left( B B^T  A_+ B B^T + B^T B A_- B^T B \right)$ \\
  \hline
  \hline
\end{tabular}
\caption{Matrix representation of the quintic basis elements}
    \label{table:3-5}
\end{table}

The above argument implies that the quintic scalars, not explicitly known yet, have totally 24
basis elements. We summarize the quintic scalar basis in Table \ref{table:3-5}.
But, as the number of independent quartic scalars has been reduced due to the existence of the linear relation \eq{19-exp}, the number of independent basis elements can be further reduced if there exists a similar linear relation among the members in Table \ref{table:3-5}.
We claim that there is such a linear relation between the basis elements,
$S^5_{1} \sim S^5_{24}$.
The linear relation \eq{19-exp} for quartic scalars was derived from Eq. (8)
in Ref. \cite{ten-poly} and such identity can be extended to quintic order
\cite{edg-hog}. So we expect the number of linearly independent quintic scalars in Table \ref{table:3-5} will be 23.
It implies that 90 quintic scalars in Ref. \cite{ten-poly} obey totally $67=90-23$ linear relations.

Although we will stop at the quintic order, it is straightforward to extend the argument to higher orders.
A nontrivial step is to find linear relations (syzygies) between the basis elements such as \eq{19-exp} whose
existence is guaranteed by the Cayley-Hamilton theorem \cite{edg-hog,sned-2,sned-3}.
In particular, it will be interesting to find a generating function
for the basis elements of scalar monomials such as Tables \ref{table:3}
and \ref{table:3-5}.

\section{Discussion}

We have verified that the irreducible decomposition of the Riemann tensor \eq{decom-riem} provides a powerful tool
for the study of the scalar invariants of the Riemann tensor. Although we have used the decomposition \eq{decom-riem}
for convenience, the true irreducible decomposition of the Riemann tensor must refer to Eq. \eq{dec-riem}, i.e.,
\begin{equation}\label{irred-riem}
  R_{abcd} = {\widetilde f}^{ij}_{(++)} \eta^i_{ab}\eta^j_{cd} + {\widetilde f}^{ij}_{(--)} \bar{\eta}^i_{ab}\bar{\eta}^j_{cd}
  + f^{ij}_{(+-)} \big( \eta^i_{ab} \bar{\eta}^j_{cd} + \bar{\eta}^j_{ab} \eta^i_{cd} \big)
   + \frac{1}{12} R (\delta_{ac} \delta_{bd} - \delta_{ad} \delta_{bc}).
\end{equation}
This decomposition has the following correspondence with the spinor representation  and
the bivector form on the complex three-dimensional space of self-dual bivectors \cite{sned-1}:
\be \la{corres-spin-biv}
\big( {\widetilde A}_+ \big)_{ij} \leftrightarrow \Psi_{ABCD} \leftrightarrow \Psi^{ij},  \qquad
\big( {\widetilde A}_- \big)_{ij} \leftrightarrow {\overline \Psi}_{\dot{A}\dot{B}\dot{C}\dot{D}} \leftrightarrow {\overline \Psi}^{ij},
\qquad \big( B + i B^T \big)_{ij} \leftrightarrow \Phi_{A B \dot{C}\dot{D}} \leftrightarrow i \Gamma^{ij},
\ee
where $\big({\widetilde A}_\pm \big)_{ij} \equiv \big( A_\pm \big)_{ij} - \frac{1}{3} \delta_{ij} \textrm{Tr} A_\pm  = {\widetilde f}^{ij}_{(\pm \pm)}$.
The corresponding rotor quantities, $\Psi$ and ${\overline \Psi}$, are symmetric and trace-free $3 \times 3$ complex matrices
and $\Gamma$ is a $3 \times 3$ Hermitian matrix. We have observed that the Riemann polynomial reveals an elegant structure under the parity transformation \eq{exchange}. The Riemann polynomials can be classified into two classes: parity even and odd polynomials.
The parity even (odd) polynomial is an eigenstate of the parity operator $P$ with eigenvalue $+1$ ($-1$).
Schematically, it can be stated as
\be \la{parity-pol}
P f(x_1, \cdots, x_n) = \pm f(x_1, \cdots, x_n)
\ee
where $x_i$'s are Riemann monomials such as the members in Table \ref{table:1} or \ref{table:2}.
A parity odd polynomial is called a pseudo-scalar or pseudo-tensor; for example, the Hirzebruch density in \eq{quadratic-r},
pseudo-tensors in Eq. \eq{4-7} and a cubic pseudo-scalar in footnote \ref{parity-odd}.
So far we have mainly considered Riemann polynomials that are parity even.

Now we will discuss how to generate a Riemann polynomial which is parity odd, from the parity even polynomials.
Since the parity operator $P$ acts on each monomial in the polynomial \eq{parity-pol}, i.e., $P f(x_1, \cdots, x_n)
=  f(P x_1, \cdots, P x_n)$, it suffices to find independent basis elements at each order which are parity odd.
Let us define the pseudo-Riemann tensor
\be \la{pseudo-riem}
{\widetilde R}_{abcd} \equiv \frac{1}{2} \varepsilon_{cdef} R_{abef}
\ee
which can be decomposed as
\begin{equation}\label{dec-psriem}
  {\widetilde R}_{abcd} = {\widetilde f}^{ij}_{(++)} \eta^i_{ab}\eta^j_{cd} - {\widetilde f}^{ij}_{(--)} \bar{\eta}^i_{ab}\bar{\eta}^j_{cd}
  + f^{ij}_{(+-)} \big( - \eta^i_{ab} \bar{\eta}^j_{cd} + \bar{\eta}^j_{ab} \eta^i_{cd} \big)
 + \frac{1}{12} R \, \varepsilon_{abcd}.
\end{equation}
Its contraction identically vanishes, e.g., ${\widetilde R}_{ab} \equiv {\widetilde R}_{acbc} = 0$,
that is due to the Bianchi identity \eq{bianchi}. Since a parity odd Riemann polynomial has to contain only
the odd power of the pseudo-Riemann tensor \eq{pseudo-riem}, it implies that the parity odd Riemann polynomial can be
generated by flipping the sign in front of ${\widetilde A}_-$ (or $A_-$) and $B$ (but keeping the sign of $B^T$)
in a parity even Riemann polynomial.

Let us first apply this rule to the scalar basis elements. We denote the set of pseudo-scalar invariants at order $n$
as ${\widetilde Q}_n$:
\bea \la{tilde-q1}
&& {\widetilde Q}_1 = 4 \big( \textrm{Tr} A_+ - \textrm{Tr} A_- \big) = 0,
\qquad \textrm{by Eq. \eq{pseudo-r}}, \\
\la{tilde-q2}
&& {\widetilde Q}_2 = \textrm{Tr} \left( A_+^2 -  A_-^2 \right), \\
\la{tilde-q3}
&& {\widetilde Q}_3 = \{ R \, \textrm{Tr} \left( A_+^2 - A_-^2 \right), \textrm{Tr} \left( A_+^3 - A_-^3 \right),
\textrm{Tr} \left( B^T A_+ B - B A_- B^T \right) \}, \\
\la{tilde-q4}
&& {\widetilde Q}_4 = \{ \widetilde{III}, \widetilde{IV}, \widetilde{V}, \widetilde{IX}, \widetilde{X},
\widetilde{XIII}, \widetilde{XIV} \} = R  {\widetilde Q}_3 \, \bigcup \, \{ \widetilde{IX}, \widetilde{X},
\widetilde{XIII}, \widetilde{XIV} \},
\eea
where the meaning of tildes in the set ${\widetilde Q}_4$ refers to a pseudo-scalar invariant obtained
by applying the above rule to the corresponding term in Table \ref{table:3}.
${\widetilde Q}_2$ corresponds to the Hirzebruch density in \eq{quadratic-r}.
The same rule can be applied to the quintic scalar invariants in Table \ref{table:3-5}.
In particular, $S^5_{16}$ generates two pseudo-scalar invariants,
$$\textrm{Tr} \left( A_+^2 - A_-^2 \right) \textrm{Tr} \left( B^T A_+ B + B A_- B^T \right), \quad
\textrm{Tr} \left( A_+^2 + A_-^2 \right) \textrm{Tr} \left( B^T A_+ B - B A_- B^T \right).$$
It is obvious that these sets thus obtained are odd under the parity transformation \eq{exchange}.
In other words, they are pseudo-scalar invariants.

Let us count the number of independent (pseudo-)scalar invariants after including the pseudo-scalars in ${\widetilde Q}_n$.
Fortunately the counting at lower orders is easy.
A notable point is that the parity even and odd polynomials are linearly independent of each other at the same order.
However, for higher-order terms, one should note the fact: $\textrm{odd} \times \textrm{odd} = \textrm{even}$ and
$\textrm{even} \times \textrm{odd} = \textrm{odd}$.
At the quadratic order, the total number of independent basis elements is now four.
Since the three elements in ${\widetilde Q}_3$ are independent of each other, the total number at the cubic order becomes
$9 = 6 + 3$. At the quartic order, new elements would be generated from the products, $Q_1 \otimes {\widetilde Q}_3$ and
${\widetilde Q}_2 \otimes {\widetilde Q}_2$. The elements from $Q_1 \otimes {\widetilde Q}_3$ are already included
in ${\widetilde Q}_4$ and the product ${\widetilde Q}_2 \otimes {\widetilde Q}_2$ does not generate new elements
because ${\widetilde Q}_2^2 = X - 2 VII$. But there is an interesting syzygy connecting the elements in ${\widetilde Q}_4$.
From the fact that Eq. \eq{19-exp} comes from a quartic polynomial of the Weyl tensor and the term $R^4$ cannot
be generated from it since ${\widetilde Q}_1 = 0$, we can infer a pseudo-scalar version of Eq. \eq{19-exp}:\footnote{Therefore, Eqs. \eq{19-exp} and \eq{odd4-0} can be rewritten as the identity
\begin{equation*}
\frac{R^4}{32} - 12 R^2 \textrm{Tr} A_\pm^2 +
128 R  \textrm{Tr} A_\pm^3 - 768  \textrm{Tr} A_\pm^4
+ 384 \textrm{Tr} \big(  A_\pm^2 \big) \textrm{Tr} \big(  A_\pm^2 \big) = 0
\end{equation*} for any symmetric $3 \times 3$ matrix $A_\pm$ with $R = 8 \textrm{Tr} A_+ = 8 \textrm{Tr} A_-.$}
\bea \la{odd4-0}
&& - 3 R^2 \textrm{Tr} \left(  A_+^2 -  A_-^2 \right) +
32 R  \textrm{Tr} \left( A_+^3 - A_-^3 \right) - 192  \textrm{Tr} \left( A_+^4  - A_-^4 \right)  \xx
&& \quad + 96 \left( \textrm{Tr} \big(  A_+^2 \big) \textrm{Tr} \big(  A_+^2 \big)
- \textrm{Tr} \big(  A_-^2 \big) \textrm{Tr} \big(  A_-^2 \big) \right) = 0.
\eea
Again it can be most easily checked in a diagonalized frame such that
$A_\pm = \textrm{diag} (a^1_\pm, a^2_\pm, a^3_\pm)$ and $a^1_+ + a^2_+ + a^3_+
= a^1_- + a^2_- + a^3_-$. Since Eq. \eq{odd4-0} provides a linear relation among the pseudo-scalar invariants in
the set ${\widetilde Q}_4$, the number of linearly independent basis elements
in the set ${\widetilde Q}_4$ is 6. Thus the total number of independent quartic (pseudo-)scalar invariants
is $19 = 13 + 6$.

For the quintic order, the situation becomes a bit complicated.
Besides the 23 independent quintic scalar invariants in Table \ref{table:3-5}, it is necessary to include
the new invariants from the products,  $Q_1 \otimes {\widetilde Q}_4, \; Q_2 \otimes {\widetilde Q}_3, \;
Q_3 \otimes {\widetilde Q}_2, \; {\widetilde Q}_2 \otimes {\widetilde Q}_3$ and ${\widetilde Q}_5^W$ where the last one is the set of quintic Weyl monomials.
It is not difficult to determine newly generated basis elements; 6 elements from $Q_1 \otimes {\widetilde Q}_4$, 7 elements from
$Q_2 \otimes {\widetilde Q}_3$ and $Q_3 \otimes {\widetilde Q}_2$ and  4 elements from ${\widetilde Q}_5^W$ while
one even element from ${\widetilde Q}_2 \otimes {\widetilde Q}_3$,
$\textrm{Tr} \left( A_+^2 - A_-^2 \right) \textrm{Tr} \left( B^T A_+ B - B A_- B^T \right)$.
Hence 17 odd elements and 1 even elememt are newly generated after including the pseudo-scalars in ${\widetilde Q}_n$.
However, the 17 odd elements may not be completely independent since it is expected that the quintic order will also give
rise to an identity similar to Eq. \eq{odd4-0} according to the Caley-Hamilton theorem \cite{edg-hog,sned-2,sned-3}.
Therefore we claim that there are totally 40 = 23 + 17 linearly independent quintic (pseudo-)scalar invariants.

It was shown \cite{ten-poly} that the quintic scalars have 90 basis elements,
among which the number of Weyl tensor monomials is 19.\footnote{Note that the result in \cite{ten-poly}
was missing the parity odd elements. We have discussed above how these can be restored.}
But we do not know the explicit form of quintic monomials since their explicit expression has not been displayed in Ref. \cite{ten-poly}.
Since we have identified 23 linearly independent quintic scalar invariants,
we may try to construct 90 basis elements from the 23 linearly independent basis elements.
The idea is to use Eq. \eq{2-2} to invert the quintic basis elements expressed in terms of the 't Hooft symbols $f^{ij}_{(\pm\pm)}$ and $f^{ij}_{(\pm\mp)}$ into homogeneous polynomials of Riemann tensors, as was done in section 4.1. To do it, it is necessary to use several identities of the 't Hooft symbols in Appendix A. Of course, a computer would be a great help
in doing this. If the 23 basis elements exhaust all 90 quintic scalars, our conjecture would be confirmed.
It will be more interesting if the parity odd elements discussed above are included.
We look forward to reporting some results of studies in this direction in the near future.

It has been well known \cite{penrose,weinberg,haskins} that, in general, there are 14 (algebraically) independent, second-order,
invariants formed from the Riemann tensor $R_{abcd}$. However, it does not provide any guidance as to how these invariants
may be constructed. An important problem is how much algebraic information that is in the Riemann tensor can be encoded
in its polynomial invariants. An ``algebraically complete" set means that it must consist of invariants of the lowest
possible degree, and also it must be the smallest set which contains the maximum number of independent invariants
for any Petrov or Segre type. It was realized \cite{carmal,sned-1,mcin} that an algebraically complete set should contain
more than 14 invariants and hence, in general, be redundant.
It was shown in \cite{zcm-algcom1} that the set of 16 invariants proposed in \cite{carmal} is not algebraically complete and
even the set of 17 invariants proposed in \cite{mcin} is missing an essential element although it is algebraically complete.
Hence it was suggested in \cite{zcm-algcom1,zcm-algcom2,zcm-algcom3} that an algebraically complete set needs at most the equivalent
of 18 real invariants with a maximum overall degree of 6. It will be interesting to have the explicit expression of these invariants in terms of the irreducible decomposition \eq{decom-riem} or \eq{irred-riem}, which may be useful to determine
whether the set of 18 invariants proposed in \cite{zcm-algcom1} is independent or not.
Any progress in this direction will be reported.

Our approach may be generalized  to the case with covariant derivatives.
The covariant derivative of the Riemann curvature tensor is defined by
\be \la{cov-riem}
\nabla_\mu R_{\nu \lambda a b} = \partial_\mu R_{\nu \lambda a b}
- \Gamma^\rho_{\mu\nu} R_{\rho \lambda a b} - \Gamma^\rho_{\mu \lambda} R_{\nu \rho a b}
+ \omega_{ac \mu}  R_{\nu \lambda c b} -  R_{\nu \lambda a c} \omega_{c b \mu}.
\ee
The spin connection can be split into a pair of $SU(2)_+$ and $SU(2)_-$ gauge fields according
to the Lie algebra splitting \eq{lie-so4} \cite{oh-yang,yang-col1,yang-col4}:
\be \la{split-spin}
\omega_{ab \mu} = A^{(+)i}_\mu \eta^i_{ab} + A^{(-)i}_\mu \bar{\eta}^i_{ab}.
\ee
Then the covariant derivative \eq{cov-riem} can be decomposed as
\begin{equation}\label{decom-coriem}
\nabla_\mu   R_{abcd} = \big( \nabla_\mu f^{ij}_{(++)} \big) \eta^i_{ab}\eta^j_{cd}
+ \big( \nabla_\mu  f^{ij}_{(+-)} \big) \eta^i_{ab} \bar{\eta}^j_{cd}
+ \big( \nabla_\mu f^{ij}_{(-+)} \big) \bar{\eta}^i_{ab} \eta^j_{cd}
+ \big( \nabla_\mu  f^{ij}_{(--)} \big) \bar{\eta}^i_{ab}\bar{\eta}^j_{cd},
\end{equation}
where the covariant derivatives are given by
\bea \la{cov-dff}
&&  \nabla_\mu f^{ij}_{(\pm\pm)} = \partial_\mu f^{ij}_{(\pm\pm)}
- 2 \left( \varepsilon^{ikl} A^{(\pm) k}_\mu f^{lj}_{(\pm\pm)}
+ \varepsilon^{jkl} A^{(\pm) k}_\mu f^{il}_{(\pm\pm)}  \right), \xx
&&  \nabla_\mu f^{ij}_{(\pm\mp)} = \partial_\mu f^{ij}_{(\pm\mp)}
- 2 \left( \varepsilon^{ikl} A^{(\pm) k}_\mu f^{lj}_{(\pm\mp)}
+ \varepsilon^{jkl} A^{(\mp) k}_\mu f^{il}_{(\pm\mp)}  \right).
\eea
The second Bianchi identity takes an interesting form:
\bea \la{2nd-bianchi}
&& \nabla_e   R_{abcd} + \nabla_a   R_{becd} + \nabla_b   R_{eacd} = 0  \xx
& \Leftrightarrow & \eta^i_{ab} \nabla_b f^{ij}_{(++)} = \bar{\eta}^i_{ab} \nabla_b f^{ij}_{(-+)}, \quad
\bar{\eta}^i_{ab} \nabla_b f^{ij}_{(--)} = \eta^i_{ab} \nabla_b f^{ij}_{(+-)}.
\eea
One can see that the decomposition of the covariant derivative \eq{decom-coriem} has essentially the same form as that of
the curvature tensor \eq{decom-riem} except that the coefficients are replaced by the covariant derivatives \eq{cov-dff}.
Therefore incorporating the covariant derivative of the curvature tensor into a Riemann polynomial would not entail great difficulties.

Although the signature of the metric is irrelevant for the linear dependence of scalar invariants of the Riemann tensor,
its dimensionality becomes important. For example, in the three-dimensional space, the Weyl tensor identically vanishes.
As a result, there is a linear relation quadratic in the Riemann tensor:
$$ R_{abcd}^2 - 4 R_{ab}^2 + R^2 = 0.$$
Moreover, the number of independent differential invariants crucially depends on the dimension of space.
In $d$-dimensions, it is kwown to be \cite{weinberg}
$$ I(d) \equiv \frac{d^2 (d^2 -1)}{12} - \left(
  \begin{array}{c}
    d \\
    2 \\
  \end{array}
\right) = \frac{d+3}{2} \left(
  \begin{array}{c}
    d \\
    3 \\
  \end{array}
\right) $$
and this number becomes $I(4) = 14, \; I(5) = 40$ and $I(6) = 90$ in four, five and six dimensions, respectively.
It was shown \cite{jack-parker} that, in six dimensions, there are eight scalar invariants cubic in the Riemann tensor
(the list is exactly the same as the four-dimensional case listed in section 4.2) but there is no linear relation
between these invariants unlike the four-dimensional case. The maximum dimension of space in which linear relation can exist
among invariants of order $n$ in the Riemann tensor is given by $d_{max} = 2n-1$ \cite{jack-parker}.
Thus it is expected that there will be linear relations for quartic $(n=4)$ scalar invariants in six dimensions.
It will be interesting to understand more closely the algebraic structure of scalar invariants in six dimensions.

In six dimensions, there also exists a global isomorphism between a six-dimensional Lorentz group and
a classical Lie group:
\be \la{so6-su4}
Spin (6) \cong SU(4).
\ee
Therefore it is possible to devise a six-dimensional version of the 't Hooft symbols using the isomorphism between
the chiral $so(6)$ Lorentz algebra and the $su(4)$ Lie algebra \cite{ya-yu}. Since the chiral and anti-chiral (or
the fundamental and anti-fundamental) representations of $so(6)$ (or $su(4)$) must be distinct, there are two kinds
of 't Hooft symbols, $\eta^{a}_{AB}$ and $\bar{\eta}^{a}_{AB}$, with $a =1, \cdots, 15, \; A, B = 1, \cdots, 6$,
and they satisfy algebraic identities similar to the four-dimensional case. The six-dimensional Riemann curvature tensor
can be classified into two classes:
\bea \la{6a-riem}
&& \mathbb{A}: R^{(+)}_{ABCD} = f^{ab}_{(++)} \eta^{a}_{AB} \eta^{b}_{CD}, \\
&& \mathbb{B}: R^{(-)}_{ABCD} = f^{ab}_{(--)} \bar{\eta}^{a}_{AB} \bar{\eta}^{b}_{CD}.
\eea
This expression may be very useful for studying the algebraic properties of six-dimensional scalar invariants.
In particular, it was argued \cite{ya-yu} that the existence of two classes and their splitting into $\mathbb{A}$ and $\mathbb{B}$
would be related to the mirror symmetry of Calabi-Yau manifolds. It will be interesting to find any relationship between
the scalar invariants of each class. We hope to address this issue in the near future too.



\section*{Acknowledgments}

This research was performed using Mathematica (www.wolfram.com) and the add-on package MathSymbolica (www.mathsymbolica.com).
This work was supported by the National Research Foundation of Korea (NRF)
with grant number NRF-2018R1D1A1B0705011314 (HSY).
We acknowledge the hospitality at APCTP where part of this work was done.

\newpage

\appendix

\section{'t Hooft symbols}

The explicit components of the 't Hooft symbols $\eta^i_{ab}$ and ${\overline \eta}^i_{ab}$
for $i = 1,2,3$ are given by
\begin{eqnarray} \label{tHooft-symbol}
\begin{array}{l}
{\eta}^i_{ab} = {\varepsilon}^{i4ab} + \delta^{ia}\delta^{4b}
- \delta^{ib}\delta^{4a}, \\
{\overline \eta}^{i}_{ab} = {\varepsilon}^{i4ab} - \delta^{ia}\delta^{4b}
+ \delta^{ib}\delta^{4a}
\end{array}
\end{eqnarray}
with ${\varepsilon}^{1234} = 1$. They satisfy the following relations
\begin{eqnarray} \label{self-eta}
&& \eta^{(\pm)i}_{ab} = \pm \frac{1}{2} {\varepsilon_{ab}}^{cd}
\eta^{(\pm)i}_{cd}, \\
\label{proj-eta}
&& \eta^{(\pm)i}_{ab} \eta^{(\pm)i}_{cd} =
\delta_{ac}\delta_{bd}
-\delta_{ad}\delta_{bc} \pm \varepsilon_{abcd}, \\
\label{self-eigen}
&& \varepsilon_{abcd} \eta^{(\pm)i}_{de} = \mp (
\delta_{ec} \eta^{(\pm)i}_{ab} + \delta_{ea} \eta^{(\pm)i}_{bc} -
\delta_{eb} \eta^{(\pm)i}_{ac} ), \\
\label{eta-etabar}
&& \eta^{(\pm)i}_{ab} \eta^{(\mp)j}_{ab}=0, \\
\label{eta^2}
&& \eta^{(\pm)i}_{ac}\eta^{(\pm)j}_{bc} =\delta^{ij}\delta_{ab} +
\varepsilon^{ijk}\eta^{(\pm)k}_{ab}, \\
\label{eta-ex}
&& \eta^{(\pm)i}_{ac}\eta^{(\mp)j}_{bc} =
\eta^{(\pm)i}_{bc}\eta^{(\mp)j}_{ac}, \\
\label{eta-o4-algebra}
&& \varepsilon^{ijk} \eta^{(\pm)j}_{ab} \eta^{(\pm)k}_{cd} =
    \delta_{ac} \eta^{(\pm)i}_{bd} - \delta_{ad} \eta^{(\pm)i}_{bc}
    - \delta_{bc} \eta^{(\pm)i}_{ad} + \delta_{bd} \eta^{(\pm)i}_{ac},
\end{eqnarray}
where $\eta^{(+)i}_{ab} \equiv \eta^i_{ab}$ and $\eta^{(-)i}_{ab} \equiv {\overline
\eta}^i_{ab}$.

If we introduce two families of $4 \times 4$ matrices defined by
\begin{equation} \label{thooft-matrix}
[\tau^i_+]_{ab} \equiv \frac{1}{2} \eta^i_{ab}, \qquad
[\tau^i_-]_{ab} \equiv \frac{1}{2} {\overline \eta}^i_{ab},
\end{equation}
the matrices in (\ref{thooft-matrix}) provide two independent spin $s=\frac{3}{2}$ representations of $su(2)$ Lie algebra.
Explicitly, they are given by
\begin{eqnarray*} \label{t+}
&& \tau^{1}_+ = \frac{1}{2}
\begin{pmatrix}
      0 & 0 & 0 & 1 \\
      0 & 0 & 1 & 0 \\
      0 & -1 & 0 & 0 \\
      -1 & 0 & 0 & 0 \\
             \end{pmatrix}, \;\;
  \tau^{2}_+ = \frac{1}{2} \begin{pmatrix}
      0 & 0 & -1 & 0 \\
      0 & 0 & 0 & 1 \\
      1 & 0 & 0 & 0 \\
      0 & -1 & 0 & 0 \\
    \end{pmatrix}, \;\;
  \tau^{3}_+ = \frac{1}{2} \begin{pmatrix}
      0 & 1 & 0 & 0 \\
      -1 & 0 & 0 & 0 \\
      0 & 0 & 0 & 1 \\
      0 & 0 & -1 & 0 \\
    \end{pmatrix}, \\
\label{t-}
&& \tau^{1}_- = \frac{1}{2} \begin{pmatrix}
      0 & 0 & 0 & -1 \\
      0 & 0 & 1 & 0 \\
      0 & -1 & 0 & 0 \\
      1 & 0 & 0 & 0 \\
    \end{pmatrix}, \;\;
  \tau^{2}_- = \frac{1}{2} \begin{pmatrix}
      0 & 0 & -1 & 0 \\
      0 & 0 & 0 & -1 \\
      1 & 0 & 0 & 0 \\
      0 & 1 & 0 & 0 \\
    \end{pmatrix}, \;\;
  \tau^{3}_- = \frac{1}{2} \begin{pmatrix}
      0 & 1 & 0 & 0 \\
      -1 & 0 & 0 & 0 \\
      0 & 0 & 0 & -1 \\
      0 & 0 & 1 & 0 \\
    \end{pmatrix}
\end{eqnarray*}
according to the definition (\ref{tHooft-symbol}).
The relations in (\ref{eta^2}) and (\ref{eta-ex}) immediately show that
$\tau^i_\pm$ satisfy $su(2)$ Lie algebras, i.e.,
\begin{equation} \label{thooft-su2}
[\tau^i_\pm, \tau^j_\pm] = - \varepsilon^{ijk} \tau^k_\pm,
\qquad [\tau^i_\pm, \tau^j_\mp] = 0.
\end{equation}

\section{Don't panic!}

Here we will analytically prove the identities \eq{dif-5} and \eq{dif-4} using the decomposition (\ref{em-riem}).
In order to simplify the calculation, it is convenient to choose pairs with the most contractions
and perform the calculation with them.
So we will show (\ref{dif-5}) by choosing the following pair:
\begin{eqnarray}\label{cal-dif-5}
  && R_{a_1 a_2 b_1 b_2} R_{a_2 b_1 b_2 e_1} \big( R_{c_1 c_2 d_1 d_2} R_{c_2 d_1 d_2 e_2} \big) R_{e_1 a_1 e_2 c_1} \nonumber \\
  &=&  R_{a_1 a_2 b_1 b_2} R_{a_2 b_1 b_2 e_1} \big( f^{i_1 i_2}_{(++)} \eta^{i_1}_{c_1 c_2} \eta^{i_2}_{d_1 d_2}
  + f^{i_1 i_2}_{(--)} \bar{\eta}^{i_1}_{c_1 c_2} \bar{\eta}^{i_2}_{d_1 d_2} \big)
  \big( f^{j_1 j_2}_{(++)} \eta^{j_1}_{c_2 d_1} \eta^{j_2}_{d_2 e_2}
  + f^{j_1 j_2}_{(--)} \bar{\eta}^{j_1}_{c_2 d_1} \bar{\eta}^{j_2}_{d_2 e_2} \big) R_{e_1 a_1 e_2 c_1} \nonumber \\
&=& R_{a_1 a_2 b_1 b_2} R_{a_2 b_1 b_2 e_1} \left( \big( \delta^{i_1 j_1} \delta_{c_1 d_1}
+ \varepsilon^{i_1 j_1 k_1} \eta^{k_1}_{c_1 d_1} \big)
\big( \delta^{i_2 j_2} \delta_{d_1 e_2}
+ \varepsilon^{i_2 j_2 k_2} \bar{\eta}^{k_2}_{d_1 e_2} \big) f^{i_1 i_2}_{(++)} f^{j_1 j_2}_{(++)} \right. \nonumber \\
&& \hspace{2.5cm} + \big( \delta^{i_1 j_1} \delta_{c_1 d_1}
+ \varepsilon^{i_1 j_1 k_1} \bar{\eta}^{k_1}_{c_1 d_1} \big)
\big( \delta^{i_2 j_2} \delta_{d_1 e_2}
+ \varepsilon^{i_2 j_2 k_2} \bar{\eta}^{k_2}_{d_1 e_2} \big) f^{i_1 i_2}_{(--)} f^{j_1 j_2}_{(--)} \nonumber \\
&& \hspace{2.5cm} \left. + [\eta^{i_1} \bar{\eta}^{j_1} \eta^{i_2} \bar{\eta}^{j_2}]_{c_1 e_2}
f^{i_1 i_2}_{(++)} f^{j_1 j_2}_{(--)}
+ [\bar{\eta}^{i_1} \eta^{j_1} \bar{\eta}^{i_2} \eta^{j_2}]_{c_1 e_2}
f^{i_1 i_2}_{(--)} f^{j_1 j_2}_{(++)} \right) R_{e_1 a_1 e_2 c_1}.
\end{eqnarray}
Note that
\begin{eqnarray*}
&& [\eta^{i_1} \bar{\eta}^{j_1} \eta^{i_2} \bar{\eta}^{j_2}]_{c_1 e_2}
=  [\eta^{i_1} \eta^{i_2} \bar{\eta}^{j_1} \bar{\eta}^{j_2}]_{c_1 e_2}
= [(\delta^{i_1 i_2} \mathbf{1} + \varepsilon^{i_1 i_2 i_3} \eta^{i_3})
(\delta^{j_1 j_2} \mathbf{1} + \varepsilon^{j_1 j_2 j_3} \bar{\eta}^{j_3} ) ]_{c_1 e_2} \\
&=& [ \delta^{i_1 i_2} \delta^{j_1 j_2} \mathbf{1} + \delta^{j_1 j_2} \varepsilon^{i_1 i_2 i_3} \eta^{i_3}
 + \delta^{i_1 i_2} \varepsilon^{j_1 j_2 j_3} \bar{\eta}^{j_3} + \varepsilon^{i_1 i_2 i_3} \varepsilon^{j_1 j_2 j_3}
 \eta^{i_3} \bar{\eta}^{j_3}  ]_{c_1 e_2}
\end{eqnarray*}
and a similar formula holds for the matrix $[\bar{\eta}^{i_1} \eta^{j_1} \bar{\eta}^{i_2} \eta^{j_2}]_{c_1 e_2}$.
Now it is easy to see that Eq. (\ref{cal-dif-5}) identically vanishes
when recalling Eq. (\ref{symm-fij}) and the fact that $[\eta^{i} \bar{\eta}^{j} ]_{c_1 e_2}$ is
a symmetric matrix by Eq. (\ref{eta-ex}). Although it is straightforward, the proof of Eq. (\ref{dif-4})
by hand requires a little algebra:
\begin{eqnarray} \label{dif-easy}
&&  2 R_{a_1 a_2 b_1 b_2} R_{c_1 c_2 a_1 d_1} R_{d_2 b_1 d_1 a_2} R_{b_2 d_2 c_1 c_2} =  \nonumber \\
&& \qquad - 2\cdot 4^3 \left( f^{i_1 i_2}_{(++)} f^{i_1 j_1}_{(++)} f^{k_1 k_2}_{(++)} f^{l_1 l_2}_{(++)}
+ f^{i_1 i_2}_{(--)} f^{i_1 j_1}_{(--)} f^{k_1 k_2}_{(--)} f^{l_1 l_2}_{(--)} \right)
\varepsilon^{j_1 l_1 k_1} \varepsilon^{i_2 l_2 k_2}, \nonumber \\
&& - R_{a_1 a_2 b_1 b_2} R_{c_1 c_2 d_1 a_1} R_{d_2 a_2 c_2 c_1} R_{b_1 b_2 d_1 d_2} =  \nonumber \\
&& \qquad - 4^3 \left( f^{i_1 i_2}_{(++)} f^{i_1 i_2}_{(++)} -
f^{i_1 i_2}_{(--)} f^{i_1 i_2}_{(--)} \right)^2  + 2\cdot 4^3
\left( f^{i_1 i_2}_{(++)} f^{i_1 j_2}_{(++)} f^{j_1 i_2}_{(++)} f^{j_1 j_2}_{(++)}
+ f^{i_1 i_2}_{(--)} f^{i_1 j_2}_{(--)} f^{j_1 i_2}_{(--)} f^{j_1 j_2}_{(--)} \right). \nonumber
\end{eqnarray}
It is convenient to arrange the remaining two terms using the identity (\ref{bianchi}):
\begin{eqnarray} \label{dif-easy2}
&&  4 R_{a_1 a_2 b_1 b_2} R_{c_1 c_2 d_1 a_1} R_{d_2 a_2 b_1 c_1}
\big( R_{d_1 b_2 d_2 c_2} + R_{d_2 d_1 b_2 c_2} \big)
= 4 R_{a_1 a_2 b_1 b_2} R_{c_1 c_2 d_1 a_1} R_{d_2 a_2 b_1 c_1} R_{b_2 d_2 c_2 d_1} \nonumber \\
&=& 4^3 \left( f^{i_1 i_2}_{(++)} f^{i_1 j_1}_{(++)} f^{k_1 k_2}_{(++)} f^{l_1 l_2}_{(++)}
+ f^{i_1 i_2}_{(--)} f^{i_1 j_1}_{(--)} f^{k_1 k_2}_{(--)} f^{l_1 l_2}_{(--)} \right)
\varepsilon^{j_1 l_1 k_1} \varepsilon^{i_2 l_2 k_2} \nonumber \\
&&  - 2\cdot 4^3  f^{i_1 i_2}_{(++)} f^{i_1 i_2}_{(++)} f^{j_1 j_2}_{(--)} f^{j_1 j_2}_{(--)}
 + 4^3 \left( f^{i_1 i_2}_{(++)} f^{i_1 i_2}_{(++)} + f^{j_1 j_2}_{(--)} f^{j_1 j_2}_{(--)} \right)
 \left( f^{k_1 k_2}_{(++)} \delta^{k_1 k_2} \right)^2 \nonumber \\
&& - 2\cdot 4^3 \left( f^{i_1 i_2}_{(++)} f^{j_1 i_2}_{(++)} f^{i_1 j_1}_{(++)}
  + f^{i_1 i_2}_{(--)} f^{j_1 i_2}_{(--)} f^{i_1 j_1}_{(--)} \right) f^{k_1 k_2}_{(++)} \delta^{k_1 k_2},
\end{eqnarray}
where we have used the identity (\ref{1-more}).
Combining all these terms leads to the identity (\ref{dif-4}).

\section{Cubic identities for a general Riemannian manifold}

The identity \eq{dxu1} becomes trivial for an Einstein manifold which satisfies the equation
$R_{ab} = \frac{1}{4} R \delta_{ab}$ and the identity \eq{dxu2} reduces to
\begin{equation*}
   R_{acbd} R_{cedf} R_{eafb} = \frac{1}{16} R^3 - \frac{3}{8} R R_{abcd} R_{abcd} + \frac{1}{2} R_{abcd} R_{cdef} R_{efab}.
\end{equation*}
Therefore, in order to prove that the identities \eq{dxu1} and \eq{dxu2} are the only linear relations existent
in the cubic terms, we will consider a general Riemannian manifold and represent the eight cubic monomials using
the decomposition \eq{decom-riem}:
\begin{eqnarray} \label{cubic-exp}
&& R R_{ab} R_{ab} = \frac{R^3}{4} + 16 R f^{ij}_{(+-)}f^{ij}_{(+-)}, \nonumber \\
&& R_{ab} R_{bc} R_{ca} = \frac{R^3}{16} + 12 R f^{ij}_{(+-)}f^{ij}_{(+-)}
- 32 \left( 2 f^{i_1 i_2}_{(+-)} f^{i_2 i_3}_{(+-)} f^{i_3 i_1}_{(+-)}
- 3 f^{i_1 i_1}_{(+-)} f^{i_2 i_3}_{(+-)} f^{i_3 i_2}_{(+-)}
+ f^{i_1 i_1}_{(+-)} f^{i_2 i_2}_{(+-)} f^{i_3 i_3}_{(+-)} \right), \nonumber \\
&& R_{ac} R_{bd} R_{abcd} = \frac{R^3}{16} + 4 R f^{ij}_{(+-)}f^{ij}_{(+-)}
+ 32 \left( 2 f^{i_1 i_2}_{(+-)} f^{i_2 i_3}_{(+-)} f^{i_3 i_1}_{(+-)}
- 3 f^{i_1 i_1}_{(+-)} f^{i_2 i_3}_{(+-)} f^{i_3 i_2}_{(+-)}
+ f^{i_1 i_1}_{(+-)} f^{i_2 i_2}_{(+-)} f^{i_3 i_3}_{(+-)} \right)  \nonumber \\
&& \hspace{2.5cm} + 32 \left( f^{i_1 i_2}_{(++)} f^{i_1 i_3}_{(+-)} f^{i_2 i_3}_{(+-)}
+ f^{i_1 i_2}_{(--)} f^{i_3 i_1}_{(+-)} f^{i_3 i_2}_{(+-)} \right),  \nonumber \\
&& R R_{abcd} R_{abcd} = 16 R \Big(f^{ij}_{(++)} f^{ij}_{(++)} + 2 f^{ij}_{(+-)}f^{ij}_{(+-)}
  + f^{ij}_{(--)}f^{ij}_{(--)} \Big),     \\
&& R_{ab} R_{acde} R_{bcde} =  4 R \Big(f^{ij}_{(++)} f^{ij}_{(++)} + 2 f^{ij}_{(+-)}f^{ij}_{(+-)}
  + f^{ij}_{(--)}f^{ij}_{(--)} \Big) \nonumber \\
&& \hspace{2.8cm}  + 64 \left( f^{i_1 i_2}_{(++)} f^{i_1 i_3}_{(+-)} f^{i_2 i_3}_{(+-)}
+ f^{i_1 i_2}_{(--)} f^{i_3 i_1}_{(+-)} f^{i_3 i_2}_{(+-)} \right),     \nonumber \\
&& R_{abcd}R_{cdef} R_{efab} = 192 \left( f^{i_1 i_2}_{(++)} f^{i_1 i_3}_{(+-)} f^{i_2 i_3}_{(+-)}
+ f^{i_1 i_2}_{(--)} f^{i_3 i_1}_{(+-)} f^{i_3 i_2}_{(+-)} \right)   \nonumber \\
&& \hspace{3.2cm} + 64 \left(  f^{i_1 i_2}_{(++)} f^{i_1 i_3}_{(++)} f^{i_2 i_3}_{(++)}
+ f^{i_1 i_2}_{(--)} f^{i_1 i_3}_{(--)} f^{i_2 i_3}_{(--)} \right),     \nonumber \\
&& R_{acbd} R_{cedf} R_{eafb} = \frac{R^3}{16}  - 6 R \Big(f^{ij}_{(++)} f^{ij}_{(++)}
+ f^{ij}_{(--)}f^{ij}_{(--)} \Big)  + 32 \left(  f^{i_1 i_2}_{(++)} f^{i_1 i_3}_{(++)} f^{i_2 i_3}_{(++)}
+ f^{i_1 i_2}_{(--)} f^{i_1 i_3}_{(--)} f^{i_2 i_3}_{(--)} \right) \nonumber \\
&& \hspace{3.1cm} + 32 \left( 2 f^{i_1 i_2}_{(+-)} f^{i_2 i_3}_{(+-)} f^{i_3 i_1}_{(+-)}
- 3 f^{i_1 i_1}_{(+-)} f^{i_2 i_3}_{(+-)} f^{i_3 i_2}_{(+-)}
+ f^{i_1 i_1}_{(+-)} f^{i_2 i_2}_{(+-)} f^{i_3 i_3}_{(+-)} \right). \nonumber
\end{eqnarray}
One can see that the above 7 monomials are expressed in terms of only 5 types of terms besides the term $R^3$,
listed below.
All the cubic monomials are symmetric under the interchange \eq{exchange} since they are scalars.
The six independent basis elements of the cubic Riemann monomials are given by
\begin{eqnarray} \label{six-cubic}
&& R^3, \quad R \Big(f^{ij}_{(++)} f^{ij}_{(++)}
  + f^{ij}_{(--)}f^{ij}_{(--)} \Big), \quad R f^{ij}_{(+-)}f^{ij}_{(+-)}, \nonumber \\
&&  \left(  f^{i_1 i_2}_{(++)} f^{i_1 i_3}_{(++)} f^{i_2 i_3}_{(++)}
+ f^{i_1 i_2}_{(--)} f^{i_1 i_3}_{(--)} f^{i_2 i_3}_{(--)} \right), \quad
\left( f^{i_1 i_2}_{(++)} f^{i_1 i_3}_{(+-)} f^{i_2 i_3}_{(+-)}
+ f^{i_1 i_2}_{(--)} f^{i_3 i_1}_{(+-)} f^{i_3 i_2}_{(+-)} \right), \\
&& \left( 2 f^{i_1 i_2}_{(+-)} f^{i_2 i_3}_{(+-)} f^{i_3 i_1}_{(+-)}
- 3 f^{i_1 i_1}_{(+-)} f^{i_2 i_3}_{(+-)} f^{i_3 i_2}_{(+-)}
+ f^{i_1 i_1}_{(+-)} f^{i_2 i_2}_{(+-)} f^{i_3 i_3}_{(+-)} \right) = 6 \, \det f^{ij}_{(+-)}, \nonumber
\end{eqnarray}
where $\det f^{ij}_{(+-)}$ is the determinant of the $3 \times 3$ matrix $f^{ij}_{(+-)}$.\footnote{Using the matrix notation \eq{def-metab}, $\det f^{ij}_{(+-)}$ can be written as $\det B = \frac{1}{3} \textrm{Tr} B^3
- \frac{1}{2} \textrm{Tr} B \, \textrm{Tr} B^2
+ \frac{1}{6} \big(\textrm{Tr} B \big)^3$. This expression appeared in Eq. (4) of Ref. \cite{sned-0} and was used to reduce independent invariants.}
The first three basis elements in \eq{six-cubic} correspond to the quadratic ones \eq{quadratic-b}
multiplied by $R$.
The last basis in \eq{six-cubic} consists of three terms which are self-symmetric under the parity
transformation \eq{exchange}. Thus they could form independent basis separately, but
they must appear with that combination because they come from the product of Ricci tensors, $R_{ab} R_{bc} R_{ca}$.
It is also clear from the fact that such combination forms $\det f^{ij}_{(+-)}$.
This proves that the cubic scalars have only six independent basis elements.
Replacing the 5 types of terms by the cubic monomials leads to the identities \eq{dxu1} and \eq{dxu2}.
There is no more linear relation. Our approach identifies the independent basis of Riemann monomials
without any ambiguity.

Now we list the result for the expansion of second-rank cubic monomials in Table \ref{table:1}.
For notational simplicity, we will use the matrix notation in \eq{def-metab}:
\begin{eqnarray*} \label{2nd-cubic:a-f}
&& A: R^2 R_{ab} = \frac{R^3}{4}  \delta_{ab} - 2 R^2 B_{i_1 i_2}
\big( \eta^{i_1} \bar{\eta}^{i_2} \big)_{ab}, \xx
&& B: R R_{ac} R_{bc} = \Big( \frac{R^3}{16} + 4 R \, \textrm{Tr} B B^T \Big) \delta_{ab}
- 2R \left( \textrm{Tr} \big( B^2 + (B^T)^2 \big)
- 2 \big( \textrm{Tr} B \big)^2 \right) \big( \eta^{i_1} \bar{\eta}^{i_1} \big)_{ab} \xx
&& \hspace{0.7cm} - \Big( R^2 B - 8 R \big( (B^T)^2  - \textrm{Tr} B B^T \big) \Big)_{i_1 i_2} \big( \eta^{i_1} \bar{\eta}^{i_2} \big)_{ab}, \xx
&& C: R_{ab} R_{cd}^2 = \left( \frac{R^3}{16} + 4 R \, \textrm{Tr} \big( B B^T \big) \right) \delta_{ab}
- \left( \frac{R^2}{2} B + 32 \textrm{Tr} \big( B B^T \big) B \right)_{i_1 i_2}
\big( \eta^{i_1} \bar{\eta}^{i_2} \big)_{ab}, \xx
&& D: R_{ac} R_{bd} R_{cd} = \left( \frac{R^3}{64} + 3 R \, \textrm{Tr} \big( B B^T \big)
-48 \, \det B \right) \delta_{ab} - \frac{3}{2} R \left( \textrm{Tr} \big( B^2 + (B^T)^2 \big) - 2 \big( \textrm{Tr} B \big)^2 \right) \big( \eta^{i_1} \bar{\eta}^{i_1} \big)_{ab} \xx
&& \qquad - \Big( \frac{3}{8} R^2 B - 6 R \big( (B^T)^2 -  \textrm{Tr} B \, B^T \big)
- 16 B B^T B  + 24 \textrm{Tr} \big( B B^T \big) B \Big)_{i_1 i_2} \big( \eta^{i_1} \bar{\eta}^{i_2} \big)_{ab}, \xx
&& E: R R_{cd} R_{acbd} = \left( \frac{R^3}{16} + 4 R \, \textrm{Tr} \big( B B^T \big) \right) \delta_{ab}
+ 2 R \left( \textrm{Tr} \big( B^2 + (B^T)^2 \big) - 2 \big( \textrm{Tr} B \big)^2 \right) \big( \eta^{i_1} \bar{\eta}^{i_1} \big)_{ab} \xx
&& \qquad - 4 R \Big(  (A_+ B + B A_-) + 2 (B^T)^2
- 2 (\textrm{Tr} B) B^T   \Big)_{i_1 i_2} \big( \eta^{i_1} \bar{\eta}^{i_2} \big)_{ab}, \xx
&& F: R_{ce} R_{de} R_{acbd} = \left( \frac{R^3}{64} + 3 R \, \textrm{Tr} \big( B B^T \big)
-48 \, \det B \right) \delta_{ab} + R \left( \textrm{Tr} \big( B^2 + (B^T)^2 \big)
- 2 \big( \textrm{Tr} B \big)^2 \right) \big( \eta^{i_1} \bar{\eta}^{i_1} \big)_{ab} \xx
&& \qquad + 4 \textrm{Tr} \Big( \big( B^2 + (B^T)^2 - 2 \textrm{Tr} B \, B \big) (A_+ + A_-) \Big)
\big( \eta^{i_1} \bar{\eta}^{i_1} \big)_{ab}  \xx
&& \qquad + \left( \frac{R^2}{8} B - 2 R \big( A_+ B + B A_- + 2 (B^T)^2
- 2 \textrm{Tr} B \, B^T \big) \right)_{i_1 i_2}
\big( \eta^{i_1} \bar{\eta}^{i_2} \big)_{ab} \xx
&& \qquad + \Big( 8 \textrm{Tr} B \, (B^T A_+ + A_- B^T) - 8  \textrm{Tr} B \, (A_+ B^T  + B^T A_-)
+ 8 \textrm{Tr} (B A_+ +  B^T A_-) B^T  \xx
&& \qquad  - 8 (B^T A_+ B^T + B^T A_- B^T) + 8 \big( A_+ (B^T)^2 + (B^T)^2 A_- - (B^T)^2 A_+ - A_- (B^T)^2 \big)
+ 8 \textrm{Tr} \big(  B B^T \big) \, B  \xx
&& \qquad + 4  \big( \textrm{Tr} B \big)^2 (A_+ + A_-) - 2 \textrm{Tr} \big(  B^2 +  (B^T)^2 \big) (A_+ + A_-)
-16 B B^T B \Big)_{i_1 i_2}  \big( \eta^{i_1} \bar{\eta}^{i_2} \big)_{ab},
\end{eqnarray*}

\begin{eqnarray*} \label{2nd-cubic:g-m}
&& G: R_{ac} R_{de} R_{becd} = \left( \frac{R^3}{64} + R \textrm{Tr} \big( B B^T \big)
+ 48 \, \det B  + 8 \textrm{Tr} \big( B^T A_+ B + B A_- B^T \big) \right) \delta_{ab} \xx
&& \qquad - \Big( \frac{R^2}{8} B + R \, \big( A_+ B + B A_- \big)
+ 4 \big( \textrm{Tr} B \big)^2 (A_+ + A_-) - 2 \textrm{Tr} \big( B^2 + (B^T)^2 \big) (A_+ + A_-)  \xx
&& \qquad  - 8 \textrm{Tr} B \big( A_+ B^T + B^T A_- \big)  + 8 \big( A_+ (B^T)^2 + (B^T)^2 A_- \big)
- 8 \textrm{Tr} \big( B B^T \big)  B + 16 BB^T B \Big)_{i_1 i_2} \big( \eta^{i_1} \bar{\eta}^{i_2} \big)_{ab} \xx
&& \qquad + 8 \varepsilon^{i_1 i_2 i_3}  \left( \big( B B^T A_+ \big)_{i_1 i_2} \eta^{i_3}_{ab}
+ \big( B^T B A_- \big)_{i_1 i_2}  \bar{\eta}^{i_3}_{ab} \right), \xx
&& H: R R_{aecd} R_{becd} = 4 R  \textrm{Tr} \big( A_+^2 + A_-^2 + 2 B B^T \big) \delta_{ab}
- 8 R \big( A_+ B + B A_- \big)_{i_1 i_2}  \big( \eta^{i_1} \bar{\eta}^{i_2} \big)_{ab}, \xx
&& I: R_{ab} R_{cdef}^2 = 4 R  \textrm{Tr} \big( A_+^2 + A_-^2 + 2 B B^T \big) \delta_{ab}
- 32 \textrm{Tr} \big( A_+^2 + A_-^2 + 2 B B^T \big) B_{i_1 i_2}  \big( \eta^{i_1} \bar{\eta}^{i_2} \big)_{ab}, \xx
&& J: R_{ac} R_{bedf} R_{cedf} = \left(  R  \textrm{Tr} \big( A_+^2 + A_-^2 + 2 B B^T \big)
+ 16 \textrm{Tr} \big( B^T A_+ B + B A_- B^T \big) \right) \delta_{ab} \xx
&& \qquad - 8 \textrm{Tr} \left( \big( B^2 + (B^T)^2 - 2  \textrm{Tr} B \, B \big) (A_+ + A_-)  \right) \big( \eta^{i_1} \bar{\eta}^{i_1} \big)_{ab} - 2 R \big( A_+ B + B A_- \big)_{i_1 i_2}  \big( \eta^{i_1} \bar{\eta}^{i_2} \big)_{ab},  \xx
&& \qquad - 8 \Big( \textrm{Tr} \big( A_+^2 + A_-^2 \big) B
+ 2 \textrm{Tr} B \big( B^T A_+ + A_- B^T \big)
- 2 \big( (B^T)^2 A_+  + A_- (B^T)^2 \big) + 2 \textrm{Tr} \big( B A_+ + B^T A_- \big) B^T \xx
&& \qquad - 2 \big( B^T A_+ B^T + B^T A_- B^T \big)
+ 2 \textrm{Tr} \big( B B^T \big) B \Big)_{i_1 i_2}  \big( \eta^{i_1} \bar{\eta}^{i_2} \big)_{ab} \xx
&& \qquad + 16 \varepsilon^{i_1 i_2 i_3}  \left( \big( B B^T A_+ \big)_{i_1 i_2} \eta^{i_3}_{ab}
+ \big( B^T B A_- \big)_{i_1 i_2}  \bar{\eta}^{i_3}_{ab} \right), \xx
&& K: R_{ef} R_{aecd} R_{bfcd} = \left(  R  \textrm{Tr} \big( A_+^2 + A_-^2 + 2 B B^T \big)
+ 16 \textrm{Tr} \big( B^T A_+ B + B A_- B^T \big) \right) \delta_{ab} \xx
&& \qquad + 8 \textrm{Tr} \Big( \big( B^2 + (B^T)^2 - 2  \textrm{Tr} B \, B \big) (A_+ + A_-)  \Big) \big( \eta^{i_1} \bar{\eta}^{i_1} \big)_{ab} - 2 R \big( A_+ B + B A_- \big)_{i_1 i_2}  \big( \eta^{i_1} \bar{\eta}^{i_2} \big)_{ab}  \xx
&& \qquad + 8 \Big( \textrm{Tr} \big( A_+^2 + A_-^2 \big) B + 2 \textrm{Tr} B \big( B^T A_+ + A_- B^T \big)
- 2 \big( (B^T)^2 A_+ + A_- (B^T)^2 \big) + 2 \textrm{Tr} \big( B A_+ + B^T A_- \big) B^T \xx
&& \qquad - 2 \big( A_+^2 B + B A_-^2 \big)
- 2 \big( B^T A_+ B^T + B^T A_- B^T \big)
+ 2 \textrm{Tr} \big( B B^T \big) B
- 4 B B^T B \Big)_{i_1 i_2}  \big( \eta^{i_1} \bar{\eta}^{i_2} \big)_{ab}, \xx
&& L: R_{ef} R_{acbd} R_{cedf} = \left( \frac{R^3}{64} + R \textrm{Tr} \big( B B^T \big)
+ 48 \, \det B  + 8 \textrm{Tr} \big( B^T A_+ B + B A_- B^T \big) \right) \delta_{ab} \xx
&& \qquad + \frac{R}{2} \left( \textrm{Tr} \big( B^2 + (B^T)^2 \big) - 2  \big(\textrm{Tr} B \big)^2 \right)
\big( \eta^{i_1} \bar{\eta}^{i_1} \big)_{ab}
- 4 \textrm{Tr} \Big( \big( B^2 + (B^T)^2 - 2  \textrm{Tr} B \, B \big) (A_+ + A_-)  \Big)
\big( \eta^{i_1} \bar{\eta}^{i_1} \big)_{ab} \xx
&& \qquad - \left( \frac{R^2}{8} B - R \big( A_+ B + B A_- \big) - R \big( B^2 + (B^T)^2 - 2 (\textrm{Tr} B) \, B^T \big) \right)_{i_1 i_2}  \big( \eta^{i_1} \bar{\eta}^{i_2} \big)_{ab}  \xx
&& \qquad - 8 \Big( \textrm{Tr} B \big( B^T A_+ + A_- B^T \big)
- \big( (B^T)^2 A_+ + A_- (B^T)^2 \big) + \textrm{Tr} \big( B A_+ + B^T A_- \big) B^T
+ \big( A_+^2 B + B A_-^2 \big) \xx
&& \qquad
- \big( B^T A_+ B^T + B^T A_- B^T \big) + 3 \textrm{Tr} \big( B B^T \big) B
- 2 B B^T B + 2 A_+ B A_- \Big)_{i_1 i_2}  \big( \eta^{i_1} \bar{\eta}^{i_2} \big)_{ab}, \xx
&& M: R_{ef} R_{acde} R_{bcde} = \left( R  \textrm{Tr} \big( A_+^2 + A_-^2 + 2 B B^T \big)
+ 16 \textrm{Tr} \big( B^T A_+ B + B A_- B^T \big) \right) \delta_{ab} \xx
&& \qquad - 2 \Big( R \big( A_+ B + B A_- \big) + 8 \big( B B^T B + A_+ B A_- \big) \Big)_{i_1 i_2}
\big( \eta^{i_1} \bar{\eta}^{i_2} \big)_{ab},
\end{eqnarray*}

\begin{eqnarray*} \label{2nd-cubic:n-p}
&& N: R_{agcd} R_{bgef} R_{cdef}  = 16 \left( \textrm{Tr} \big( A_+^3 + A_-^3 \big)
+ 3 \textrm{Tr} \big( B^T A_+ B + B A_- B^T \big) \right) \delta_{ab} \xx
&& \qquad -32 \Big(  A_+^2 B + B A_-^2 + B B^T B + A_+ B A_- \Big)_{i_1 i_2}
\big( \eta^{i_1} \bar{\eta}^{i_2} \big)_{ab}, \xx
&& O: R_{aecg} R_{bfdg} R_{cdef} = \left( \frac{R^3}{64} - \frac{3}{2} R \textrm{Tr} \big( A_+^2 + A_-^2 \big)
+ 48 \, \det B  + 8 \textrm{Tr} \big( A^3_+ + A^3_-\big) \right) \delta_{ab} \xx
&& \qquad - \left( \frac{R^2}{8} B - R \big( A_+ B + B A_- \big) - 4 \textrm{Tr} \big( A_+^2 + A_-^2 \big) B
+ 4 \big( \textrm{Tr} B \big)^2 (A_+ + A_-) \right)_{i_1 i_2}  \big( \eta^{i_1} \bar{\eta}^{i_2} \big)_{ab}  \xx
&& \qquad + 4 \Big( \frac{1}{2} \textrm{Tr} \big( B^2 + (B^T)^2 \big) (A_+ + A_-) + 2 \textrm{Tr} B
\big( A_+ B^T + B^T A_- \big) \xx
&& \hspace{1.5cm} - 2 \big( A_+^2 B + B A_-^2 + A_+ (B^T)^2 + (B^T)^2 A_- \big) \Big)_{i_1 i_2}
\big( \eta^{i_1} \bar{\eta}^{i_2} \big)_{ab}, \xx
&& P: R_{aebg} R_{cdef} R_{cdfg} = - \left( R  \textrm{Tr} \big( A_+^2 + A_-^2 + 2 B B^T \big)
+ 16 \textrm{Tr} \big( B^T A_+ B + B A_- B^T \big) \right) \delta_{ab} \xx
&& \qquad - 8 \textrm{Tr} \Big( \big( B^2 + (B^T)^2 - 2  \textrm{Tr} B \, B \big) (A_+ + A_-)  \Big)
\big( \eta^{i_1} \bar{\eta}^{i_1} \big)_{ab} - 2 R \left( A_+ B + B A_- \right)_{i_1 i_2}
\big( \eta^{i_1} \bar{\eta}^{i_2} \big)_{ab}, \xx
&& \qquad + 8 \Big( \textrm{Tr} \big( A_+^2 + A_-^2 \big) B  - 2 \textrm{Tr} B
\big( B^T A_+ + A_- B^T \big) - 2 \textrm{Tr} \big( B A_+ +  B^T A_- \big) B^T + 2 \textrm{Tr} \big( B B^T \big) B  \xx
&& \qquad + 2 \big(  A_+^2 B + B A_-^2 + (B^T)^2 A_+ + A_- (B^T)^2 + B^T A_+ B^T + B^T A_- B^T
+ 2 A_+ B A_- \big) \Big)_{i_1 i_2}
\big( \eta^{i_1} \bar{\eta}^{i_2} \big)_{ab}.
\end{eqnarray*}

Note that the second-rank cubic tensors in Table \ref{table:1} are parity even, so they must be invariant under the parity transformation \eq{parity-mat}.
One can quickly check it by noting that the parity transformation acts on
$P: \big( \eta^{i_1} \bar{\eta}^{i_2} \big)_{ab} \to \big( \bar{\eta}^{i_1} \eta^{i_2} \big)_{ab}
= \big( \eta^{i_2} \bar{\eta}^{i_1} \big)_{ab}$. The second-rank cubic monomials in Table \ref{table:1} are symmetric with respect to $(a \leftrightarrow b)$ except in $G$ and $J$ which contain an anti-symmetric part as was shown above. Although the expansion is much more complicated compared to the scalar case, it turns out that terms that would appear in the scalar case reappear with the factor $\frac{1}{4} \delta_{ab}$, and novel terms that do not appear in the scalar case (or should disappear in the scalar case) appear with the factor $(\eta^i \bar{\eta}^j)_{ab}$. Indeed,
$\delta_{ab}$ and $(\eta^i \bar{\eta}^j)_{ab}$ are the only symmetric tensors constructed from the product of 't Hooft symbols. The latter is trace-free due to Eq. \eq{eta-etabar}. In particular, $(\eta^i \bar{\eta}^i)_{ab}$ is a diagonal matrix given by
\be \la{diag-eebar}
(\eta^i \bar{\eta}^i)_{ab} =  - \eta^i_{ac} \bar{\eta}^i_{bc}
= - \big(\delta_{ab} - 4 \delta_{a4} \delta_{b4} \big) = - \left(
  \begin{array}{cccc}
    1 & 0 & 0 & 0 \\
    0 & 1 & 0 & 0 \\
    0 & 0 & 1 & 0 \\
    0 & 0 & 0 & -3 \\
  \end{array}
\right).
\ee

We found that there are two algebraic relations for the second-rank cubic tensors in Table \ref{table:1}. We list them below:
\bea \la{2nd-rank-syzy1}
&& R^2 R_{ab} - 4 R_{ab} R_{cd}^2  - 8 R_{ac} R_{de} R_{becd}
-2 R R_{acde} R_{becd} + R_{ab} R_{cdef}^2 + 4 R_{ef} R_{aecd} R_{bfcd} \xx
&& + 8 R_{ef} R_{aecd} R_{bcdf} + 4 R_{ac} R_{bedf} R_{cedf} + 8 R_{aecg} R_{bfdg} R_{cdef}
- 4 R_{agcd} R_{bgef} R_{cdef} = 0, \\
\la{2nd-rank-syzy2}
&& R R_{ac} R_{bc} - 2 R_{ac} R_{bd} R_{cd} + R R_{cd} R_{acbd} - 2 R_{ce} R_{de} R_{acbd} - 6 R_{ac} R_{de} R_{becd} - R R_{aecd} R_{becd} \xx
&& + 4 R_{ef} R_{acde} R_{bcdf} + 2 R_{ef} R_{aecd} R_{bfcd} - 2 R_{ef} R_{acbd} R_{cedf}
+ 3 R_{ac} R_{bedf} R_{cedf} \xx
&& + 4 R_{aecg} R_{bfdg} R_{cdef} -2 R_{agcd} R_{bgef} R_{cdef} - R_{aebg} R_{cdef} R_{cdfg} = 0.
\eea
It is natural to expect that there exist more than two linear relations between the second-rank cubic tensors in Table \ref{table:1} since contracting the free indices $a$
and $b$ has to reproduce the eight cubic scalars which obey the syzygy relations \eq{dxu1} and \eq{dxu2}. However we found that there exist 14 linearly independent second-rank cubic tensors using a computer algorithm.\footnote{The expression for these 14 basis elements is quite complicated and not illuminating, so will not be displayed here. But we can e-mail the result to those who request it.} This means that Eqs. \eq{2nd-rank-syzy1} and \eq{2nd-rank-syzy2} exhaust all possible linear relations for the second-rank cubic tensors in Table \ref{table:1}. It also implies that Eqs. \eq{2nd-rank-syzy1} and \eq{2nd-rank-syzy2}
have to reproduce the syzygies \eq{dxu1} and \eq{dxu2} when contracting $a$
and $b$. It can be checked by eliminating $R_{a e c g} R_{b f d g} R_{c d e f}$ and $R_{a c} R_{b e d f} R_{c e d f}$ from Eqs. \eq{2nd-rank-syzy1} and \eq{2nd-rank-syzy2},  respectively, and then combining them together to obtain
\begin{eqnarray} \la{3-421}
&& R^2 R_{a b} - 2 R R_{a c} R_{b c} + 4 R_{a c} R_{b d} R_{c d}
- 4 R_{a b} R_{c d}^2 - 2 R R_{c d} R_{a c b d} + 4 R_{c e} R_{d e} R_{a c b d} \xx
&& + 4 R_{a c} R_{d e} R_{b e c d} + R_{a b} R_{c d e f}^2
+ 4 R_{e f} R_{a c b d} R_{c e d f}
- 2 R_{a c} R_{b e d f} R_{c e d f}
+ 2 R_{a e b g} R_{c d e f} R_{c d f g} = 0, \\
 \la{3-422}
&& 3 R^2 R_{a b} - 4 R R_{a c} R_{b c} + 8 R_{a c} R_{b d} R_{c d}
- 12 R_{a b} R_{c d}^2 - 4 R R_{c d} R_{a c b d} + 8 R_{c e} R_{d e} R_{a c b d} \xx
&& - 2 R R_{a e c d} R_{b e c d} + 8 R_{e f} R_{a c d e} R_{b c d f}
+ 4 R_{e f} R_{a e c d} R_{b f c d} + 3 R_{a b} R_{c d e f}^2
+ 8 R_{e f} R_{a c b d} R_{c e d f}  \xx
&& - 4 R_{a g c d} R_{b g e f} R_{c d e f} + 8 R_{a e c g} R_{b f d g} R_{c d e f}
+ 4 R_{a e b g} R_{c d e f} R_{c d f g} = 0.
\end{eqnarray}
Combining the above equations after contracting $a$ and $b$ reproduces the cubic identities \eq{dxu1} and \eq{dxu2}.

\section{Quartic identities for a general Riemannian manifold}

We present the expansion of the 26 quartic monomials in Table \ref{table:2} using
the decomposition \eq{decom-riem} for a general Riemannian manifold:
\begin{eqnarray*} \label{quartic:a-j}
&& A: R^4=256 \left(  f_{(++)}^{i_1 i_1} + f_{(--)}^{i_1 i_1} \right)^4, \nonumber \\
&& B: R^2 R_{a b} R_{a b}=\frac{R^4}{4} + 16 R^2 \left( f_{(+-)}^{i_1 i_2}\right)^2, \\
&& C: R R_{a b} R_{b c} R_{c a} = \frac{R^4}{16} + 12 R^2 \left(f_{(+-)}^{i_1 i_2}\right)^2
- 32 R \left( 2 f_{(+-)}^{i_1 i_2} f_{(+-)}^{i_2 i_3} f_{(+-)}^{i_3 i_1} - 3 f_{(+-)}^{i_1 i_1} f_{(+-)}^{i_2 i_3} f_{(+-)}^{i_3 i_2} + f_{(+-)}^{i_1 i_1} f_{(+-)}^{i_2 i_2} f_{(+-)}^{i_3 i_3} \right), \\
&& D: \left(R_{a b} R_{a b} \right)^2 = \frac{R^4}{16} + 8 R^2 \left( f_{(+-)}^{i_1 i_2} \right)^2 + 256 \left( f_{(+-)}^{i_1 i_2} \right)^2 \left(f_{(+-)}^{i_3 i_4} \right)^2, \\
&& E: R_{a b} R_{b c} R_{c d} R_{d a} \\
&& \quad \; = \frac{R^4}{64} + 6 R^2 \left(f_{(+-)}^{i_1 i_2}\right)^2
- 32 R\left(2f_{(+-)}^{i_1 i_2} f_{(+-)}^{i_2 i_3}f_{(+-)}^{i_3 i_1} - 3f_{(+-)}^{i_1 i_1} f_{(+-)}^{i_2 i_3} f_{(+-)}^{i_3 i_2} + f_{(+-)}^{i_1 i_1} f_{(+-)}^{i_2 i_2} f_{(+-)}^{i_3 i_3}\right) \\
&& \qquad + 192 \left(f_{(+-)}^{i_1 i_2}\right)^2 \left(f_{(+-)}^{i_3 i_4} \right)^2
- 128 f_{(+-)}^{i_1 i_2} f_{(+-)}^{i_1 i_3} f_{(+-)}^{i_4 i_2} f_{(+-)}^{i_4 i_3}, \\
&& F: R R_{a b} R_{c d} R_{a c b d} \\
&& \quad \; = \frac{R^4}{16} + 4R^2 \left(f_{(+-)}^{i_1 i_2}\right)^2 + 32 R \left( \left(2f_{(+-)}^{i_1 i_2}f_{(+-)}^{i_2 i_3} f_{(+-)}^{i_3 i_1} - 3 f_{(+-)}^{i_1 i_1} f_{(+-)}^{i_2 i_3} f_{(+-)}^{i_3 i_2}
+ f_{(+-)}^{i_1 i_1} f_{(+-)}^{i_2 i_2} f_{(+-)}^{i_3 i_3} \right) \right. \\
&& \qquad \left. + \left( f_{(+-)}^{i_1 i_2} f_{(+-)}^{i_3 i_2} f_{(++)}^{i_1 i_3}
+ f_{(+-)}^{i_2 i_1} f_{(+-)}^{i_2 i_3} f_{(--)}^{i_1 i_3}  \right) \right), \\
&& G: R_{a b} R_{c e} R_{e d} R_{a c b d} \\
&& \quad \; = \frac{R^4}{64} + 2 R^2 \left(f_{(+-)}^{i_1 i_2} \right)^2
+ 8 R \left( \left( 2 f_{(+-)}^{i_1 i_2}f_{(+-)}^{i_2 i_3} f_{(+-)}^{i_3 i_1}
- 3 f_{(+-)}^{i_1 i_1} f_{(+-)}^{i_2 i_3} f_{(+-)}^{i_3 i_2}
+ f_{(+-)}^{i_1 i_1} f_{(+-)}^{i_2 i_2} f_{(+-)}^{i_3 i_3} \right) \right. \\
&& \qquad \left. + 2 \left( f_{(+-)}^{i_1 i_2} f_{(+-)}^{i_3 i_2} f_{(++)}^{i_1 i_3}
+ f_{(+-)}^{i_2 i_1} f_{(+-)}^{i_2 i_3} f_{(--)}^{i_1 i_3} \right) \right)
-64 \left(f_{(+-)}^{i_1 i_2} \right)^2 \left(f_{(+-)}^{i_3 i_4}\right)^2
+ 128 f_{(+-)}^{i_1 i_2} f_{(+-)}^{i_1 i_3} f_{(+-)}^{i_4 i_2} f_{(+-)}^{i_4 i_3} \\
&& \qquad
- 32 \left(f_{(+-)}^{i_1 i_2} f_{(+-)}^{i_2 i_1} f_{(+-)}^{i_3 i_4} f_{(++)}^{i_3 i_4} - f_{(+-)}^{i_1 i_1} f_{(+-)}^{i_2 i_2} f_{(+-)}^{i_3 i_4} f_{(++)}^{i_3 i_4} \right. \\
&& \qquad \qquad \left.
+ 2 f_{(+-)}^{i_1 i_1} f_{(+-)}^{i_2 i_3} f_{(+-)}^{i_4 i_2} f_{(++)}^{i_3 i_4}
-2 f_{(+-)}^{i_1 i_2} f_{(+-)}^{i_2 i_3} f_{(+-)}^{i_4 i_1} f_{(++)}^{i_3 i_4} \right)  \\
&& \qquad  - 32 \left( f_{(+-)}^{i_1 i_2} f_{(+-)}^{i_2 i_1} f_{(+-)}^{i_3 i_4} f_{(--)}^{i_3 i_4}
- f_{(+-)}^{i_1 i_1} f_{(+-)}^{i_2 i_2} f_{(+-)}^{i_3 i_4} f_{(--)}^{i_3 i_4} \right. \\
&& \qquad \qquad \left.
+ 2 f_{(+-)}^{i_1 i_1} f_{(+-)}^{i_3 i_2} f_{(+-)}^{i_2 i_4} f_{(--)}^{i_3 i_4}  - 2 f_{(+-)}^{i_2 i_1} f_{(+-)}^{i_3 i_2} f_{(+-)}^{i_1 i_4} f_{(--)}^{i_3 i_4}  \right), \xx
&& H: R^2 R_{a b c d} R_{a b c d} = 16 R^2 \left( \left( f_{(++)}^{i_1 i_2} \right)^2 + 2 \left( f_{(+-)}^{i_1 i_2} \right)^2
+ \left( f_{(--)}^{i_1 i_2} \right)^2 \right), \\
&& I: R R_{a b} R_{a c d e} R_{b c d e} \\
&& \quad \; = 4 R^2 \left( \left( f_{(++)}^{i_1 i_2} \right)^2
+ 2 \left( f_{(+-)}^{i_1 i_2} \right)^2 + \left( f_{(--)}^{i_1 i_2} \right)^2 \right)
+ 64R \left( f_{(+-)}^{i_1 i_2} f_{(+-)}^{i_3 i_2} f_{(++)}^{i_1 i_3}
+ f_{(+-)}^{i_2 i_1} f_{(+-)}^{i_2 i_3} f_{(--)}^{i_1 i_3} \right), \\
&& J: R_{a b} R_{a b} R_{c d e f} R_{c d e f} \\
&& \quad \; = 4 R^2 \left( \left( f_{(++)}^{i_1 i_2}\right)^2 + 2\left( f_{(+-)}^{i_1 i_2} \right)^2
+  \left(f_{(--)}^{i_1 i_2} \right)^2  \right)
+ 256 \left( f_{(+-)}^{i_3 i_4} \right)^2 \left( \left( f_{(++)}^{i_1 i_2} \right)^2
+ 2 \left( f_{(+-)}^{i_1 i_2} \right)^2
+ \left( f_{(--)}^{i_1 i_2} \right)^2 \right),
\end{eqnarray*}

\begin{eqnarray*} \label{quartic:k-m}
&& K: R_{a b} R_{b c} R_{d e f a} R_{d e f c} \\
&& \quad \; = R^2 \left( \left( f_{(++)}^{i_1 i_2} \right)^2 + 2 \left( f_{(+-)}^{i_1 i_2} \right)^2 + \left( f_{(--)}^{i_1 i_2} \right)^2 \right)
+ 32 R \left( f_{(+-)}^{i_1 i_2}f_{(+-)}^{i_3 i_2} f_{(++)}^{i_1 i_3}
+ f_{(+-)}^{i_2 i_1} f_{(+-)}^{i_2 i_3} f_{(--)}^{i_1 i_3} \right) \\
&& \qquad \: + 64 \left( f_{(+-)}^{i_3 i_4} \right)^2 \left( \left( f_{(++)}^{i_1 i_2} \right)^2
+ 2 \left( f_{(+-)}^{i_1 i_2}\right)^2 + \left(f_{(--)}^{i_1 i_2} \right)^2 \right) \\
&& \qquad \;
+ 128 \left( f_{(+-)}^{i_1 i_2} f_{(+-)}^{i_2 i_1}
f_{(+-)}^{i_3 i_4} f_{(++)}^{i_3 i_4} - f_{(+-)}^{i_1 i_1} f_{(+-)}^{i_2 i_2} f_{(+-)}^{i_3 i_4} f_{(++)}^{i_3 i_4} \right. \\
&& \qquad \qquad \left.
+ 2 f_{(+-)}^{i_1 i_1} f_{(+-)}^{i_2 i_3} f_{(+-)}^{i_4 i_2} f_{(++)}^{i_3 i_4}
- 2 f_{(+-)}^{i_1 i_2} f_{(+-)}^{i_2 i_3} f_{(+-)}^{i_4 i_1} f_{(++)}^{i_3 i_4} \right) \\
&& \qquad \;
+ 128 \left( f_{(+-)}^{i_1 i_2} f_{(+-)}^{i_2 i_1} f_{(+-)}^{i_3 i_4} f_{(--)}^{i_3 i_4} - f_{(+-)}^{i_1 i_1} f_{(+-)}^{i_2 i_2} f_{(+-)}^{i_3 i_4} f_{(--)}^{i_3 i_4} \right. \\
&& \qquad \qquad \left.
+ 2 f_{(+-)}^{i_1 i_1} f_{(+-)}^{i_3 i_2} f_{(+-)}^{i_2 i_4} f_{(--)}^{i_3 i_4}
-2 f_{(+-)}^{i_2 i_1} f_{(+-)}^{i_3 i_2} f_{(+-)}^{i_1 i_4} f_{(--)}^{i_3 i_4} \right),  \\
&& L: R_{a b} R_{cd} R_{a c e f} R_{b d e f} \\
&& \quad \; = R^2 \left( \left( f_{(++)}^{i_1 i_2} \right)^2 + 2 \left( f_{(+-)}^{i_1 i_2} \right)^2 + \left( f_{(--)}^{i_1 i_2} \right)^2 \right)
+ 32 R \left( f_{(+-)}^{i_1 i_2}f_{(+-)}^{i_3 i_2} f_{(++)}^{i_1 i_3}
+ f_{(+-)}^{i_2 i_1} f_{(+-)}^{i_2 i_3} f_{(--)}^{i_1 i_3} \right) \\
&& \qquad \: - 64 \left( f_{(+-)}^{i_3 i_4} \right)^2 \left( \left( f_{(++)}^{i_1 i_2} \right)^2
+ 2 \left( f_{(+-)}^{i_1 i_2}\right)^2 + \left(f_{(--)}^{i_1 i_2} \right)^2 \right) \\
&& \qquad \;
- 128 \left( f_{(+-)}^{i_1 i_2} f_{(+-)}^{i_2 i_1} f_{(+-)}^{i_3 i_4} f_{(++)}^{i_3 i_4}
- f_{(+-)}^{i_1 i_1} f_{(+-)}^{i_2 i_2} f_{(+-)}^{i_3 i_4} f_{(++)}^{i_3 i_4} \right. \\
&& \qquad \qquad \left. + 2 f_{(+-)}^{i_1 i_1} f_{(+-)}^{i_2 i_3} f_{(+-)}^{i_4 i_2} f_{(++)}^{i_3 i_4}
- 2 f_{(+-)}^{i_1 i_2} f_{(+-)}^{i_2 i_3} f_{(+-)}^{i_4 i_1} f_{(++)}^{i_3 i_4} \right) \\
&& \qquad \;
- 128 \left( f_{(+-)}^{i_1 i_2} f_{(+-)}^{i_2 i_1} f_{(+-)}^{i_3 i_4} f_{(--)}^{i_3 i_4} - f_{(+-)}^{i_1 i_1} f_{(+-)}^{i_2 i_2} f_{(+-)}^{i_3 i_4} f_{(--)}^{i_3 i_4} \right. \\
&& \qquad \qquad \left.+ 2 f_{(+-)}^{i_1 i_1} f_{(+-)}^{i_3 i_2} f_{(+-)}^{i_2 i_4} f_{(--)}^{i_3 i_4}
-2 f_{(+-)}^{i_2 i_1} f_{(+-)}^{i_3 i_2} f_{(+-)}^{i_1 i_4} f_{(--)}^{i_3 i_4} \right) \\
&& \qquad \; + 128 \left( f_{(+-)}^{i_1 i_2} f_{(+-)}^{i_3 i_2} f_{(++)}^{i_1 i_4} f_{(++)}^{i_3 i_4}
+ 2 f_{(+-)}^{i_1 i_2} f_{(+-)}^{i_1 i_3} f_{(+-)}^{i_4 i_2} f_{(+-)}^{i_4 i_3}
+ f_{(+-)}^{i_2 i_1} f_{(+-)}^{i_2 i_3} f_{(--)}^{i_1 i_4} f_{(--)}^{i_3 i_4} \right),  \\
&& M: R_{ab} R_{cd} R_{a e b f} R_{c e d f} \\
&& \quad \; = \frac{R^4}{64} + 2 R^2 \left( f_{(+-)}^{i_1 i_2} \right)^2
+ 12 R \left( 2 f_{(+-)}^{i_1 i_2}f_{(+-)}^{i_2 i_3} f_{(+-)}^{i_3 i_1}
- 3 f_{(+-)}^{i_1 i_1} f_{(+-)}^{i_2 i_3} f_{(+-)}^{i_3 i_2}
+ f_{(+-)}^{i_1 i_1} f_{(+-)}^{i_2 i_2} f_{(+-)}^{i_3 i_3} \right) \\
&& \qquad \: + 192 \left( f_{(+-)}^{i_1 i_2} \right)^2 \left( f_{(+-)}^{i_3 i_4} \right)^2
+ 16 \left( f_{(+-)}^{i_1 i_2} f_{(+-)}^{i_2 i_1} f_{(+-)}^{i_3 i_4} f_{(++)}^{i_3 i_4}
- f_{(+-)}^{i_1 i_1} f_{(+-)}^{i_2 i_2} f_{(+-)}^{i_3 i_4} f_{(++)}^{i_3 i_4} \right. \\
&& \qquad \qquad \left. + 2 f_{(+-)}^{i_1 i_1} f_{(+-)}^{i_2 i_3} f_{(+-)}^{i_4 i_2} f_{(++)}^{i_3 i_4}
- 2 f_{(+-)}^{i_1 i_2} f_{(+-)}^{i_2 i_3} f_{(+-)}^{i_4 i_1} f_{(++)}^{i_3 i_4} \right) \\
&& \qquad \;
+ 16 \left( f_{(+-)}^{i_1 i_2} f_{(+-)}^{i_2 i_1} f_{(+-)}^{i_3 i_4} f_{(--)}^{i_3 i_4} - f_{(+-)}^{i_1 i_1} f_{(+-)}^{i_2 i_2} f_{(+-)}^{i_3 i_4} f_{(--)}^{i_3 i_4} \right. \\
&& \qquad \qquad \left.+ 2 f_{(+-)}^{i_1 i_1} f_{(+-)}^{i_3 i_2} f_{(+-)}^{i_2 i_4} f_{(--)}^{i_3 i_4}
-2 f_{(+-)}^{i_2 i_1} f_{(+-)}^{i_3 i_2} f_{(+-)}^{i_1 i_4} f_{(--)}^{i_3 i_4} \right) \\
&& \qquad \;
+ 64 \left( f_{(+-)}^{i_1 i_2} f_{(+-)}^{i_3 i_2} f_{(++)}^{i_1 i_4} f_{(++)}^{i_3 i_4}
+ 2 f_{(+-)}^{i_1 i_3} f_{(+-)}^{i_2 i_4} f_{(++)}^{i_1 i_2} f_{(--)}^{i_3 i_4} \right. \\
&& \qquad \qquad \left. - 2 f_{(+-)}^{i_1 i_2} f_{(+-)}^{i_1 i_3} f_{(+-)}^{i_4 i_2} f_{(+-)}^{i_4 i_3}
+ f_{(+-)}^{i_2 i_1} f_{(+-)}^{i_2 i_3} f_{(--)}^{i_1 i_4} f_{(--)}^{i_3 i_4} \right),
\end{eqnarray*}

\begin{eqnarray*} \label{quartic:n-r}
&& N: R_{a b} R_{c d} R_{a e c f} R_{b e d f} \\
&& \quad \; = R^2 \left( \left( f_{(++)}^{i_1 i_2} \right)^2
+ 2 \left( f_{(+-)}^{i_1 i_2} \right)^2 + \left( f_{(--)}^{i_1 i_2} \right)^2 \right)
+ 32 R \left( f_{(+-)}^{i_1 i_2} f_{(+-)}^{i_3 i_2} f_{(++)}^{i_1 i_3}
+ f_{(+-)}^{i_2 i_1} f_{(+-)}^{i_2 i_3} f_{(--)}^{i_1 i_3} \right) \\
&& \qquad \;
+ 128 \left( f_{(++)}^{i_1 i_2} f_{(--)}^{i_3 i_4} f_{(+-)}^{i_1 i_3} f_{(+-)}^{i_2 i_4}
+ f_{(+-)}^{i_1 i_2} f_{(+-)}^{i_1 i_3} f_{(+-)}^{i_4 i_2} f_{(+-)}^{i_4 i_3} \right), \\
&& O: R R_{a b c d} R_{c d e f} R_{e f a b} \\
&& \quad \; = 64 R \left( f_{(++)}^{i_1 i_2} f_{(++)}^{i_1 i_3} f_{(++)}^{i_2 i_3}
+ 3 f_{(++)}^{i_1 i_3} f_{(+-)}^{i_1 i_2} f_{(+-)}^{i_3 i_2}
+ 3 f_{(--)}^{i_1 i_3} f_{(+-)}^{i_2 i_1} f_{(+-)}^{i_2 i_3}
+ f_{(--)}^{i_1 i_2} f_{(--)}^{i_1 i_3} f_{(--)}^{i_2 i_3} \right), \\
&& P: R R_{a c b d} R_{a e b f} R_{c e d f}  = \frac{R^4}{16} - 6 R^2 \left( \left( f_{(++)}^{i_1 i_2} \right)^2
+ \left( f_{(--)}^{i_1 i_2} \right)^2 \right) \\
&& \qquad \: + 32 R \left( \left( 2 f_{(+-)}^{i_1 i_2}f_{(+-)}^{i_2 i_3} f_{(+-)}^{i_3 i_1}
- 3 f_{(+-)}^{i_1 i_1} f_{(+-)}^{i_2 i_3} f_{(+-)}^{i_3 i_2}
+ f_{(+-)}^{i_1 i_1} f_{(+-)}^{i_2 i_2} f_{(+-)}^{i_3 i_3} \right) \right. \\
&& \qquad \: + \left. \left( f_{(++)}^{i_1 i_2} f_{(++)}^{i_1 i_3} f_{(++)}^{i_2 i_3}
+ f_{(--)}^{i_1 i_2} f_{(--)}^{i_1 i_3} f_{(--)}^{i_2 i_3} \right) \right), \\
&& Q: R_{a b} R_{a c b d} R_{e f g c} R_{e f g d} \\
&& \quad \; = R^2 \left( \left( f_{(++)}^{i_1 i_2} \right)^2 + 2 \left( f_{(+-)}^{i_1 i_2} \right)^2 + \left( f_{(--)}^{i_1 i_2} \right)^2 \right)
+ 64 \left( f_{(+-)}^{i_3 i_4} \right)^2 \left( \left( f_{(++)}^{i_1 i_2} \right)^2
+ 2 \left( f_{(+-)}^{i_1 i_2}\right)^2 + \left(f_{(--)}^{i_1 i_2} \right)^2 \right) \\
&& \qquad \;
- 128 \left( f_{(+-)}^{i_1 i_2} f_{(+-)}^{i_2 i_1}
f_{(+-)}^{i_3 i_4} f_{(++)}^{i_3 i_4} - f_{(+-)}^{i_1 i_1} f_{(+-)}^{i_2 i_2} f_{(+-)}^{i_3 i_4} f_{(++)}^{i_3 i_4} \right. \\
&& \qquad \qquad \left. + 2 f_{(+-)}^{i_1 i_1} f_{(+-)}^{i_2 i_3} f_{(+-)}^{i_4 i_2} f_{(++)}^{i_3 i_4}
- 2 f_{(+-)}^{i_1 i_2} f_{(+-)}^{i_2 i_3} f_{(+-)}^{i_4 i_1} f_{(++)}^{i_3 i_4} \right) \\
&& \qquad \;
- 128 \left( f_{(+-)}^{i_1 i_2} f_{(+-)}^{i_2 i_1} f_{(+-)}^{i_3 i_4} f_{(--)}^{i_3 i_4} - f_{(+-)}^{i_1 i_1} f_{(+-)}^{i_2 i_2} f_{(+-)}^{i_3 i_4} f_{(--)}^{i_3 i_4} \right. \\
&& \qquad \qquad \left.+ 2 f_{(+-)}^{i_1 i_1} f_{(+-)}^{i_3 i_2} f_{(+-)}^{i_2 i_4} f_{(--)}^{i_3 i_4}
-2 f_{(+-)}^{i_2 i_1} f_{(+-)}^{i_3 i_2} f_{(+-)}^{i_1 i_4} f_{(--)}^{i_3 i_4} \right)  \\
&& \qquad \; + 128 \left( f_{(+-)}^{i_1 i_2} f_{(+-)}^{i_3 i_2} f_{(++)}^{i_1 i_4} f_{(++)}^{i_3 i_4}
+ 2 f_{(++)}^{i_1 i_2} f_{(--)}^{i_3 i_4} f_{(+-)}^{i_1 i_3} f_{(+-)}^{i_2 i_4}
+ f_{(+-)}^{i_2 i_1} f_{(+-)}^{i_2 i_3} f_{(--)}^{i_1 i_4} f_{(--)}^{i_3 i_4} \right),  \\
&& R: R_{ab} R_{c d e f} R_{a g e f} R_{b g c d} \\
&& \quad \; = 16 R \left( f_{(++)}^{i_1 i_2} f_{(++)}^{i_1 i_3} f_{(++)}^{i_2 i_3}
+ 3 f_{(++)}^{i_1 i_3} f_{(+-)}^{i_1 i_2} f_{(+-)}^{i_3 i_2}
+ 3 f_{(--)}^{i_1 i_3} f_{(+-)}^{i_2 i_1} f_{(+-)}^{i_2 i_3}
+ f_{(--)}^{i_1 i_2} f_{(--)}^{i_1 i_3} f_{(--)}^{i_2 i_3} \right) \\
&& \qquad \; + 256 \left( f_{(++)}^{i_1 i_2} f_{(++)}^{i_2 i_3} f_{(+-)}^{i_3 i_4} f_{(+-)}^{i_1 i_4}
+ f_{(++)}^{i_1 i_2} f_{(--)}^{i_3 i_4} f_{(+-)}^{i_1 i_3} f_{(+-)}^{i_2 i_4} \right. \\
&& \qquad \qquad \left. +  f_{(+-)}^{i_1 i_2} f_{(+-)}^{i_1 i_3} f_{(+-)}^{i_4 i_2} f_{(+-)}^{i_4 i_3}
+ f_{(--)}^{i_1 i_2} f_{(--)}^{i_2 i_3} f_{(+-)}^{i_4 i_3} f_{(+-)}^{i_4 i_1}  \right), \\
\end{eqnarray*}

\begin{eqnarray*} \label{quartic:s-w}
&& S: R_{a b} R_{c e d f} R_{e g f a} R_{g c b d}
= \frac{R^4}{64} - \frac{R^2}{2} \left( 3 \left( f_{(++)}^{i_1 i_2} \right)^2
- 2 \left( f_{(+-)}^{i_1 i_2} \right)^2 + 3 \left( f_{(--)}^{i_1 i_2} \right)^2 \right) \\
&& \qquad \; + 8 R \left( \left( 2 f_{(+-)}^{i_1 i_2}f_{(+-)}^{i_2 i_3} f_{(+-)}^{i_3 i_1}
- 3 f_{(+-)}^{i_1 i_1} f_{(+-)}^{i_2 i_3} f_{(+-)}^{i_3 i_2}
+ f_{(+-)}^{i_1 i_1} f_{(+-)}^{i_2 i_2} f_{(+-)}^{i_3 i_3} \right) \right. \\
&& \qquad \qquad \left. +
\left( f_{(++)}^{i_1 i_2} f_{(++)}^{i_1 i_3} f_{(++)}^{i_2 i_3}
- f_{(++)}^{i_1 i_3} f_{(+-)}^{i_1 i_2} f_{(+-)}^{i_3 i_2}
- f_{(--)}^{i_1 i_3} f_{(+-)}^{i_2 i_1} f_{(+-)}^{i_2 i_3}
+ f_{(--)}^{i_1 i_2} f_{(--)}^{i_1 i_3} f_{(--)}^{i_2 i_3}   \right) \right)\\
&& \qquad \;
- 32 \left( f_{(+-)}^{i_3 i_4} \right)^2
\left(  \left( f_{(++)}^{i_1 i_2} \right)^2 + \left( f_{(--)}^{i_1 i_2} \right)^2 \right)
+ 64 \left( f_{(+-)}^{i_1 i_2} f_{(+-)}^{i_3 i_2} f_{(++)}^{i_1 i_4} f_{(++)}^{i_3 i_4}
+ f_{(+-)}^{i_2 i_1} f_{(+-)}^{i_2 i_3} f_{(--)}^{i_1 i_4} f_{(--)}^{i_3 i_4} \right) \\
&& \qquad \;
- 32 \left( f_{(+-)}^{i_1 i_2} f_{(+-)}^{i_2 i_1} f_{(+-)}^{i_3 i_4} f_{(++)}^{i_3 i_4}
- f_{(+-)}^{i_1 i_1} f_{(+-)}^{i_2 i_2} f_{(+-)}^{i_3 i_4} f_{(++)}^{i_3 i_4} \right. \\
&& \qquad \qquad \left. + 2 f_{(+-)}^{i_1 i_1} f_{(+-)}^{i_2 i_3} f_{(+-)}^{i_4 i_2} f_{(++)}^{i_3 i_4}
- 2 f_{(+-)}^{i_1 i_2} f_{(+-)}^{i_2 i_3} f_{(+-)}^{i_4 i_1} f_{(++)}^{i_3 i_4} \right) \\
&& \qquad \;
- 32 \left( f_{(+-)}^{i_1 i_2} f_{(+-)}^{i_2 i_1} f_{(+-)}^{i_3 i_4} f_{(--)}^{i_3 i_4} - f_{(+-)}^{i_1 i_1} f_{(+-)}^{i_2 i_2} f_{(+-)}^{i_3 i_4} f_{(--)}^{i_3 i_4} \right. \\
&& \qquad \qquad \left.+ 2 f_{(+-)}^{i_1 i_1} f_{(+-)}^{i_3 i_2} f_{(+-)}^{i_2 i_4} f_{(--)}^{i_3 i_4}
-2 f_{(+-)}^{i_2 i_1} f_{(+-)}^{i_3 i_2} f_{(+-)}^{i_1 i_4} f_{(--)}^{i_3 i_4} \right), \\
&& T: \big( R_{a b c d} R_{a b c d} \big)^2 = 256 \left( \left( f_{(++)}^{i_1 i_2} \right)^2
+ 2 \left( f_{(+-)}^{i_1 i_2} \right)^2 + \left( f_{(--)}^{i_1 i_2} \right)^2 \right)^2, \\
&& U: R_{a b c d} R_{a b c e} R_{f g h d} R_{f g h e}
= 256 \left( f_{(+-)}^{i_3 i_4} \right)^2 \left( \left( f_{(++)}^{i_1 i_2} \right)^2
+ \left( f_{(+-)}^{i_1 i_2}\right)^2 + \left(f_{(--)}^{i_1 i_2} \right)^2  \right) \\
&& \qquad \; + 64 \left( \big( f^{i_1 i_2}_{(++)} \big)^2  \big( f^{i_3 i_4}_{(++)} \big)^2
+ 2 \big( f^{i_1 i_2}_{(++)} \big)^2 \big( f^{i_3 i_4}_{(--)} \big)^2
+  \big( f^{i_1 i_2}_{(--)} \big)^2  \big( f^{i_3 i_4}_{(--)} \big)^2  \right) \\
&& \qquad \; + 256 \left( f_{(++)}^{i_1 i_2} f_{(++)}^{i_2 i_3} f_{(+-)}^{i_3 i_4} f_{(+-)}^{i_1 i_4}
+ 2 f_{(++)}^{i_1 i_2} f_{(--)}^{i_3 i_4} f_{(+-)}^{i_1 i_3} f_{(+-)}^{i_2 i_4}
+ f_{(--)}^{i_1 i_2} f_{(--)}^{i_2 i_3} f_{(+-)}^{i_4 i_3} f_{(+-)}^{i_4 i_1} \right), \\
&& V: R_{a b c d} R_{c d e f} R_{e f g h} R_{g h a b} \\
&& \quad \; = 256 \left( f_{(++)}^{i_1 i_2} f_{(++)}^{i_1 i_3} f_{(++)}^{i_2 i_4} f_{(++)}^{i_3 i_4}
+ 4 f_{(++)}^{i_1 i_2} f_{(++)}^{i_2 i_4} f_{(+-)}^{i_1 i_3} f_{(+-)}^{i_4 i_3}
+ 4 f_{(++)}^{i_1 i_2} f_{(--)}^{i_3 i_4} f_{(+-)}^{i_1 i_3} f_{(+-)}^{i_2 i_4} \right. \\
&& \qquad \; \left. + 2 f_{(+-)}^{i_1 i_2} f_{(+-)}^{i_1 i_3} f_{(+-)}^{i_4 i_2} f_{(+-)}^{i_4 i_3}
+ 4 f_{(--)}^{i_1 i_2} f_{(--)}^{i_2 i_4} f_{(+-)}^{i_3 i_1} f_{(+-)}^{i_3 i_4}
+ f_{(--)}^{i_1 i_2} f_{(--)}^{i_1 i_3} f_{(--)}^{i_2 i_4} f_{(--)}^{i_3 i_4} \right), \\
&& W: R_{a b c d} R_{a b e f} R_{c e g h} R_{d f g h} \\
&& \quad \; = - 64 \left( \big( f^{i_1 i_2}_{(++)} \big)^2  \big( f^{i_3 i_4}_{(++)} \big)^2
- 2 \big( f^{i_1 i_2}_{(++)} \big)^2 \big( f^{i_3 i_4}_{(--)} \big)^2
+  \big( f^{i_1 i_2}_{(--)} \big)^2  \big( f^{i_3 i_4}_{(--)} \big)^2  \right) \\
&& \qquad + 128 \left( f_{(++)}^{i_1 i_2} f_{(++)}^{i_1 i_3} f_{(++)}^{i_2 i_4} f_{(++)}^{i_3 i_4}
+ 4 f_{(++)}^{i_1 i_2} f_{(++)}^{i_2 i_4} f_{(+-)}^{i_1 i_3} f_{(+-)}^{i_4 i_3}
+ 4 f_{(++)}^{i_1 i_2} f_{(--)}^{i_3 i_4} f_{(+-)}^{i_1 i_3} f_{(+-)}^{i_2 i_4} \right. \\
&& \qquad \quad \left. + 2 f_{(+-)}^{i_1 i_2} f_{(+-)}^{i_1 i_3} f_{(+-)}^{i_4 i_2} f_{(+-)}^{i_4 i_3}
+ 4 f_{(--)}^{i_1 i_2} f_{(--)}^{i_2 i_4} f_{(+-)}^{i_3 i_1} f_{(+-)}^{i_3 i_4}
+ f_{(--)}^{i_1 i_2} f_{(--)}^{i_1 i_3} f_{(--)}^{i_2 i_4} f_{(--)}^{i_3 i_4} \right), \\
\end{eqnarray*}

\begin{eqnarray*} \label{quartic:x-z}
&& X: R_{a b c d} R_{ef a b} R_{g c h e} R_{g d h f}
= R^2 \left( \left( f_{(++)}^{i_1 i_2} \right)^2
+ 2 \left( f_{(+-)}^{i_1 i_2}\right)^2 + \left(f_{(--)}^{i_1 i_2} \right)^2  \right) \\
&& \qquad \; - 16 R \left( f_{(++)}^{i_1 i_2} f_{(++)}^{i_1 i_3} f_{(++)}^{i_2 i_3}
+  f_{(++)}^{i_1 i_3} f_{(+-)}^{i_1 i_2} f_{(+-)}^{i_3 i_2}
+  f_{(--)}^{i_1 i_3} f_{(+-)}^{i_2 i_1} f_{(+-)}^{i_2 i_3}
+ f_{(--)}^{i_1 i_2} f_{(--)}^{i_1 i_3} f_{(--)}^{i_2 i_3} \right) \\
&& \qquad \; - 64 \left( \big( f^{i_1 i_2}_{(++)} \big)^2  \big( f^{i_3 i_4}_{(++)} \big)^2
+ \big( f^{i_1 i_2}_{(++)} \big)^2 \big( f^{i_3 i_4}_{(+-)} \big)^2
+ \big( f^{i_1 i_2}_{(--)} \big)^2 \big( f^{i_3 i_4}_{(+-)} \big)^2
+  \big( f^{i_1 i_2}_{(--)} \big)^2  \big( f^{i_3 i_4}_{(--)} \big)^2  \right) \\
&& \qquad \;
- 128 \left( f_{(+-)}^{i_1 i_2} f_{(+-)}^{i_2 i_1}
f_{(+-)}^{i_3 i_4} f_{(++)}^{i_3 i_4} - f_{(+-)}^{i_1 i_1} f_{(+-)}^{i_2 i_2} f_{(+-)}^{i_3 i_4} f_{(++)}^{i_3 i_4} \right. \\
&& \qquad \qquad \left. + 2 f_{(+-)}^{i_1 i_1} f_{(+-)}^{i_2 i_3} f_{(+-)}^{i_4 i_2} f_{(++)}^{i_3 i_4}
- 2 f_{(+-)}^{i_1 i_2} f_{(+-)}^{i_2 i_3} f_{(+-)}^{i_4 i_1} f_{(++)}^{i_3 i_4} \right) \\
&& \qquad \;
- 128 \left( f_{(+-)}^{i_1 i_2} f_{(+-)}^{i_2 i_1} f_{(+-)}^{i_3 i_4} f_{(--)}^{i_3 i_4} - f_{(+-)}^{i_1 i_1} f_{(+-)}^{i_2 i_2} f_{(+-)}^{i_3 i_4} f_{(--)}^{i_3 i_4} \right. \\
&& \qquad \qquad \left.+ 2 f_{(+-)}^{i_1 i_1} f_{(+-)}^{i_3 i_2} f_{(+-)}^{i_2 i_4} f_{(--)}^{i_3 i_4}
-2 f_{(+-)}^{i_2 i_1} f_{(+-)}^{i_3 i_2} f_{(+-)}^{i_1 i_4} f_{(--)}^{i_3 i_4} \right)  \\
&& \qquad \; + 128 \left( f_{(++)}^{i_1 i_2} f_{(++)}^{i_1 i_3} f_{(++)}^{i_2 i_4} f_{(++)}^{i_3 i_4}
+  f_{(++)}^{i_1 i_2} f_{(++)}^{i_2 i_4} f_{(+-)}^{i_1 i_3} f_{(+-)}^{i_4 i_3} \right. \\
&& \qquad \qquad \left. +  f_{(--)}^{i_1 i_2} f_{(--)}^{i_2 i_4} f_{(+-)}^{i_3 i_1} f_{(+-)}^{i_3 i_4}
+ f_{(--)}^{i_1 i_2} f_{(--)}^{i_1 i_3} f_{(--)}^{i_2 i_4} f_{(--)}^{i_3 i_4} \right),  \\
&& Y: R_{acbd}R_{cedf}R_{egfh}R_{gahb}
 = 64 \big( f^{i_3 i_4}_{(+-)} \big)^2 \left(
\big( f^{i_1 i_2}_{(++)} \big)^2 + 3 \big( f^{i_1 i_2}_{(+-)} \big)^2 + \big( f^{i_1 i_2}_{(--)} \big)^2 \right) \\
&& \qquad  - 32 \left( f^{i_1 i_2}_{(++)} f^{i_1 i_3}_{(++)} f^{i_2 i_4}_{(++)} f^{i_3 i_4}_{(++)}
- 4 f^{i_2 i_1}_{(+-)} f^{i_3 i_1}_{(+-)} f^{i_2 i_4}_{(++)} f^{i_3 i_4}_{(++)}
- 12 f^{i_1 i_2}_{(++)} f^{i_1 i_3}_{(+-)} f^{i_2 i_4}_{(+-)} f^{i_3 i_4}_{(--)} \right. \\
&& \qquad  \left. + 2 f^{i_1 i_2}_{(+-)} f^{i_1 i_3}_{(+-)} f^{i_4 i_2}_{(+-)} f^{i_4 i_3}_{(+-)}
- 4 f^{i_1 i_2}_{(+-)} f^{i_1 i_3}_{(+-)} f^{i_2 i_4}_{(--)} f^{i_3 i_4}_{(--)}
+ f^{i_1 i_2}_{(--)} f^{i_1 i_3}_{(--)} f^{i_2 i_4}_{(--)} f^{i_3 i_4}_{(--)} \right) \\
&& \qquad  + 48 \left( \big(f^{i_1 i_2}_{(++)} \big)^2 \big( f^{i_3 i_4}_{(++)} \big)^2
+ 2 \big( f^{i_1 i_2}_{(++)} \big)^2 \big(f^{i_3 i_4}_{(--)} \big)^2
+ \big(f^{i_1 i_2}_{(--)} \big)^2 \big( f^{i_3 i_4}_{(--)} \big)^2 \right), \\
&& Z: R_{acbd} R_{eafb} R_{gehc} R_{fgdh}
= \frac{7}{512} R^4 - \frac{1}{8} R^2 \left(
13 \big( f^{i_1 i_2}_{(++)} \big)^2 - 16 \big( f^{i_1 i_2}_{(+-)} \big)^2 + 13 \big( f^{i_1 i_2}_{(--)} \big)^2 \right) \\
&& \qquad + R \left( 12 f^{i_1 i_2}_{(++)} f^{i_1 i_3}_{(++)} f^{i_2 i_3}_{(++)}
- 16 f^{i_2 i_1}_{(+-)} f^{i_3 i_1}_{(+-)} f^{i_2 i_3}_{(++)}
+ 7 f^{i_1 i_1}_{(+-)} f^{i_2 i_2}_{(+-)} f^{i_3 i_3}_{(+-)}
- 21 f^{i_1 i_1}_{(+-)} f^{i_2 i_3}_{(+-)} f^{i_3 i_2}_{(+-)} \right. \\
&& \qquad  \left. + 14 f^{i_1 i_2}_{(+-)} f^{i_2 i_3}_{(+-)} f^{i_3 i_1}_{(+-)}
- 16 f^{i_1 i_2}_{(+-)} f^{i_1 i_3}_{(+-)} f^{i_2 i_3}_{(--)}
+ 12 f^{i_1 i_2}_{(--)} f^{i_1 i_3}_{(--)} f^{i_2 i_3}_{(--)} \right) \\
&& \qquad  - 44 \left( f^{i_1 i_2}_{(+-)} f^{i_2 i_1}_{(+-)} f^{i_3 i_4}_{(+-)} f^{i_3 i_4}_{(++)}
- f^{i_1 i_1}_{(+-)} f^{i_2 i_2}_{(+-)} f^{i_3 i_4}_{(+-)} f^{i_3 i_4}_{(++)} \right. \\
&& \qquad \left. + 2  f^{i_1 i_1}_{(+-)} f^{i_2 i_3}_{(+-)} f^{i_4 i_2}_{(+-)} f^{i_3 i_4}_{(++)}
- 2  f^{i_1 i_2}_{(+-)} f^{i_2 i_3}_{(+-)} f^{i_4 i_1}_{(+-)} f^{i_3 i_4}_{(++)} \right) \\
&& \qquad  - 44 \left( f^{i_1 i_2}_{(+-)} f^{i_2 i_1}_{(+-)} f^{i_3 i_4}_{(+-)} f^{i_3 i_4}_{(--)}
- f^{i_1 i_1}_{(+-)} f^{i_2 i_2}_{(+-)} f^{i_3 i_4}_{(+-)} f^{i_3 i_4}_{(--)} \right. \\
&& \qquad \left. + 2  f^{i_1 i_1}_{(+-)} f^{i_3 i_2}_{(+-)} f^{i_2 i_4}_{(+-)} f^{i_3 i_4}_{(--)}
- 2  f^{i_2 i_1}_{(+-)} f^{i_3 i_2}_{(+-)} f^{i_1 i_4}_{(+-)} f^{i_3 i_4}_{(--)} \right) \\
&& \qquad - 8 \left( 5 f^{i_1 i_2}_{(++)} f^{i_1 i_3}_{(++)} f^{i_2 i_4}_{(++)} f^{i_3 i_4}_{(++)}
- 16  f^{i_1 i_2}_{(+-)} f^{i_3 i_2}_{(+-)} f^{i_1 i_4}_{(++)} f^{i_3 i_4}_{(++)}
- 32  f^{i_1 i_2}_{(++)} f^{i_1 i_3}_{(+-)} f^{i_2 i_4}_{(+-)} f^{i_3 i_4}_{(--)} \right. \\
&& \qquad  \left. -16 f^{i_2 i_1}_{(+-)} f^{i_2 i_3}_{(+-)} f^{i_1 i_4}_{(--)} f^{i_3 i_4}_{(--)}
+ 5 f^{i_1 i_2}_{(--)} f^{i_1 i_3}_{(--)} f^{i_2 i_4}_{(--)} f^{i_3 i_4}_{(--)} \right)  \\
&& \qquad  + 4 \left( 5 \big( f^{i_1 i_2}_{(++)} \big)^2  \big( f^{i_3 i_4}_{(++)} \big)^2
+ 16 \big( f^{i_1 i_2}_{(++)} \big)^2 \big( f^{i_3 i_4}_{(--)} \big)^2
+ 16 \big( f^{i_1 i_2}_{(+-)} \big)^2 \big( f^{i_3 i_4}_{(+-)} \big)^2
+ 5 \big( f^{i_1 i_2}_{(--)} \big)^2  \big( f^{i_3 i_4}_{(--)} \big)^2  \right).
\end{eqnarray*}
Note that all terms are symmetric under the interchange \eq{exchange}
since scalar quantities are parity even.

For some terms in $G, K, L, M, Q, S, X$ and $Z$, we have the identity
\begin{eqnarray} \label{id-trdet}
  && \big(f^{i_3 i_4}_{(++)} - f^{i_3 i_4}_{(--)} \big) \left( f^{i_1 i_2}_{(+-)} f^{i_2 i_1}_{(+-)} f^{i_3 i_4}_{(+-)}
- f^{i_1 i_1}_{(+-)} f^{i_2 i_2}_{(+-)} f^{i_3 i_4}_{(+-)}
+ 2  f^{i_1 i_1}_{(+-)} f^{i_2 i_3}_{(+-)} f^{i_4 i_2}_{(+-)}
- 2  f^{i_1 i_2}_{(+-)} f^{i_2 i_3}_{(+-)} f^{i_4 i_1}_{(+-)} \right) \nonumber \\
&=& - 2 \Big(f^{i j}_{(++)} \delta^{ij} - f^{ij}_{(--)} \delta^{ij} \Big) \det f^{ij}_{(+-)} = 0.
\end{eqnarray}
Since \eq{id-trdet} is odd under the parity transformation \eq{exchange},
its vanishing can be easily understood from this property.
This means that such terms can be written as $- \frac{1}{4} R \, \det f^{ij}_{(+-)}$.
It is interesting to compare this expression for $\det f^{ij}_{(+-)}$ with the result in
Eq. \eq{six-cubic}.
Using the identity \eq{id-trdet}, it is straightforward to identity the
basis elements for the quartic monomials in Table \ref{table:2} which superficially look independent. The result is summarized in Table \ref{table:4}.
\begin{table} [h!]
    \centering
\begin{tabular}{|l|l|}
  \hline
  \hline
  I: $R^4$ &
  VIII: $\big(f_{(+-)}^{i_1 i_2} \big)^2 \big(f_{(+-)}^{i_3 i_4} \big)^2$ \\
  II: $R^2 \left( f^{i_1 i_2}_{(+-)} \right)^2$ &
  IX: $ \big( f_{(+-)}^{i_3 i_4} \big)^2 \left( \big(f_{(++)}^{i_1 i_2} \big)^2 + \big(f_{(--)}^{i_1 i_2} \big)^2 \right) $\\
  III: $R^2 \left( \big( f_{(++)}^{i_1 i_2} \big)^2 + \big( f_{(--)}^{i_1 i_2} \big)^2 \right) $ &
  X: $\big(f_{(++)}^{i_1 i_2} \big)^2 \big(f_{(++)}^{i_3 i_4} \big)^2
  + \big(f_{(--)}^{i_1 i_2} \big)^2 \big(f_{(--)}^{i_3 i_4} \big)^2 $ \\
  IV: $ R \left( f_{(++)}^{i_1 i_2} f_{(++)}^{i_1 i_3} f_{(++)}^{i_2 i_3}
+ f_{(--)}^{i_1 i_2} f_{(--)}^{i_1 i_3} f_{(--)}^{i_2 i_3} \right)$  &
  XI: $f_{(+-)}^{i_1 i_2} f_{(+-)}^{i_1 i_3} f_{(+-)}^{i_4 i_2} f_{(+-)}^{i_4 i_3}$  \\
  V: $ R \left( f_{(++)}^{i_1 i_2} f_{(+-)}^{i_1 i_3} f_{(+-)}^{i_2 i_3}
+ f_{(--)}^{i_1 i_2} f_{(+-)}^{i_3 i_1} f_{(+-)}^{i_3 i_2} \right)$ &
  XII: $f_{(++)}^{i_1 i_2} f_{(--)}^{i_3 i_4} f_{(+-)}^{i_1 i_3} f_{(+-)}^{i_2 i_4}$ \\
  VI: $R  \,  \det f^{ij}_{(+-)}  $ &
  XIII: $f_{(++)}^{i_1 i_2} f_{(++)}^{i_2 i_4} f_{(+-)}^{i_1 i_3} f_{(+-)}^{i_4 i_3}
  + f_{(--)}^{i_1 i_2} f_{(--)}^{i_2 i_4} f_{(+-)}^{i_3 i_1} f_{(+-)}^{i_3 i_4} $ \\
  VII: $ \big(f_{(++)}^{i_1 i_2} \big)^2 \big(f_{(--)}^{i_3 i_4} \big)^2$ &
  XIV: $ f_{(++)}^{i_1 i_2} f_{(++)}^{i_1 i_3} f_{(++)}^{i_2 i_4} f_{(++)}^{i_3 i_4}
  + f_{(--)}^{i_1 i_2} f_{(--)}^{i_1 i_3} f_{(--)}^{i_2 i_4} f_{(--)}^{i_3 i_4} $ \\
  \hline
  \hline
\end{tabular}
\caption{The 14 quartic basis elements}
    \label {table:4}
\end{table}
The first 6 basis elements, from $I$ to $VI$, are simply coming from the cubic ones \eq{six-cubic} multiplied by $R$.
The remaining 8 elements are newly generated in the quartic order.

We will show that the 26 quartic monomials in Table \ref{table:2} can be represented by
using only 13 linearly independent basis elements. This means that the 14 basis elements listed in Table \ref{table:4} are not completely independent. We will explain later how one linear relation among the quartic scalars arises.
This implies that there must be 13 linear relations among the quartic monomials.
A. Harvey found 6 such relations using the fact \cite{harvey} that an $(n+1)$ index object anti-symmetrized on an $n$-dimensional manifold vanishes identically.
However his results contain several errors and we will correct them here.
We find that there are 7 more linear relations in addition to the 6 relations found by Harvey.
We can find such linear relations by replacing the quartic basis elements in Table \ref{table:4}
in favor of the corresponding quartic monomials in Table \ref{table:2}.
We first list the 12 linear relations between the Riemann monomials in Table \ref{table:2}.
\begin{eqnarray}
&& (a): \; -R^4 + 9 R^2 R_{ab} R_{ab} - 14 R R_{ab} R_{bc} R_{ca}
 - 6 \big( R_{ab} R_{ab} \big)^2
 + 12 R_{ab} R_{bc} R_{cd} R_{da} \xx
&& \qquad  - 6 R R_{ab} R_{cd} R_{acbd}
+ 12 R_{ab} R_{ce} R_{ed} R_{acbd} = 0, \xx
&& (b): \; - \frac{R^4}{4} + 2 R^2 R_{a b} R_{a b} -2 R R_{a b} R_{b c} R_{c a}
- 2 R R_{a b} R_{c d} R_{a c b d} - \frac{1}{4} R^2 R_{a b c d} R_{a b c d}
+ R R_{a b} R_{a c d e} R_{b c d e} = 0, \xx
&& (c): \; - 2 R^4 + 15 R^2 R_{a b} R_{a b} - 16 R R_{a b} R_{b c} R_{c a} -12 R R_{a b} R_{c d} R_{a c b d}
-3 R_{a b} R_{a b} R_{c d e f} R_{c d e f} \xx
&& \qquad + 12 R_{a b} R_{b c} R_{a d e f} R_{c d e f} = 0, \nonumber
\end{eqnarray}

\begin{eqnarray}
&& (d): \; -\frac{11}{2}  R^4  + \frac{87}{2} R^2 R_{a b} R_{a b}
- 56 R R_{a b} R_{b c} R_{c a}
-18 \big(R_{ab} R_{ab} \big)^2
+ 36 R_{a b} R_{b c} R_{c d}  R_{d a} \xx
&& \qquad
- 18 R R_{a b} R_{c d} R_{a c b d}
- \frac{3}{2} R^2 R_{a b c d} R_{a b c d}
+ \frac{3}{2} R_{a b} R_{a b} R_{c d e f} R_{c d e f}
 + 12 R_{a b} R_{c d} R_{a e c f} R_{b e d f} \xx
&& \qquad - 12 R_{a b} R_{c d} R_{a e b f} R_{c e d f}
+ 6 R_{a b} R_{c d} R_{a c e f} R_{b d e f} = 0, \xx
&& (e): \; \frac{5}{4}  R^4 - 9 R^2 R_{a b} R_{a b}
+ 8 R R_{a b} R_{b c} R_{c a}
+ 6 R R_{a b} R_{c d} R_{a c b d}
+ \frac{3}{4} R^2 R_{a b c d} R_{a b c d} \xx
&& \qquad
- R R_{a b c d} R_{c d e f}  R_{e f a b}
+ 2 R R_{a b c d} R_{a e c f} R_{b e d f} = 0, \xx
&& (f): \; -2 R^4 + 15 R^2 R_{a b} R_{a b} -28 R R_{a b} R_{b c} R_{c a}
+24 R_{a b} R_{b c} R_{c d} R_{d a}
-3 R_{a b} R_{a b} R_{c d e f} R_{c d e f}  \xx
&& \qquad
- 24 R_{a b} R_{c d} R_{a e b f} R_{c e d f}
+ 12 R_{a b} R_{a c b d} R_{c e f g} R_{d e f g} = 0, \xx
&& (g): \; -\frac{R^4}{2} + 3 R^2 R_{a b} R_{a b}  - 8 R R_{a b} R_{b c} R_{c a}
 + 4 R R_{a b} R_{c d} R_{a c b d} + 8 R_{a b} R_{b c} R_{c d} R_{d a}
 + \frac{1}{2} R^2 R_{a b c d} R_{a b c d}\xx
&& \qquad
-8 R_{a b} R_{c d} R_{a e b f} R_{c e d f}
- R_{a b} R_{a b} R_{c d e f} R_{c d e f}
- 4 R_{a b} R_{c d} R_{a c e f} R_{b d e f}
- R R_{a b c d} R_{c d e f} R_{e f a b} \xx
&& \qquad
 + 4 R_{a b} R_{c d e f} R_{c d g a} R_{e f g b}  = 0, \xx
&& (h): \; 4 R^4 - 30 R^2 R_{a b} R_{a b}
+32 R R_{a b} R_{b c} R_{c a} + 6 \left(R_{a b} R_{a b}\right)^2
-12 R_{a b} R_{b c} R_{c d} R_{d a} \xx
&& \qquad + 18 R R_{a b} R_{c d} R_{a c b d}
+ \frac{3}{2} R^2 R_{a b c d} R_{a b c d}
- 6 R_{a b} R_{c d} R_{a c e f} R_{b d e f}  \xx
&& \qquad
-\frac{3}{2} R R_{a b c d} R_{c d e f} R_{e f a b}
+ 12 R_{a b} R_{c e d f} R_{a d c g} R_{b f e g}  = 0, \xx
&& (i): \;-5 R^4 + 36 R^2 R_{a b} R_{a b}
- 64 R R_{a b} R_{b c} R_{c a}
+ 48 R_{a b} R_{b c} R_{c d} R_{d a}
- 48 R_{a b} R_{c d} R_{a e b f} R_{c e d f} \xx
&& \qquad -3 \left(R_{a b c d} R_{a b c d}\right)^2
+ 12 R_{a b c d} R_{a b c e} R_{d h f g } R_{e h f g} = 0, \xx
&& (j): \; R^4 - 6 R^2 R_{a b} R_{a b} + 16 R R_{a b} R_{b c} R_{c a}
- 16 R_{a b} R_{b c} R_{c d} R_{d a}
- 8 R R_{a b} R_{c d} R_{a c b d}
- \frac{1}{2} R^2 R_{a b c d} R_{a b c d}  \xx
&& \qquad
+4 R_{a b} R_{c d} R_{a c e f} R_{b d e f}
+16 R_{a b} R_{c d} R_{a e b f} R_{c e d f}
+ R R_{a b c d} R_{c d e f} R_{e f a b}
+ \frac{1}{2} \left(R_{a b c d} R_{a b c d}\right)^2 \xx
&& \qquad + 4 R_{a b c d} R_{a b e f} R_{c g e h} R_{d g f h}
- R_{a b c d} R_{c d e f} R_{e f g h} R_{g h a b}
- 2 R_{a b c d} R_{a b e f} R_{c e g h} R_{g h d f} = 0, \xx
&& (k): \; -\frac{23}{6}  R^4 + 28 R^2 R_{a b} R_{a b}
-\frac{128}{3} R R_{a b} R_{b c} R_{c a}
-8 \left(R_{a b} R_{a b}\right)^2
+ 32 R_{a b} R_{b c} R_{c d} R_{d a} \xx
&& \qquad
- R^2 R_{a b c d} R_{a b c d}
+ 4 R_{a b} R_{a b} R_{c d e f} R_{c d e f}
- 32 R_{a b} R_{c d} R_{a e b f} R_{c e d f}
-\frac{3}{2} \left(R_{a b c d} R_{a b c d}\right){}^2 \xx
&& \qquad
+ 8 R_{a c b d} R_{c e d f} R_{e g f h} R_{a g b h}
+ R_{a b c d} R_{c d e f} R_{e f g h} R_{g h a b} = 0, \xx
&& (l): \; \frac{17}{12} R^4 - 12 R^2 R_{a b} R_{a b}
- \frac{128}{3} R R_{a b} R_{b c} R_{c a}
- 6 \left(R_{a b} R_{a b}\right)^2
+ 84 R_{a b} R_{b c} R_{c d} R_{d a}  \xx
&& \qquad
+ 72 R R_{a b} R_{c d} R_{a c b d}
+ \frac{5}{2} R^2 R_{a b c d} R_{a b c d}
-20 R_{a b} R_{c d} R_{a c e f} R_{b d e f}
+ 8 R_{a b} R_{a b} R_{c d e f} R_{c d e f} \xx
&& \qquad
- 104 R_{a b} R_{c d} R_{a e b f} R_{c e d f}
- 6 R R_{a b c d} R_{c d e f}  R_{e f a b}
-\frac{13}{4} \left(R_{a b c d} R_{a b c d}\right)^2  \xx
&& \qquad
+ \frac{13}{2} R_{a b c d} R_{c d e f} R_{e f g h}  R_{g h a b}
- 3 R_{a b c d} R_{a b e f} R_{c e g h} R_{g h d f}
+ 32 R_{a c b d} R_{a e b f} R_{c h e g } R_{d h f g}  = 0. \nonumber
\end{eqnarray}

Our results $(a), \, (b)$ and $(c)$ precisely reproduce Eqs. (11), (9), and (10) in Ref. \cite{harvey}, respectively.\footnote{Eq. (11) in Ref. \cite{harvey} contains a sign error: $\cdots - 9 R^2 R_{ab} R_{ab} + R^4 \to \cdots + 9 R^2 R_{ab} R_{ab} - R^4$.}
Unfortunately, Eqs. (15), (17), (18), and (19) in Ref. \cite{harvey} contain errors too.\footnote{We used Mathematica (www.wolfram.com) and the add-on package MathSymbolica (www.mathsymbolica.com) for calculation
of the quartic Riemann monomials, $A \sim Z$, and the linear relations, $(a) \sim (l)$.
We verified the results by numerical tests using randomly generated numbers
for the coefficients $f^{ij}_{(\pm\pm)}$ and $f^{ij}_{(\pm\mp)}$,
and hence, we believe that our results are error-free.}
For example, Eq. (15) in Ref. \cite{harvey} should read as
\begin{eqnarray} \label{harvey-15}
&& -\frac{5}{4}  R^4  + \frac{39}{4} R^2 R_{a b} R_{a b}
- 12 R R_{a b} R_{b c} R_{c a}
- 3 \big(R_{ab} R_{ab} \big)^2
+ 6 R_{a b} R_{b c} R_{c d}  R_{d a} \xx
&& \quad
- 5 R R_{a b} R_{c d} R_{a c b d}
- \frac{1}{4} R^2 R_{a b c d} R_{a b c d}
+ \frac{1}{4} R_{a b} R_{a b} R_{c d e f} R_{c d e f}
+ 2 R_{a b} R_{b c} R_{a d e f} R_{c d e f} \\
&& \quad + 2 R_{a b} R_{c d} R_{a e c f} R_{b e d f}
- 2 R_{a b} R_{c d} R_{a e b f} R_{c e d f}
+ R_{a b} R_{c d} R_{a c e f} R_{b d e f} = 0. \nonumber
\end{eqnarray}
In particular, Eq. (15) in Ref. \cite{harvey} was missing the term,
$- \frac{1}{4} R^2 R_{a b c d} R_{a b c d}$.
The correct equation \eq{harvey-15} can simply be obtained by adding $(c)$ and $(d)$
and dividing by 6.
Since Eq. (17) in Ref. \cite{harvey} used the incorrect equation (15), it is  incorrect too. The correct forms of Eqs. (17) and (19) are given by, respectively,\footnote{Eq. (19) in Ref. \cite{harvey} was derived using Eq. (18), but Eq. (18) contains typographic errors.
The rectified version must read as $- 2 R^2 R_{abcd}^2 \to + 2 R^2 R_{abcd}^2$ and $-\frac{64}{3} R R_{ab} R_{bc} R_{ca} \to + \frac{64}{3} R R_{ab} R_{bc} R_{ca}$.}
\begin{eqnarray} \label{harvey-17}
&& \textrm{(17)} : \;  R^4 - 6 R^2 R_{a b} R_{a b}
+ 5 R R_{a b} R_{b c} R_{c a}
+ 6 R_{a b} R_{c e} R_{e d}  R_{a c b d}
+ \frac{3}{2} R R_{a b} R_{c d e a} R_{c d e b} \xx
&& \qquad
+ 3 R_{a b} R_{b c a d} R_{e f g c} R_{e f g d}
- 3 R_{a b} R_{c d e f} R_{e f g a} R_{g b c d}
+ 6 R_{a b} R_{c e d f} R_{e g f a} R_{g c b d} = 0, \\
\label{harvey-19}
&& \textrm{(19)} : \; - \frac{5}{48} R^4 + \frac{1}{2} R^2 R^2_{a b}
-\frac{1}{3} R R_{a b} R_{b c} R_{c a} + \frac{1}{16} R_{abcd}^4
- R_{a b} R_{c d e f} R_{e f g a}  R_{g b c d} \xx
&& \qquad \;
+ 2 R_{ab} R_{c e d f} R_{e g f a} R_{g c b d}
+ R_{a b} R_{b c a d} R_{e f g c} R_{g d e f}
+ R_{a c b d} R_{c e d f} R_{e g f h} R_{a g b h}  \xx
&& \qquad  \;
- 2 R_{a b c d} R_{a b e f} R_{c g e h} R_{d g f h}
- 2 R_{a c b d} R_{a e b f} R_{c h e g} R_{d h f g}
+ \frac{1}{8} R_{a b c d} R_{c d e f} R_{e f g h} R_{g h a b} \xx
&& \qquad \;
+ R_{a b c d} R_{a b e f} R_{c e g h} R_{g h d f}
- R_{a b c d} R_{a b c e} R_{d h f g} R_{f g e h}  = 0.
\end{eqnarray}
Eq. \eq{harvey-17} can be reproduced from our results by considering the combination,
$\frac{1}{4} (f) - \frac{3}{4} (g) + \frac{1}{2} (h) + \frac{1}{2} (a) + \frac{3}{2}(b)$.
Using 12 linear equations $(a) \sim (l)$, Eq. \eq{harvey-19} can be further reduced as
\begin{eqnarray} \label{reduced-19}
&& - \frac{1}{4} R^4 + R^2 R_{a b} R_{a b} + R_{ab}^4
- 2 R_{a b} R_{b c} R_{c d} R_{d a}
- 2 R_{a b} R_{c d} R_{e f a c}  R_{e f b d}
+ 2 R_{a c b d} R_{c e d f} R_{e g f h} R_{g a h b}  \xx
&& \qquad
- 4 R_{a c b d} R_{e a f b} R_{f g d h} R_{g e h c}
+ \frac{3}{2} R_{a b c d} R_{a b e f} R_{c e g h} R_{g h d f}
- R_{a b c d} R_{a b c e} R_{f g h d} R_{h e f g} = 0.
\end{eqnarray}
Using the notation \eq{def-metab}, Eqs. \eq{harvey-19} and \eq{reduced-19} are equally written as
\bea \la{19-exp}
&& \frac{R^4}{16} - 12 R^2 \textrm{Tr} \left(  A_+^2 +  A_-^2 \right) +
128 R  \textrm{Tr} \left( A_+^3 + A_-^3 \right) - 768  \textrm{Tr} \left( A_+^4  + A_-^4 \right) \xx
&& + 384 \left( \textrm{Tr} \big(  A_+^2 \big) \textrm{Tr} \big(  A_+^2 \big) +
\textrm{Tr} \big(  A_-^2 \big) \textrm{Tr} \big(  A_-^2 \big) \right) = 0.
\eea
The reason why the matrix $B$ does not appear in \eq{19-exp} is that Eqs. \eq{harvey-19} and \eq{reduced-19} originated from a quartic polynomial of Weyl tensors. The validity of the linear relation \eq{19-exp} may be understood
by the fact that the four elements, $III, \, IV, \, X$ and $XIV$, in Table \ref{table:4} can be shown to be implicitly
connected by this relationship.
We have confirmed that Eq. \eq{harvey-19} cannot be derived as a linear combination of the 12
equations, $(a) \sim (l)$, by checking all possible cases out of 26 terms.
This means that Eq. \eq{harvey-19} must be regarded as a new linear relation in addition
to the 12 relations, $(a) \sim (l)$. It is quite revealing to see that Eq. \eq{19-exp} is an identity
for general symmetric $3 \times 3$ matrices $A_\pm$ satisfying the property $\textrm{Tr} A_+
= \textrm{Tr} A_- = \frac{R}{8}$. It can be most easily checked in a diagonalized frame such that
$A_\pm = \textrm{diag} (a^1_\pm, a^2_\pm, a^3_\pm)$ and $a^1_+ + a^2_+ + a^3_+
= a^1_- + a^2_- + a^3_-$. Since Eq. (D.6) provides one more linear relation among the quartic
basis elements in Table \ref{table:4}, the number of linearly independent basis elements
for quartic scalars is 13.
\\

{\bf Note added.} The above linear relations,  \eq{harvey-19}, \eq{reduced-19}, and \eq{19-exp}, have been derived
from the identity ${R^{ab}}_{[ab} {R^{cd}}_{cd} {R^{ef}}_{ef} {R^{gh}}_{gh]} = 0$
whose expansion is very complicated \cite{harvey}.
But we found that Eq. \eq{19-exp} can be rewritten as $P- 6S + 3X = R R_{abcd} R_{cedf} R_{eafb}
- 6 R_{ab} R_{cedf} R_{egfa} R_{gcbd} + 3 R_{abcd} R_{efab} R_{gche} R_{gdhf} =0$ using only the three members
in Table \ref{table:2}.

\end{document}